\documentclass[a4paper,11pt]{article}
\pdfoutput=1
\usepackage{jheppub}
\usepackage[T1]{fontenc}
\usepackage{amsmath,amssymb,graphicx}
\usepackage{orcidlink}
\usepackage{subcaption}
\usepackage{braket}
\usepackage{placeins}
\usepackage{afterpage}
\usetikzlibrary{decorations.pathmorphing} 
\usetikzlibrary{arrows.meta}     
\usepackage{slashed}
\usepackage{tabularray}
\usetikzlibrary{patterns,snakes}
\newcommand{\midarrow}{\tikz \draw[-latex] (0,0) -- +(.1,0);}
\tikzstyle{spring}=[line width=0.8,blue!7!black!80,snake=coil,segment amplitude=5,segment length=5,line cap=round]
\usepackage{cleveref}
\Crefname{equation}{Eq.}{Eq.}
\Crefname{figure}{Fig.}{Fig.}
\Crefname{tabular}{Tab.}{Tab.}
\Crefname{table}{Tab.}{Tab.}
\Crefname{section}{Sec.}{Sec.}

\title{Nucleon tomography from a calibrated diquark spectator model}
\author[a]{Jakub Terlecki}
\author[b]{Shohini Bhattacharya \orcidlink{0000-0001-8536-082X}}
\author[a]{Krzysztof Cichy \orcidlink{0000-0002-5705-3256}}

\affiliation[a]{Faculty of Physics and Astronomy, Adam Mickiewicz University, ul. Uniwersytetu Pozna\'nskiego 2, 61-614 Pozna\'n, Poland}
\affiliation[b]{Department of Physics, University of Connecticut, Storrs, Connecticut 06269, USA}

\emailAdd{jakter1@st.amu.edu.pl, shohinib@uconn.edu, kcichy@amu.edu.pl}

\abstract{Generalized parton distributions (GPDs) provide a unified description of the longitudinal-momentum and transverse spatial structure of the nucleon. We calculate the twist-2 vector GPDs \(H^{q}\) and \(E^{q}\) in a spectator model containing both scalar and axial-vector diquark configurations. A central new result is the explicit analytic expression for the axial-vector-diquark contribution to the helicity-flip GPD \(E^{q}\). To regulate the transverse-momentum integrations while retaining a physically sensible dependence on the momentum transfer, we introduce a momentum-transfer-dependent quasi-hard cutoff.
The model parameters are determined through two complementary calibrations: one using proton observables only and another simultaneously including proton and neutron observables. Electromagnetic moments and radii, together with selected constraints on gravitational form factors, determine the quark and diquark masses, the relative scalar and axial-vector contributions, and the transverse-momentum scale. The two calibrations yield compatible parameter ranges, while the inclusion of neutron observables reduces the parameter uncertainties and more strongly constrains the flavor structure.
From the resulting zero-skewness GPDs \(H^{u,d}(x,0,t)\) and \(E^{u,d}(x,0,t)\), we calculate the valence-quark PDFs and the impact-parameter distributions of unpolarized up and down quarks in both unpolarized and transversely polarized nucleons. The transverse distributions exhibit the expected localization of large-\(x\) quarks near the transverse center of momentum and a flavor-dependent distortion when the nucleon is transversely polarized. Our results provide a unified phenomenological description of the longitudinal-momentum and transverse-position structure of valence quarks in the nucleon.}

\begin{document}
\maketitle

\section{Introduction}
Parton distribution functions (PDFs) are key quantities for describing the
quark and gluon structure of hadrons. They encode the probability of finding
a parton carrying a longitudinal momentum fraction \(x\) of a fast-moving
hadron and enter the description of a wide range of high-energy scattering
processes through QCD factorization theorems. However, PDFs provide only a
one-dimensional picture of hadron structure and contain no direct information
about the spatial distribution of partons inside the hadron.

Generalized parton distributions (GPDs) \cite{Mueller:1998fv, Ji:1996ek,Ji:1996nm,Radyushkin:1996nd} extend ordinary PDFs by allowing a
nonzero momentum transfer between the initial and final hadron states
\cite{Diehl:2003ny,Guo:2026swz}. At leading twist, the vector GPDs \(H^{q}(x,\xi,t)\) and
\(E^{q}(x,\xi,t)\) depend on the average longitudinal momentum fraction \(x\),
the skewness \(\xi\), and the invariant momentum transfer \(t\). In the forward
limit, the GPD \(H^{q}\) reduces to the usual unpolarized PDF,
\begin{equation}
    H^{q}(x,0,0)=q(x),
\end{equation}
whereas \(E^{q}(x,0,0)\) has no ordinary forward-PDF counterpart. The Mellin
moments of \(H^{q}\) and \(E^{q}\) are related to electromagnetic and
gravitational form factors. GPDs therefore provide a common framework for
connecting parton distributions, form factors, and the spatial structure of
the nucleon.

A particularly transparent interpretation emerges at zero skewness. In this
case, a two-dimensional Fourier transform of \(H^{q}(x,0,t)\) with respect to
the transverse momentum transfer gives the impact-parameter distribution
\cite{Burkardt:2000za,Ralston:2001xs,Diehl:2002he,Burkardt:2002hr},
\begin{equation}
    q\bigl(x,\vec{b}_{\perp}\bigr)
    =
    \int \frac{d^{2}\vec{\Delta}_{\perp}}{(2\pi)^{2}}\,
    e^{-i\vec{b}_{\perp}\cdot\vec{\Delta}_{\perp}}
    H^{q}\bigl(x,0,-\vec{\Delta}_{\perp}^{2}\bigr).
\end{equation}
The variable \({b}_{\perp}\) specifies the transverse position of
the active quark relative to the transverse center of momentum of the
nucleon. Consequently,
\(q(x,\vec{b}_{\perp})\) correlates the longitudinal momentum carried
by a quark with its transverse spatial position and provides a tomographic
picture of the nucleon.
For a transversely polarized nucleon, the distribution is no longer
azimuthally symmetric. If the nucleon is polarized along the \(x\) direction,
the density of unpolarized quarks takes the form
\cite{Burkardt:2002hr}
\begin{equation}
    q_{X}\bigl(x,\vec{b}_{\perp}\bigr)
    =
    q\bigl(x,\vec{b}_{\perp}\bigr)
    -
    \frac{1}{2M}
    \frac{\partial}{\partial b_{y}}\,
    e^{q}\bigl(x,\vec{b}_{\perp}\bigr),
\end{equation}
where
\begin{equation}
    e^{q}\bigl(x,\vec{b}_{\perp}\bigr)
    =
    \int \frac{d^{2}\vec{\Delta}_{\perp}}{(2\pi)^{2}}\,
    e^{-i\vec{b}_{\perp}\cdot\vec{\Delta}_{\perp}}
    E^{q}\bigl(x,0,-\vec{\Delta}_{\perp}^{2}\bigr).
\end{equation}
The GPD \(E^{q}\) therefore controls the sideways distortion of the
unpolarized-quark distribution induced by transverse nucleon polarization \cite{Burkardt_2004, Brodsky_2002, Pasquini_2019}.
A simultaneous description of \(H^{q}\) and \(E^{q}\) is thus required to
study both the azimuthally symmetric transverse density and its
spin-dependent deformation.

Although GPDs can in principle be accessed through hard exclusive processes,
including deeply virtual Compton scattering \cite{Mueller:1998fv,Ji:1996nm,Radyushkin:1996nd,Collins:1998be} and deeply virtual meson
production \cite{Radyushkin:1996ru,Collins:1996fb,Mankiewicz:1997uy}, determining their full dependence on \(x\), \(\xi\), and \(t\)
from experimental data remains challenging. Phenomenological determinations of GPDs have been pursued using complementary experimental constraints. At zero skewness, analyses combining forward parton distributions with nucleon electromagnetic form factors have constrained the valence-quark GPDs \(H^q\) and \(E^q\), allowing the reconstruction of their transverse spatial distributions~\cite{Diehl:2004cx,Guidal:2004nd,Diehl:2013xca}. Fits to deeply virtual Compton scattering observables have provided complementary constraints on GPD parametrizations and Compton form factors~\cite{Kumericki:2009uq,Goldstein:2010gu,Moutarde:2018kwr}. These studies have established a phenomenological basis for nucleon tomography, while also demonstrating the importance of quantifying experimental uncertainties and the dependence on the assumed parametrizations.

On the other hand, Lattice QCD provides
complementary information through Mellin moments extracted from local matrix elements, see, e.g., Refs.~\cite{Shintani:2018ozy,Bali:2018zgl,Alexandrou:2018sjm,Alexandrou:2019olr,Alexandrou:2019ali,Alexandrou:2020okk,Alexandrou:2021wzv,Alexandrou:2023qbg,PhysRevLett.132.251904,Alexandrou:2025vto,Alexandrou:2026tjs} and, more recently, through
methods designed to access the $x$ dependence of partonic distributions and/or higher moments, such as quasi- and pseudo-distributions \cite{Ji:2013dva,Ji:2014gla,Radyushkin:2017cyf}, see, e.g., Refs.~\cite{Chen:2019lcm,Alexandrou:2020zbe,Alexandrou:2021bbo,Bhattacharya:2023nmv,Lin:2020rxa,Lin:2021brq,Lin:2023gxz,Holligan:2023jqh,Bhattacharya:2022aob,Bhattacharya:2023ays,Bhattacharya:2023jsc,Bhattacharya:2024qpp,HadStruc:2024rix,Ding:2024saz,Bhattacharya:2024wtg,Bhattacharya:2025yba,Chu:2025kew,Chu:2025jsi,Gao:2025inf,Dutrieux:2026grg}.
The quasi-distribution approach has also been extended to GPDs and examined
in diquark spectator models (DSM)
\cite{Bhattacharya:2018zxi,Bhattacharya:2020cen,Ma:2019agv}.
Recent work combining
lattice-QCD information with elastic-scattering data has demonstrated the
potential of such complementary constraints for extracting valence-quark
nucleon tomography \cite{Cichy:2024afd}. 
Also, first global analyses of lattice and experimental data have been attempted in the framework of GPDs through universal moment parametrization (GUMP) \cite{Guo:2022upw,Guo:2023ahv,Guo:2024wxy,Guo:2025muf}.
Nevertheless, obtaining continuous GPDs from the available information still requires parametrizations or model assumptions and a laborious quantification of several sources of systematic effects.

Spectator models provide a useful framework for investigating the relationships among PDFs, GPDs, form factors, and transverse spatial distributions. In these models, the active quark participates in the interaction, while the remaining degrees of freedom are represented by an effective spectator. Early studies established the use of scalar and axial-vector diquark spectators for describing quark distribution and fragmentation functions~\cite{Jakob_1997}, and subsequent calculations extended this framework to transverse-momentum-dependent distributions~\cite{Bacchetta:2008af}. Spectator-model calculations have also explored GPDs and their relations to transverse-momentum-dependent distributions~\cite{Mei_ner_2007}, while light-front overlap constructions have addressed the simultaneous implementation of positivity and polynomiality~\cite{Hwang:2007tb}. Furthermore, spectator-based GPD parametrizations have been constrained using hadronic observables and deeply virtual Compton scattering data~\cite{Ahmad:2006gn,Goldstein:2010gu}. More recently, diquark spectator models have been used to investigate quasi-GPDs and their approach to the light-cone limit~\cite{Bhattacharya:2018zxi,Bhattacharya:2020cen,tan2024chiraleventwist3gpdsproton, Ma:2019agv, Bacchetta_2017} and to generalized form factors and angular momentum densities \cite{Liu_2014, Ma:2014jla,Kumar_2017}. Their relatively simple structure makes it possible to derive analytic results while retaining nontrivial correlations among quark momentum, momentum transfer, and nucleon spin.

A scalar-diquark configuration alone, however, is insufficient for a
flavor-dependent description of both up and down quarks in the nucleon \cite{Mezrag_2018, Kumar_2017, Bednar_2018}. In
the present work, we therefore include both scalar and axial-vector diquark
configurations. The up-quark GPDs receive contributions from both sectors,
whereas the down-quark GPDs are generated by the axial-vector sector. This
structure makes the axial-vector contribution especially important for the
flavor decomposition and for the helicity-flip GPD \(E^{q}\). A central new
result of this work is an explicit analytic expression for the
axial-vector-diquark contribution to \(E^{q}\). As an independent check, we
also derive the corresponding quasi-GPD and verify that it reduces to the
standard light-cone GPD in the infinite-momentum limit.

A practical issue in spectator-model calculations is the regulation of the
transverse-momentum integration. A fixed hard cutoff is straightforward to
implement, but we find that it can produce an unphysical growth of the
axial-vector GPDs at large
\(\lvert\vec{\Delta}_{\perp}\rvert\). Smooth vertex form factors avoid
a sharp integration boundary, but they can also modify the distributions
throughout the integration region rather than acting predominantly as
ultraviolet regulators. We therefore introduce a momentum-transfer-dependent
quasi-hard cutoff,
\begin{equation}
\label{eq:quasihardcutoff}
    k_{\perp}^{2}
    \leq
    \beta^{2}
    -
    \frac{\vec{\Delta}_{\perp}^{2}}{4}.
\end{equation}
This prescription treats the transverse momenta entering the initial and
final nucleon--quark--diquark vertices symmetrically. It reduces to an
ordinary hard cutoff in the forward limit and progressively restricts the
available transverse-momentum region as the momentum transfer increases.

The predictive content of a spectator model depends on how its parameters
are determined. In this work, the quark mass, the scalar- and
axial-vector-diquark masses, the relative scalar--axial contribution, and the
transverse-momentum scale are constrained by known nucleon observables. We
consider two complementary calibrations. The first uses proton
electromagnetic moments and radii together with selected constraints on
gravitational form factors. Neutron observables then constitute predictions
of the proton-calibrated model. The second calibration simultaneously
includes proton and neutron observables and therefore provides a more
restrictive test of the flavor decomposition. Comparing the two
calibrations allows us to examine the stability of the extracted model
parameters and the sensitivity of the resulting tomography to the data used
as input.

Using the calibrated parameter sets, we calculate the flavor-separated zero-skewness GPDs \(H^{\text{up},\text{down}}(x,0,t)\) and \(E^{\text{up},\text{down}}(x,0,t)\), their forward limits, and the corresponding impact-parameter distributions. These distributions are not directly included in the calibration; they follow from the model constrained by nucleon observables and allow us to examine how those constraints translate into valence-quark momentum and spatial structure.
We also examine their dependence on the active-quark and spectator masses.
In particular, we show that finite constituent masses lead to finite
valence-quark distributions as \(x\to0\), although a light active quark and
a heavier spectator can generate a pronounced enhancement at small \(x\).
At the opposite endpoint, we verify analytically that
\begin{equation}
    \lim_{x\to1}
    \frac{\partial H^{q}(x,0,t)}{\partial t}
    =
    \lim_{x\to1}
    \frac{\partial E^{q}(x,0,t)}{\partial t}
    =
    0.
\end{equation}
This property ensures that a quark carrying the full longitudinal momentum
of the nucleon becomes localized at the transverse center of momentum.

The paper is organized as follows. In \Cref{Sec:Theoretical formalism}, we introduce the GPD
kinematics, derive the scalar- and axial-vector-diquark contributions to
\(H^{q}\) and \(E^{q}\), and review the relations between their Mellin
moments and electromagnetic and gravitational form factors. In \Cref{Sec:numerical results}, we
discuss the transverse-momentum regulator, study the analytic and numerical
properties of the distributions, and determine the model parameters through
the proton-only and simultaneous proton--neutron calibrations. We then
present the resulting form factors, valence-quark PDFs, zero-skewness GPDs,
and impact-parameter distributions for unpolarized quarks in unpolarized and
transversely polarized nucleons. We summarize our findings and discuss
possible extensions in \Cref{Sec:summary and prospects}. Additional analytic expressions are collected
in the Appendix.

\section{Theoretical formalism}
\label{Sec:Theoretical formalism}
\subsection{General definitions}

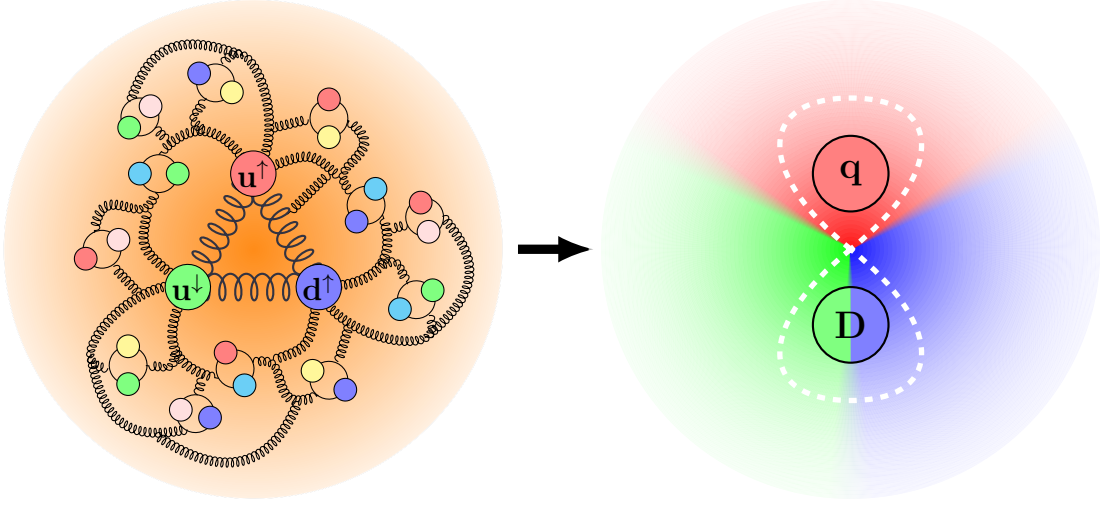
\begin{figure}
    \centering
\begin{tikzpicture}[
    decoration={coil,amplitude=0.6mm,segment length=1.95}
                    ]
\begin{scope}[rotate around={35:(0,0)}]

\filldraw[draw=white,even odd rule,inner color=orange!90!white,outer color=white] (0,0) circle (3.3);

\end{scope}

\draw[decorate] (0,1) .. controls (-0.42,1.7) and (-0.84,1.7) .. (-1.25,1.25);
\draw[thin] (-1.25,1) circle (0.25);
\filldraw[thin, fill=green!50] (-1,1) circle (0.15);
\filldraw[thin, fill=cyan!50] (-1.5,1) circle (0.15);

\draw[decorate] (-0.4,2.45) ..controls (0,3) and (0.5,1.5) .. (0,1);

\draw[decorate] (-0.636,1.55) .. controls (-0.82,1.7) .. (-0.621,2);
\draw[thin] (-0.75+0.25,2.2) circle (0.25);
\begin{scope}[rotate around={-30:(-0.5,2.2)}]
\filldraw[thin, fill=blue!50] (-0.75,2.2) circle (0.15);
\filldraw[thin, fill=yellow!50] (-0.75+0.5,2.2) circle (0.15);
\end{scope}

\draw[decorate] (-1.625,1.95) .. controls (-1.4,2.6) and (-0,4,2.8) .. (-0.12,2.55);
\draw[decorate] (-1.3,1.6) -- (-1.1,1.4);

\draw[decorate] (-1.3,0.75) .. controls (-1.45,0.2) and (-2,1) .. (-2.1,0.22);

\draw[decorate] (-1.4,0.5) .. controls (-1.5,0) .. (-0.866,-0.5);

\draw[decorate] (-1.85,-0.23) -- (-1.5,-0.6);

\draw[thin] (-2,0) circle (0.25);
\begin{scope}[rotate around={30:(-2,0)}]
\filldraw[thin, fill=red!50] (-2.25,0) circle (0.15);
\filldraw[thin, fill=pink!50] (-1.75,0) circle (0.15);
\end{scope}

\draw[thin] (-1.5,1.75) circle (0.25);
\begin{scope}[rotate around={45:(-1.5,1.75)}]
\filldraw[thin, fill=pink!50] (-1.3,1.75) circle (0.15);
\filldraw[thin, fill=green!50] (-1.7,1.75) circle (0.15);
\end{scope}

\begin{scope}[rotate around={-120:(0,0)}]

\draw[decorate] (0,1) .. controls (-0.42,1.7) and (-0.84,1.7) .. (-1.25,1.25);
\draw[thin] (-1.25,1) circle (0.25);
\filldraw[thin, fill=blue!50] (-1,1) circle (0.15);
\filldraw[thin, fill=cyan!50] (-1.5,1) circle (0.15);

\draw[decorate] (-0.4,2.45) ..controls (0,3) and (0.5,1.5) .. (0,1);

\draw[decorate] (-0.636,1.55) .. controls (-0.82,1.7) .. (-0.621,2);
\draw[thin] (-0.75+0.25,2.2) circle (0.25);
\begin{scope}[rotate around={-30:(-0.5,2.2)}]
\filldraw[thin, fill=green!50] (-0.75,2.2) circle (0.15);
\filldraw[thin, fill=cyan!50] (-0.75+0.5,2.2) circle (0.15);
\end{scope}

\draw[decorate] (-1.625,1.95) .. controls (-1.4,2.6) and (-0,4,2.8) .. (-0.12,2.55);
\draw[decorate] (-1.3,1.6) -- (-1.1,1.4);

\draw[decorate] (-1.3,0.75) .. controls (-1.45,0.2) and (-2,1) .. (-2.1,0.22);

\draw[decorate] (-1.4,0.5) .. controls (-1.5,0) .. (-0.866,-0.5);

\draw[decorate] (-1.85,-0.23) -- (-1.5,-0.6);

\draw[thin] (-2,0) circle (0.25);
\begin{scope}[rotate around={30:(-2,0)}]
\filldraw[thin, fill=red!50] (-2.25,0) circle (0.15);
\filldraw[thin, fill=yellow!50] (-1.75,0) circle (0.15);
\end{scope}

\draw[thin] (-1.5,1.75) circle (0.25);
\begin{scope}[rotate around={45:(-1.5,1.75)}]
\filldraw[thin, fill=pink!50] (-1.3,1.75) circle (0.15);
\filldraw[thin, fill=red!50] (-1.7,1.75) circle (0.15);
\end{scope}

\end{scope}

\begin{scope}[rotate around={120:(0,0)}]
[
    decoration={coil,amplitude=0.5mm,segment length=1mm}
                    ]

\draw[decorate] (0,1) .. controls (-0.42,1.7) and (-0.84,1.7) .. (-1.25,1.25);
\draw[thin] (-1.25,1) circle (0.25);
\filldraw[thin, fill=red!50] (-1,1) circle (0.15);
\filldraw[thin, fill=cyan!50] (-1.5,1) circle (0.15);

\draw[decorate] (-0.4,2.45) ..controls (0,3) and (0.5,1.5) .. (0,1);

\draw[decorate] (-0.636,1.55) .. controls (-0.82,1.7) .. (-0.621,2);
\draw[thin] (-0.75+0.25,2.2) circle (0.25);
\begin{scope}[rotate around={-30:(-0.5,2.2)}]
\filldraw[thin, fill=green!50] (-0.75,2.2) circle (0.15);
\filldraw[thin, fill=yellow!50] (-0.75+0.5,2.2) circle (0.15);
\end{scope}

\draw[decorate] (-1.625,1.95) .. controls (-1.4,2.6) and (-0,4,2.8) .. (-0.12,2.55);
\draw[decorate] (-1.3,1.6) -- (-1.1,1.4);

\draw[decorate] (-1.3,0.75) .. controls (-1.45,0.2) and (-2,1) .. (-2.1,0.22);

\draw[decorate] (-1.4,0.5) .. controls (-1.5,0) .. (-0.866,-0.5);

\draw[decorate] (-1.85,-0.23) -- (-1.5,-0.6);

\draw[thin] (-2,0) circle (0.25);
\begin{scope}[rotate around={30:(-2,0)}]
\filldraw[thin, fill=blue!50] (-2.25,0) circle (0.15);
\filldraw[thin, fill=yellow!50] (-1.75,0) circle (0.15);
\end{scope}

\draw[thin] (-1.5,1.75) circle (0.25);
\begin{scope}[rotate around={45:(-1.5,1.75)}]
\filldraw[thin, fill=pink!50] (-1.3,1.75) circle (0.15);
\filldraw[thin, fill=blue!50] (-1.7,1.75) circle (0.15);
\end{scope}

\end{scope}

\draw[decorate] (-1.3,-2.45) .. controls (-0.5,-3) and (0.2,-3) .. (0.6,-2);
\draw[decorate] (1,1) -- (0.45,0.462);

\draw[spring,segment length=6] (0.866,-0.5) -- (0,1);   
\draw[spring,segment length=6] (0,1) -- (-0.866,-0.5);   
\draw[spring,segment length=6] (-0.866,-0.5) --  (0.866,-0.5);   

\filldraw[fill=red!50] (0,1) circle (0.3) node {\large{\textbf{u}}$^\uparrow$};
\filldraw[fill=green!50,rotate around={120:(0,0)}] (0,1) circle (0.3) node {\large{\textbf{u}}$^\downarrow$};
\filldraw[fill=blue!50,rotate around={-120:(0,0)}] (0,1) circle (0.3) node {\large{\textbf{d}}$^\uparrow$};

\draw[line width=3.5pt, -latex] (3.5,0) -- (4.5,0);

\begin{scope}[xshift=7.9cm] 
\begin{scope}[rotate around={35:(0,0)}]

\def\angleA{0}    
\def\angleB{120}  
\def\angleC{120}  
\def\angleD{240}  
\def\angleE{240}  
\def\angleF{360}  
\def\angleG{360}  

\foreach \i in {0,0.25,0.5,...,110} {
    \draw[red!90] (0,0) -- (\i:3.3);
}

\foreach \i in {110,110.25,...,120} {
    \pgfmathsetmacro{\ratio}{100*(\i-110)/10}
    \draw[green!\ratio!red!90] (0,0) -- (\i:3.3);
}

\foreach \i in {120.25,120.5,...,229.75} {
    \draw[green!90] (0,0) -- (\i:3.3);
}

\foreach \i in {230,230.25,...,240} {
    \pgfmathsetmacro{\ratio}{100*(\i-230)/10}
    \draw[blue!\ratio!green!90] (0,0) -- (\i:3.3);
}

\foreach \i in {240.25,240.5,...,349.75} {
    \draw[blue!90] (0,0) -- (\i:3.3);
}

\foreach \i in {350,350.25,...,360} {
    \pgfmathsetmacro{\ratio}{100*(\i-350)/10}
    \draw[red!\ratio!blue!90] (0,0) -- (\i:3.3);
}

\foreach \i in {0,1,...,60} {  
    \pgfmathsetmacro{\rInner}{\i/60*3.3 }  
    \pgfmathsetmacro{\rOuter}{(\i+3)/60*3.3 }  
    \pgfmathsetmacro{\opacity}{(\i+1)/90}
    \fill[white, opacity=\opacity, even odd rule] 
         (0,0) circle (\rOuter) (0,0) circle (\rInner);
}

\end{scope}

\filldraw[fill=red!50, thick] (0,1) circle (0.5) node{\Large{\textbf{q}}};

    \coordinate (center) at (0, -1);

    \fill[green!50] (center) -- +(0, 0.5) arc (90:270:0.5) -- cycle;

    \fill[blue!50] (center) -- +(0, -0.5) arc (-90:90:0.5) -- cycle;

    \draw[black, thick] (center) circle (0.5) node {\Large{\textbf{D}}};

\begin{scope}[rotate around={90:(0,0)}]
  \draw[white, dashed, thick, line width=2pt] 
    (0,0) .. controls (1,1.25) and (2,1.25) .. (2,0) 
          .. controls (2,-1.25) and (1,-1.25) .. (0,0)
          .. controls (-1,-1.25) and (-2,-1.25) .. (-2,0) 
          .. controls (-2,1.25) and (-1,1.25) .. (0,0);
\end{scope}

\end{scope}

\end{tikzpicture}
\caption{A schematic transition from a physical model of the proton, in which, in addition to the valence quarks, the internal dynamics of the particle involves a gluon-quark sea, to a DSM, in which the two valence quarks are combined into a single effective diquark $D$, and the dynamics of the gluon-quark sea is accounted for within the quark-diquark pair itself.}
\label{fig:quark_to_diquark}
\end{figure}

The standard twist-2 vector GPDs of quarks for a spin $\tfrac{1}{2}$ hadron, such as a proton or neutron, are defined through the light-cone correlation function \cite{Diehl:2003ny}

\begin{equation}
    \label{eq:correlation_function}
    F^{[\Gamma]} (x,\Delta) = \frac{1}{2} \int 
    \frac{dz^-}{2\pi} e^{i k \cdot z}
    \bra{p'}\bar{\psi}\left(\scriptstyle  -\frac{z}{2}\right)
    \Gamma \mathcal{W} \left(\scriptstyle -\frac{z}{2}, \frac{z}{2}\right)
    \psi\left(\scriptstyle \frac{z}{2}\right)\ket{p} \bigg|_{z^+=0,\,\vec{z}_\bot = \vec{0}_\bot},
\end{equation}
where $\psi$ is a quark (Dirac) field, $z$ is a light-cone coordinates four-vector and $\Gamma$ denotes a Dirac gamma matrix. Wilson line $\mathcal{W}$ in the form of \Cref{eq:Wilson_line} is inserted to ensure gauge invariance of the quark-quark correlator,
\begin{equation}
    \label{eq:Wilson_line}
     \mathcal{W}\left(\scriptstyle-\frac{z}{2},\,\frac{z}{2} \right) 
     \bigg|_{z^+=0,\,\vec{z}_\bot = \vec{0}_\bot} = \mathcal{P}\, \text{exp} \left( -ig_s \int^{\frac{z^-}{2}}_{-\frac{z^-}{2}} 
      dy^- A^+\left( 0^+, y^-, \vec{0}_\bot \right)\right),
\end{equation}
with $\mathcal{P}$ being the path-ordering operator, $g_s$ denotes the strong coupling constant and $A^+$ is a plus component (in light-cone coordinates) of the gluon field. Throughout this work, we employ light-cone coordinates. In this convention, the components of a four-vector are defined as
\begin{equation}
    \label{eq:four_vectors}
    v^\pm = \frac{1}{\sqrt{2}}\left(v^0 \pm v^3 \right), \hspace{2cm} \vec{v}_\bot = \left(v^1,v^2\right),
\end{equation}
where the Minkowski components of a four-vector are denoted by $v = \left(v^0,v^1,v^2,v^3 \right)$, while its light-cone representation is written as $v = (v^+,v^-,\vec{v}_\bot)$. If the initial four-momentum of a hadron is equal to $p$ and the final four-momentum is equal to $p'$, then the following kinematic variables can be defined:
\begin{equation}
    \label{eq:kinematic_variables}
    P=\frac{p+p'}{2}, \qquad \Delta = p'-p, \qquad t=\Delta^2, \qquad \xi = - \frac{p'^+-p^+}{p'^+ + p^+} = - \frac{\Delta^+}{2P^+},
\end{equation}
where $P$ denotes an average four-momentum of a hadron, $\Delta$ denotes a four-momentum transfer, $t$ is a Mandelstam variable for t-channel and $\xi$ is the skewness. The skewness parameter provides information about the momentum transfer relative to the average momentum along the $+$ direction on the light-cone. The system is analyzed in a symmetric frame where $\vec{P}_\bot=0$ ($\vec{p}_\bot = -\vec{p'}_\bot=\vec{\Delta}_\bot/2$) and $P^3 \gg0$. The following quantities are used to express the Mandelstam variable in a more useful way:
\begin{equation}
    \label{eq:mandelstam_quantities}
    \begin{split}
   &\hspace{-1.2cm} p^- = \frac{p^2_\bot +M^2}{2p^+}    , \quad p'^- = \frac{p'^2_\bot +M^2}{2p'^+}, \quad 
   p^+ = P^+(1+\xi), \quad p'^+ = P^+(1-\xi) \\
   &\Delta^- = \frac{\xi}{1-\xi^2} \frac{M^2 + \vec{\Delta}^2_\bot/4}{P^+}, \qquad 
   P^- = \frac{1}{2(1-\xi^2)}\frac{M^2 + \vec{\Delta}^2_\bot/4}{P^+},
    \end{split}
\end{equation}
where $M$ is the proton mass. After using these equations and deriving $\Delta^\pm = \mp 2\xi P^\pm$ from \Cref{eq:kinematic_variables}, $t$ can be expressed as
\begin{equation}
    \label{eq:mandelstam}
    t = - \frac{1}{1-\xi^2}\left(4\xi^2M^2 + \vec{\Delta}^2_\bot\right).
\end{equation}
We also define $x$, which denotes momentum fraction of a parton along the direction of the hadron's motion and is equal to $x = \frac{\text{ parton longitudinal momentum }}{\text{  hadron longitudinal momentum}} = \frac{k^+}{P^+}$.
A useful consequence of the kinematic relations presented above is $P \cdot \Delta=0$.
For the light-cone coordinates, we use the metric tensor defined as follows,
\begin{equation}
    \label{eq:metric_tensor}
    g^{\mu\nu}=g_{\mu\nu} = \begin{pmatrix}
0\quad & 1\quad & 0\quad& 0 \\
1\quad & 0\quad & 0\quad & 0 \\
0\quad & 0\quad & -1\quad & 0 \\
0\quad & 0\quad& 0\quad &-1
\end{pmatrix}.
\end{equation}
After using $P^2 = M^2-t/4$, one can define a variable $\delta$ to simplify further notations, $\delta = \sqrt{1+\frac{M^2-t/4}{(P^3)^2}}$ \cite{Bhattacharya:2018zxi}. For $\Gamma = \gamma^+$, \Cref{eq:correlation_function} defines the twist-2 vector GPDs $H$ and $E$, \cite{Diehl:2003ny, Mei_ner_2007}
\begin{equation}
    \label{eq:twist_2_vector_H_and_E}
    F^{[\gamma^+]} (x, \Delta)
     = \frac{1}{2 P^+} \bar{u}(p') \left[\gamma^+
     H(x,\xi,t) + \frac{i \sigma^{+\mu}\Delta_\mu}{2M}
     E(x,\xi,t)
     \right]u(p),
\end{equation}
where $u(p)$ and $u(p')$ are 4-component Dirac spinors for the incoming and the outgoing hadron, and $\sigma ^{\mu\nu}$ is defined through a commutator of gamma matrices as $\sigma^{\mu\nu} = \frac{i}{2} (\gamma^\mu \gamma^\nu - \gamma^\nu \gamma^\mu)$. In the DSM (graphical representation of which is shown in  \Cref{fig:quark_to_diquark}), the proton wave function can be written as
\begin{equation}
    \label{eq:proton_wave_function}
    \ket{p,\,\uparrow} = 
   c_{S}^{+} \ket{S^0_0u^\uparrow} + 
    c_{A}^+ \ket{A^0_0u^\uparrow} - 
   c_{A}^- \ket{A^1_0u^\downarrow} - 
   c_{A}'^+ \ket{A^0_1 d^\uparrow} + 
   c_A'^- \ket{A^1_1 d^\downarrow},
\end{equation}
where $A$ stands for the axial-vector diquark and $S$ stands for the scalar diquark. The lower index next to $A(S)$ indicates the isospin value, $0$ for the diquark built with up and down quarks and $1$ for the diquark built with two up quarks. The upper index is the value of diquark spin projection onto the $z$ axis. $c(c')_{S/A}^{+/-}$ are constants describing contributions of given quark-diquark configurations to the whole proton wave function, where from the normalization condition one has

\begin{equation}
    (c_S^+)^2 + 
(c_A^+)^2 + (c_A^-)^2+ (c_A'^+)^2 + (c_A'^-)^2 =1.
\end{equation}
It is important to note that there is no (down quark) -- (scalar diquark) combination, which is a direct consequence of the Pauli exclusion principle.
If the value of the spin projection onto the $z$ axis is not important (like in this work) the above equation looks as follows,
\begin{equation}
    \label{eq_proton_wave_function_without_spin}
    \ket{p} = c_S\ket{S_0 u} -c_A \ket{A_0 u} + c'_A\ket{A_1 d },
\end{equation}
which gives the normalization $c_S^2 + c_A^2 + c_A'^2=1$.
By convention, the PDF is normalized to $2$ for the valence up quark and to $1$ for the valence down quark,
\begin{equation}
    \label{eq:normalization_of_valence_quarks}
    \int_0^1 f^{u_v}_1(x)dx=3c_S^2 + 3c_A^2 = 2, \hspace{1.5cm} \int_0^1 f_1^{d_v}(x)dx= 3c_A'^2 = 1.
\end{equation}
So it is possible to rewrite $c(c')_{S/A}$ constants as $3 c_{S}^2 = \lambda$, $3 c_A^2 = 2-\lambda$ and $3c'^2_A = 1$, where $\lambda$ is called the decomposition constant and describes the contributions of the scalar and axial-vector diquarks to the state with up quark. Because of this, the up quark GPDs $H(E)^{\text{up}}$ can be expressed as the sum of the scalar diquark GPDs $H(E)^S$ and the axial-vector diquark GPDs $H(E)^A$, and the down quark GPDs $H(E)^{\text{down}}$ are the same as the pure axial-vector diquark GPDs,
\begin{equation}
\label{eq:H_quark_and_E_quark}
\begin{split}
        &H^{\text{up}} =\lambda H^{S}+(2-\lambda)H^A,  
       \hspace{1cm} H^{\text{down}} = H^A,\\
        &\hspace{0.1cm}E^{\text{up}} =\lambda E^{S}+(2-\lambda)E^A,  
       \hspace{1cm} E^{\text{down}} = E^A.
       \end{split}
\end{equation}
If $q$ stands for up or down quark, then the forward limits ($\xi=0$ and $t=0$) of $H^q$ and $E^q$ are 
\begin{equation}
        H^q(x,0,0) = q(x) , \hspace{2cm}
        E^q(x,0,0) = e_q(x),
\end{equation}
where $q(x)$ is the unpolarized PDF and $e_q(x)$ has no PDF counterpart. In this work, we investigate the nucleon tomography, which requires consideration of only zero-skewness GPDs. For an unpolarized hadron in position space, it is given by the Fourier transform of the momentum space $H$,
\begin{figure}
    \centering
\begin{tikzpicture}
    \filldraw[black, fill = orange!40, very thick] (0,0) circle (1.5);
    \draw[->, -{Latex}, line width=4pt, orange!50!black] (-2,0) -- (-5,0); 
    \draw[ultra thick, -latex] (1,0) arc (0:300:1);

    \draw[ultra thick, -latex] (2.2,-2) -- (2.2,0);
    \draw[ultra thick, -latex] (2.2,-2) -- (0.2,-2);
    \filldraw[ultra thick, fill=white] (2.2,-2) circle (0.25);
    \filldraw[fill=black] (2.2,-2) circle (0.1);

    \node at (2.5,0) {$y$};
    \node at (0.1,-2.25) {$z$};
    \node at (2.65,-1.7) {$x$};

    \draw[latex-, ultra thick, red!70] (0.3,1.2) -- (1.8,1.2) -- (3,0.75);

    \draw[latex-, ultra thick, blue!70] (-0.3,-1.2) -- (-1.8,-1.2) -- (-3,-0.75);

    \node [red!80] at (3.4,0.5) {Higher momentum};
    \node [blue!80] at (-3.4,-0.5) {Lower momentum};

\end{tikzpicture}
    \caption{A schematic view of a hadron polarized along the $x$-axis and moving in the $z$-direction, depicted as a sphere rotating about the $x$-axis. It can be seen that, as a result of the rotation about the $x$-axis, for one half of the sphere (from the center upwards), the rotational momentum is greater than the total momentum of the particle, whereas for the other half of the sphere, the momentum is correspondingly smaller. Similar interpretation can be found in \cite{Burkardt:2002hr}.}
        \label{fig:polarized_hadron}
\end{figure}
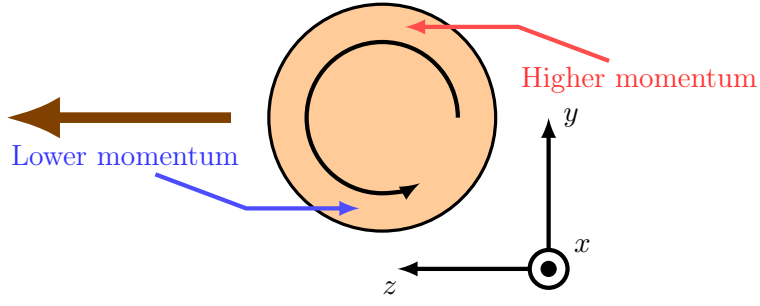
\begin{equation}
    \label{eq:unpolarized_tomography}
    q\left(x,\vec{b}_\bot\right) = \int \frac{d^2 \vec{\Delta}_\bot}{(2\pi)^2}
    e^{-i\vec{b}_\bot \cdot \vec{\Delta}_\bot}
    H^q(x,0,-\vec{\Delta}^2_\bot),
\end{equation}
where $\vec{b}_\bot=(b_x,b_y)$ is called the impact parameter, and provides information about the spatial distribution of partons in the plane perpendicular to the direction of hadron motion, in the frame where the center of mass is at the origin \cite{Burkardt:2002hr}. If the functions are normalized, $q(x,\vec{b}_\bot)$ can be interpreted as the density of partons at a given position. If the hadron is polarized transversely to its direction of motion, the density of partons will be distorted by terms coming from the Fourier transform of the $E$ GPD \cite{Burkardt_2004, Brodsky_2002, Pasquini_2019},
\begin{equation}
    \label{eq:E_GPD_tomography}
        e_q\left(x,\vec{b}_\bot\right) = \int \frac{d^2 \vec{\Delta}_\bot}{(2\pi)^2}
    e^{-i\vec{b}_\bot \cdot \vec{\Delta}_\bot}
    E^q(x,0,-\vec{\Delta}^2_\bot).
\end{equation}
In a situation where the hadron is polarized along the $x$-axis, the total parton distribution function is given by the following equation,
\begin{equation}
    \label{eq:polarized_tomography}
    q_X  \left(x,\vec{b}_\bot\right) = q(x,\vec{b}_\bot) - \frac{1}{2 M} \frac{\partial }{\partial b_y}e_q(x,\vec{b}_\bot) \, .
\end{equation}
It is important to note that for the hadron polarized along the $x$ axis, $e_q$ must be differentiated along the axis perpendicular to both the direction of motion and the axis of polarization, as shown in \Cref{fig:polarized_hadron}. 
\subsection{Results for GPDs}
\begin{figure}[!h]
\centering
\begin{tikzpicture}
\begin{scope}[very thick, every node/.style={sloped,allow upside down}]

 \draw[draw=blue!40, decorate, decoration={zigzag, segment length=10pt, amplitude=3pt}, shorten >=3pt]  (0,-1) -- (0,3.3);
 
 \filldraw[line width=0.4mm,dashed] (-2.5,0) -- node[scale=2]  {\midarrow}  (2.5,0);

\draw[line width=1mm] (-5,-1) -- node[scale=3]  {\midarrow}  (-2.5,0);
\draw[line width=1mm] (2.5,0) -- node[scale=3]  {\midarrow}  (5,-1);

\draw[] (-2.5,0) -- node[scale=2]  {\midarrow}  (-2.5,3);
\draw[] (2.5,3) -- node[scale=2]  {\midarrow}  (2.5,0);

\fill[] (-2.5,0) circle (0.15);
\fill[] (2.5,0) circle (0.15);

\node[] at (-2,0.4) {\large $\mu$};
\node[] at (2,0.4) {\large $\nu$};

\node[] at (-5.7,-0.25) {\large $ p=P-\frac{\Delta}{2}$};
\node[] at (5.5,-0.25) {\large $p'=P+\frac{\Delta}{2}$};

\node[] at (-3.5,1.5) {\large $k-\frac{\Delta}{2}$};
\node[] at (3.5,1.5) {\large $k+\frac{\Delta}{2}$};

\node[] at (-1,-0.5) {\large $P-k$};

\node[blue!50] at (0.6,2.8) {\Large $\gamma^+$};

\end{scope}

\end{tikzpicture}
\caption{The Feynman diagram for the leading-twist cut vertex GPD. The vertical blue zigzag line represents the cut vertex $\gamma^+$, which effectively replaces the interaction between quarks and photons. The thick solid line represents the incoming (outgoing) hadron with momentum $p$($p'$), the solid thin line represents the quark interacting with the photon, and the dashed line is the spectator diquark.}
\label{cut DVCS}
\end{figure}
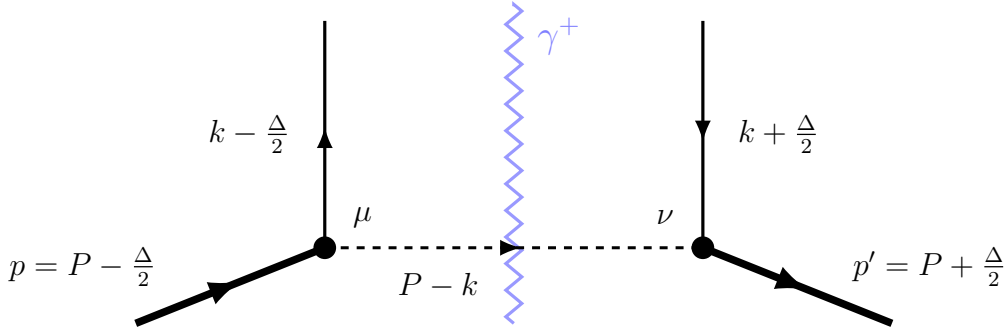
We begin by recalling the known results for the scalar diquark GPDs (for more information, see \cite{Bhattacharya:2018zxi,Mei_ner_2009}). For the lowest nontrivial order shown in the Feynman diagram in \Cref{cut DVCS}, the correlator \Cref{eq:correlation_function} takes the form:
\begin{equation}
    \label{eq:GPD_from_Fyenmann_for_scalar}
    F^{[\gamma^+
    ]}(x,\Delta) = \frac{i g^2}{2(2\pi)^4}
    \int dk^- d^2\vec{k}_\bot 
    \frac{\bar{u}(p') 
    \left(\slashed{k} +\slashed{\Delta}/2 +m_q
    \right)\gamma^+
    \left( \slashed{k} -\slashed{\Delta}/2 +m_q
    \right)u(p)
    }{D_{\text{scalar}
    }}\, .
\end{equation}
In this equation, we omit the spin of the incoming and the outgoing hadron. The denominator $D_{\text{scalar}}$ is given by
\begin{equation}
    \label{eq:D_scalar}
    D_{\text{scalar}} = 
    \left[ \left(k+\Delta/2 \right)^2 - m_q^2 +i\varepsilon
    \right]
        \left[ 
        (P-k)^2 - m_s^2+i\varepsilon
    \right]
    \left[
    (k-\Delta/2)^2-m_q^2
    +i\varepsilon
    \right].
\end{equation}
In the above equations, $m_q$ denotes the valence quark mass, $m_s$ the scalar diquark mass, and $g$ denotes the coupling constant of the hadron-quark-diquark vertex.
After using the Gordon identities\footnote{In this paper we use the following Gordon identities: $2M\bar{u}(p')\Gamma u(p)=\bar{u}(p') 
\left[\{\Gamma, \slashed{P} \} - \frac{1}{2}[\Gamma, \slashed{\Delta}]   \right]
u(p)$ and $0=\bar{u}(p') 
\left[\{\Gamma, \slashed{P} \} + \frac{1}{2}[\Gamma, \slashed{\Delta}]   \right]
u(p)$, for $\Gamma = I, \gamma^\mu,\gamma_5,\gamma^\mu\gamma_5, i\sigma^{\mu\nu}, i\sigma^{\mu\nu} \gamma_5$.}
and performing the contour integral over $k^-$, and changing integral coordinates of $\vec{k}_\bot$ to polar ones, one obtain:
  \begin{equation}
  \label{eq:H_GPD_scalar}
H(x,\xi,t) =
\begin{cases}
    0 & -1 \leq x \leq -\xi, \\
  \displaystyle  \frac{g^2 (x+\xi)(1+\xi)(1-\xi^2)}{4(2\pi)^3} \int _0^{2\pi}d\theta \int_0^{\Lambda} d {k}_\perp\;  k_\bot \dfrac{N_H^S}{D_{1S} D_{2S}^{-\xi\leq x \leq \xi}} & -\xi \leq x \leq \xi, \\
 \displaystyle   \frac{g^2 (1-x)(1-\xi^2)}{2(2\pi)^3} \int _0^{2\pi}d\theta \int_0^{\Lambda} d {k}_\perp\; k_\bot \frac{N_H^S}{ D_{1S} D_{2S}^{x \geq \xi}} & x \geq \xi,
\end{cases}
\end{equation}
where the numerator $N_H^S$ is the same in every $x$ region and is given by:
\begin{equation}
    \label{eq:N_H_scalar}
    N_H^S = \vec{k}_\bot^2 + (m_q+xM)^2+(1+x)^2\frac{t}{4} - (1-x)\xi t \frac{\vec{k}_\bot \cdot \vec{\Delta}_\bot}{\vec{\Delta^2}_\bot}.
\end{equation}
The denominators, on the other hand, are given by:
\begin{equation}
\label{eq:D_scalar12}
\begin{split}
&D_{1S}=(1+\xi)^2\vec{k}^2_\bot+\frac{1}{4}(1-x)^2\vec{\Delta}^2_\bot-(1-x)(1+\xi)\vec{k}_\bot\cdot \vec{\Delta}_\bot+(1-x)(1+\xi)m_q^2+\\
&\hspace{8cm}+(x+\xi)(1+\xi)m_s^2-(1-x)(x+\xi)M^2 ,\\
&D_{2S}^{-\xi\leq x\leq \xi}=\xi(1-\xi^2)\vec{k}_\bot^2+\frac{1}{4}(1-x^2)\xi\vec{\Delta}^2_\bot+x(1-\xi^2)\vec{k}_\bot\cdot \vec{\Delta}_\bot+\xi(1-\xi^2)m_q^2\\
&\hspace{12cm}-\xi(x^2-\xi^2)M^2,\\
&D_{2S}^{x \geq\xi}=(1-\xi)^2\vec{k}_\bot^2+\frac{1}{4}(1-x)^2\vec{\Delta}^2_\bot+(1-x)(1-\xi)\vec{k}_\bot\cdot \vec{\Delta}_\bot+(1-x)(1-\xi)m_q^2+\\
&\hspace{8cm}+(x-\xi)(1-\xi)m_s^2-(1-x)(x-\xi)M^2 .
\end{split}
\end{equation}
$\Lambda$ is the cutoff of integration over $k_\bot$ (integration radius); it will be discussed in greater detail below, in Appendix \ref{NR}.
For the $E$ GPD, the formalism is analogous, with the difference that the numerator in \Cref{eq:H_GPD_scalar}, $N_H^S$, must be replaced by $N_E^S$,
\begin{equation}
    \label{eq:E_GPD_nominator}
    N_E^S = 2(1-x)M\left[m_q+\left(x+2\xi \frac{\vec{k}_\bot\cdot\vec{\Delta}_\bot}{\vec{\Delta}^2_\bot}\right)M \right]\, .
\end{equation}
The denominator remains the same as for the scalar $H$ GPD. 
\\
\\
Similarly, the GPDs can be calculated for the axial-vector diquark case, where both the numerator and the denominator change with respect to \Cref{eq:H_GPD_scalar}. The difference stems from the fact that the contribution from the interaction vertex is different, due to the diquark transitioning from a scalar to an axial-vector in the hadron. For this reason, the axial-vector diquark propagator takes a more complex form and, with respect to the propagator for the scalar diquark, additionally contains in the numerator a polarization tensor, $d^{\mu\nu} = -g^{\mu\nu} + \frac{P^\mu P^\nu}{m_a^2}$. One can also use a different form of the interaction vertex for the axial-vector diquark; for further information, we refer, for example, to Ref \cite{Bacchetta:2008af}. Using the Feynman rules for the axial case, one gets:
\begin{equation}
    \label{eq:GPD_from_Fyenmann_for_axial}
    F^{[\gamma^+
    ]}(x,\Delta) = \frac{i g^2}{2(2\pi)^4}
    \int dk^- d^2\vec{k}_\bot d_{\mu\nu}
    \frac{\bar{u}(p') \gamma^\mu\gamma_5
    \left(\slashed{k} +\frac{\slashed{\Delta}}{2} +m_q
    \right)\gamma^+
    \left( \slashed{k} -\frac{\slashed{\Delta}}{2} +m_q
    \right)
    \gamma^\nu\gamma_5
    u(p)
    }{D_{\text{axial}
    }}\, .
\end{equation}
The change in the denominator is minimal; all that is required is to replace the scalar diquark mass with the axial-vector diquark mass, $m_s \to m_a$, to get $D_{\text{axial}}$.
However, due to the additional tensor factor in the numerator, the resulting numerators for the axial-vector diquark case are more complex and are given by the following equations for the $H$ and $E$ GPDs cases:
    \begin{equation}
    \label{eq:NHa}
    \begin{split}
2N_H^A = &
\left(2 + \frac{M^2}{m_a^2} + \frac{t}{4m_a^2}\right)
(\vec{k}_\perp^2 + m_q^2 + x^2 M^2)
+ 2\left[4x - x\frac{M^2}{m_a^2} + (2 + x)\frac{t}{4m_a^2}\right] m_q M\\
&\hspace{4cm}+ (1 + x)\left[2(1 + x) + (1 - 3x)\frac{M^2}{m_a^2} + (1 + x)\frac{t}{4m_a^2}\right] \frac{t}{4} \\
&\hspace{6cm}+ (1 + x)\left[2 - \frac{M^2}{m_a^2} + \frac{t}{4m_a^2}\right] \xi t
\frac{\vec{k}_\perp \cdot \vec{\Delta}_\perp}{\vec{\Delta}_\perp^2},
\end{split}
\end{equation}

    \begin{align}
    \label{eq:NEA}
        2 N_E^A&=-2\frac{m_q^2}{m_a^2}M^2+2m_qM\left[\frac{M^2}{m_a^2}(-1+x)-\frac{t}{4m_a^2}(1+x)-4x \right]-4 M^2x(x+1)\nonumber\\
        &\hspace{1.5cm}-8\xi M^2 (x+1)\frac{\vec{k}_\bot\cdot \vec{\Delta}_\bot}{\vec{\Delta}^2_\bot} 
         -2\frac{M^2}{m_a^2}\vec{k}^2_\bot+2x\frac{M^2}{m_a^2}\left(M^2-\frac{t}{4}\right)-\frac{M^2}{2m_a^2}t\nonumber\\
         &\hspace{6cm}+4\frac{M^2}{m_a^2}\xi \left(M^2-\frac{t}{4} \right)(x+1)\frac{\vec{k}_\bot\cdot \vec{\Delta}_\bot}{\vec{\Delta}^2_\bot}.
    \end{align}
\Cref{eq:NEA} is a new result previously not reported in the literature; for this reason, we have also derived equations describing the quasi-E GPD for an axial diquark and demonstrated the convergence of quasi-GPD $\to$ GPD; this analysis can be found in  Appendix \ref{app:convergence}. QGPDs (quasi-GPDs) are defined through spatial correlations rather than on the light-cone. For this model, the convergence of the QGPD to the GPD for a large average hadron momentum has previously been demonstrated in \cite{Bhattacharya:2018zxi} for the scalar diquark, and in \cite{Bhattacharya:2019cme} for the $H$ GPD for the axial-vector diquark.

\subsection{Form factors}
 The Mellin moments of GPDs, $A^q_{n+1,i}(t)$ (for the $H$ GPD) and $B^q_{n+1,i}(t)$ (for the $E$ GPD), are referred to as generalized form factors. 
In this paper, we are primarily interested in the zero-skewness case. For this reason, we consider only skewness-independent form factors with $i=0$,
\begin{equation}
    \label{eq:FF_non_xi}
    \int^1_{-1} dx\, x^n H^q (x,t) =  A^{q}_{n+1,0}(t)\,, \hspace{1cm}
        \int^1_{-1} dx\, x^n E^q (x,t) =  B^{q}_{n+1,0}(t)\,.
\end{equation}
 For $n=0$, these terms describe the contribution of a given quark to the Dirac, $F_1^q(t)$, and Pauli, $F_2^q(t)$, form factors:
\begin{equation}
    \label{eq:Dirrac_Pauli}
    \int _{-1}^1 dx \,H^q(x,t) = A^q_{1,0}(t)=F_1^q(t), \hspace{1.5cm} \int _{-1}^1 dx \,E^q(x,t)=B^q_{1,0}(t) = F_2^q(t).
\end{equation}
This result does not depend on $\xi$, so it is convenient to set $\xi = 0$ in further calculations. Since the charge of a single up quark is $+2/3$, while that of a down quark is $-1/3$, these form factors for the whole proton can be written as
\begin{equation}
        \label{eq:Dirac_Pauli_whole_proton}
\begin{split}
    F_{1/2}^{\text{proton}}(t)=\frac{2}{3}F_{1/2}^{\text{up}}(t) - \frac{1}{3}F_{1/2}^{\text{down}}(t) &= \frac{2}{3}\lambda F_{1/2}^S(t) + \frac{2}{3} (2-\lambda) F_{1/2}^A(t) - \frac{1}{3}F_{1/2}^A(t) \\
     &= \frac{2}{3}\left[\lambda F_{1/2}^S(t)+  \left(\frac{3}{2}-\lambda\right)F_{1/2}^A(t)\right].
\end{split}
\end{equation}
It is worth noting that for $\lambda=1.5$ (which is often the standard choice (see, for instance \cite{Jakob:1997wg,Bacchetta:2008af}), made with the assumption that the proton wavefunction has equal contributions from the scalar and axial-vector diquarks), the contribution from the axial-vector diquark to the Dirac and Pauli FFs vanishes entirely. It is also important to note that, under the normalization condition of the 
$H$ GPD in the forward limit, $\int_0^1dx H^{S/A}(x,0,0)=1$, the value of $F^{S/A}_1$ at $t=0$ is well defined and equals $1$. Consequently
\begin{equation}
    \label{eq:F1_FF_t=0}
    F_1^{\text{proton}}(t=0) = \frac{2}{3}\left[
    \lambda + \left(\frac{3}{2}-\lambda\right)
    \right] = 1,
\end{equation}
which means that this value does not depend on the decomposition constant $\lambda$. This quantity is interpreted as the total charge of the proton. From $F_{1/2}^q$, one can calculate the electromagnetic form factors (EMFFs), defined as
\begin{equation}
    \label{eq:Electro_Magnetic_FFs}
    G_E^q(t)=F_1^q(t) + \frac{t}{4M^2}F^q_2(t), \hspace{2cm} G_M^q = F_1^q(t) + F_2^q(t),
\end{equation}
where proton EMFFs can be written as:
\begin{equation}
    \label{eq:Electro_Magnetic_FFs_proton}
    G_{E/M}^{\text{proton}}(t) = \frac{2}{3}G_{E/M}^{\text{up}}(t) - \frac{1}{3} G_{E/M}^{\text{down}}(t) = \frac{2}{3}\left[
    \lambda G_{E/M}^S(t) + \left(\frac{3}{2} - \lambda\right)G_{E/M}^A(t) \right].
\end{equation}
Again, for $\lambda = 1.5$, the results for the proton do not depend on axial-vector diquark terms\footnote{However, analogous calculations for the neutron do not lead to the isolation of purely scalar or axial-vector diquark factors. For the neutron, one can get $F^{\text{neutron}}(t) = \frac{2}{3}F^{\text{down}}(t) -\frac{1}{3}F^{\text{up}}(t)  =  \frac{\lambda}{3}\left(F^A(t) - F^S(t) \right)$. From the normalization condition $F_1^{\text{neutron}}(t=0)=\frac{\lambda}{3}(1-1)=0$, so, regardless of the value of $\lambda$, the total neutron charge is always zero.}. The EMFFs allow the electric ($r_E$) and magnetic ($r_M$) radii of a hadron to be defined,
\begin{equation}
    \label{eq:electro_magnetic_radii}
    \left<r_E^2\right> = 6 \frac{\text{d}G^{\text{hadron}}_E}{\text{d}t}\bigg|_{t=0},
    \hspace{2cm}
    \left<r^2_M \right> = \frac{6}{G_M^{\text{hadron}}(t=0)}\frac{\text{d}G^{\text{hadron}}_M}{\text{d}t}\bigg|_{t=0}.
\end{equation}
The $n=1$ Mellin moments describe the gravitational form factors (GFFs), $A^q$ and $B^q$. Here, $A^q$ is defined as $A^q_{2,0}$ and $B^q$ as $B^q_{2,0}$ (see, for instance, \cite{burkert2024colloquiumgravitationalformfactors,Liu_2014,Kumar_2017}),
\begin{equation}
    \label{eq:A_and_B_GGFs}
    \int^{1}_{-1} dx\cdot xH^q(x,t) = A^q(t),\hspace{1.5cm} \int^1_{-1}dx\cdot xE(x,t) = B^q(t).
\end{equation}
For these equations, the relations of the proton (and of the up and down quarks) to the scalar and axial-vector diquarks stated in \Cref{eq:Dirac_Pauli_whole_proton} are also valid. 
The form factors $A^q(t)$ and $B^q(t)$ can be understood as quark contributions to the matrix element of the symmetric energy-momentum tensor $T^{\mu\nu}_q$. Following Ji's decomposition \cite{Ji:1996ek,Lorc__2019}, this matrix element is parametrized as
\begin{equation}
    \label{eq:EMT_decomposition}
    \begin{split}
    \langle p'|T^{\mu\nu}_q|p\rangle = \bar{u}(p')\Bigg[A^q(t)\,\gamma^{(\mu}\bar{P}^{\nu)} + B^q(t)\,\frac{i\bar{P}^{(\mu}\sigma^{\nu)\alpha}\Delta_\alpha}{2M} + C^q(t)\,\frac{\Delta^\mu \Delta^\nu - g^{\mu\nu}\Delta^2}{M}\\
    + \bar{C}^q(t)\,Mg^{\mu\nu}
    \Bigg]u(p)\,,
    \end{split}
\end{equation}
where the round brackets around the indices denote symmetrization, $a^{(\mu}b^{\nu)}\equiv \tfrac{1}{2}(a^\mu b^\nu + a^\nu b^\mu)$.
$A^q(t)$ and $B^q(t)$ couple to $T^{\mu\nu}$ in the same way that $F_1^q(t)$ and $F_2^q(t)$ couple to the electromagnetic current. There is also a third form factor, $C^q(t)$ in this decomposition, closely related to the so-called $D$ term. At $\xi=0$ its contribution to the polynomiality expansion vanishes, so it does not enter the moments we consider here, and we do not discuss it further in this work. It is nevertheless an important quantity, since it is connected to the pressure and shear-stress distributions inside the proton \cite{Polyakov:2002yz, burkert2024colloquiumgravitationalformfactors}.
A fourth form factor, $\bar{C}^q(t)$, also appears in the decomposition of the single-flavor EMT $T^{\mu\nu}_q$, reflecting the fact that it is not separately conserved; only the total $T^{\mu\nu}=\sum_q T^{\mu\nu}_q + T^{\mu\nu}_g$ is. It cancels upon summing over quark flavors and gluons. Similarly to $C^q(t)$, it is not discussed further in this work.
It is also important to recall Ji's sum rule, defined by adding both skewness-independent Mellin moments (for $t=0$) with $n=1$ \cite{Adhikari_2014},
\begin{equation}
    \label{eq:Ji_sum}
    2J^q = A^q(0)+B^q(0)\, .
\end{equation}
 So Ji's sum can be strictly defined in terms of forward-limit GPDs as: $2J^q = \int ^1_{-1} dx\, x(q^{}(x)+e_q^{}(x))$. It is worth noting that unlike the EMFFs, the GFFs for proton and neutron are the same. This is because going from the proton to the neutron requires the exchange $\text{up} \leftrightarrow \text{down}$, which is symmetric for GFFs in the absence of electric charge in the GFF description. One can also show it by explicitly rewriting the above equations in terms of diquark contribution:
 \begin{equation}
    A^\text{proton} =  A^{\text{up}} + A^\text{down} = \lambda A^S + (2-\lambda)A^A + A^A  = \lambda A^S + A^A(3-\lambda)\, , 
 \end{equation}
  \begin{equation}
    A^\text{neutron} =  A^{\text{down}} + A^\text{up} = \lambda A^S + (2-\lambda)A^A + A^A  = \lambda A^S + A^A(3-\lambda)\, .
 \end{equation}
\section{Numerical results}
\label{Sec:numerical results}

In this work, we determine the integration limit $\Lambda$ from
\begin{equation}
    \Lambda = \sqrt{\beta^2-\vec\Delta^2_\bot/4},
\end{equation}
and leave $\beta$ as a model parameter.
We call this type of cutoff a quasi-hard cutoff.
A broader discussion of this cutoff is given in Appendix~\ref{NR}. 

\subsection{General characteristics of GPDs}
\label{sec:General characteristics of GPDs}
\begin{figure}[htbp]
    \centering
    \begin{subfigure}[b]{0.51\linewidth}
        \centering
        \includegraphics[width=\linewidth]{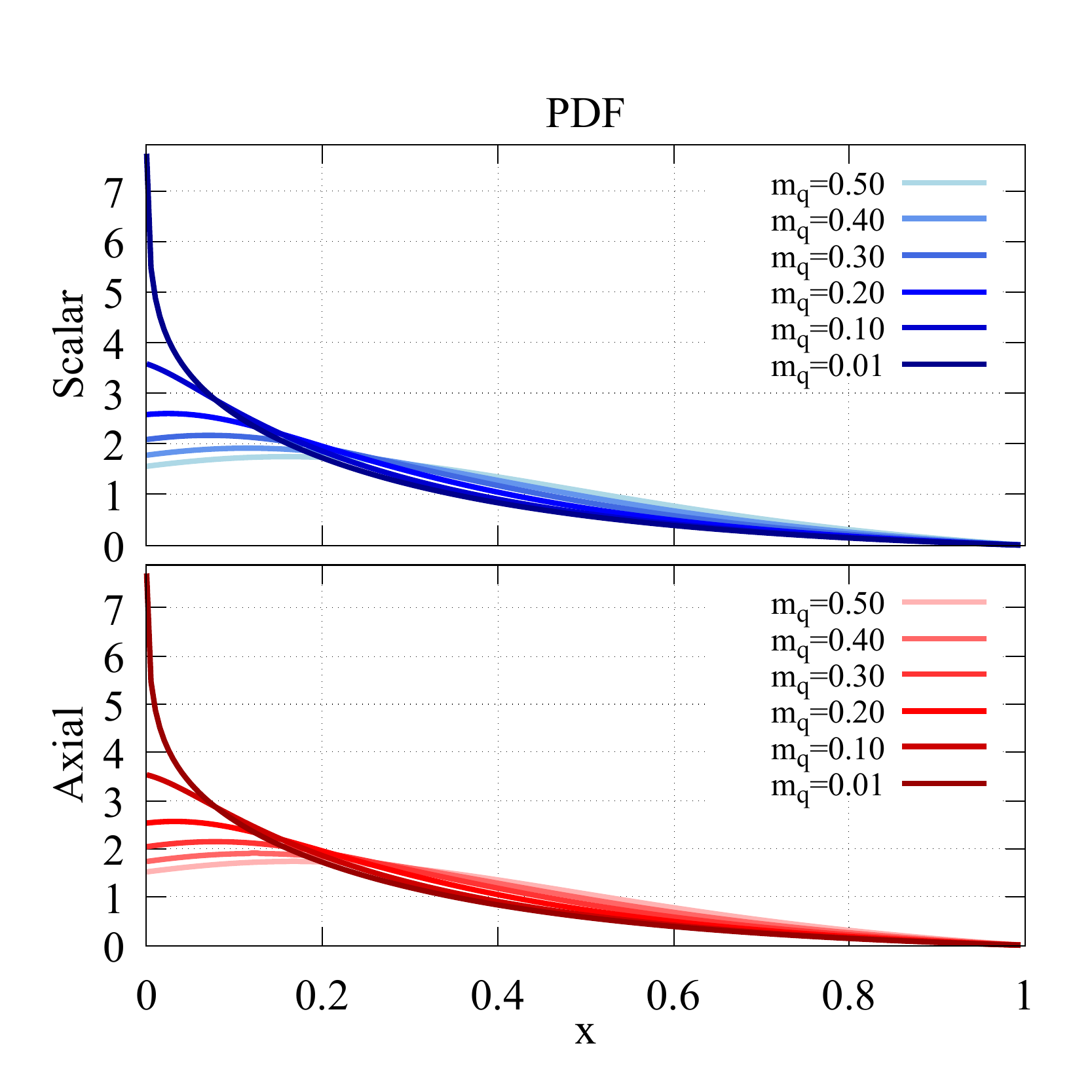}
    \end{subfigure}
\hspace{-0.7cm}
    \begin{subfigure}[b]{0.51\linewidth}
        \centering
        \includegraphics[width=\linewidth]{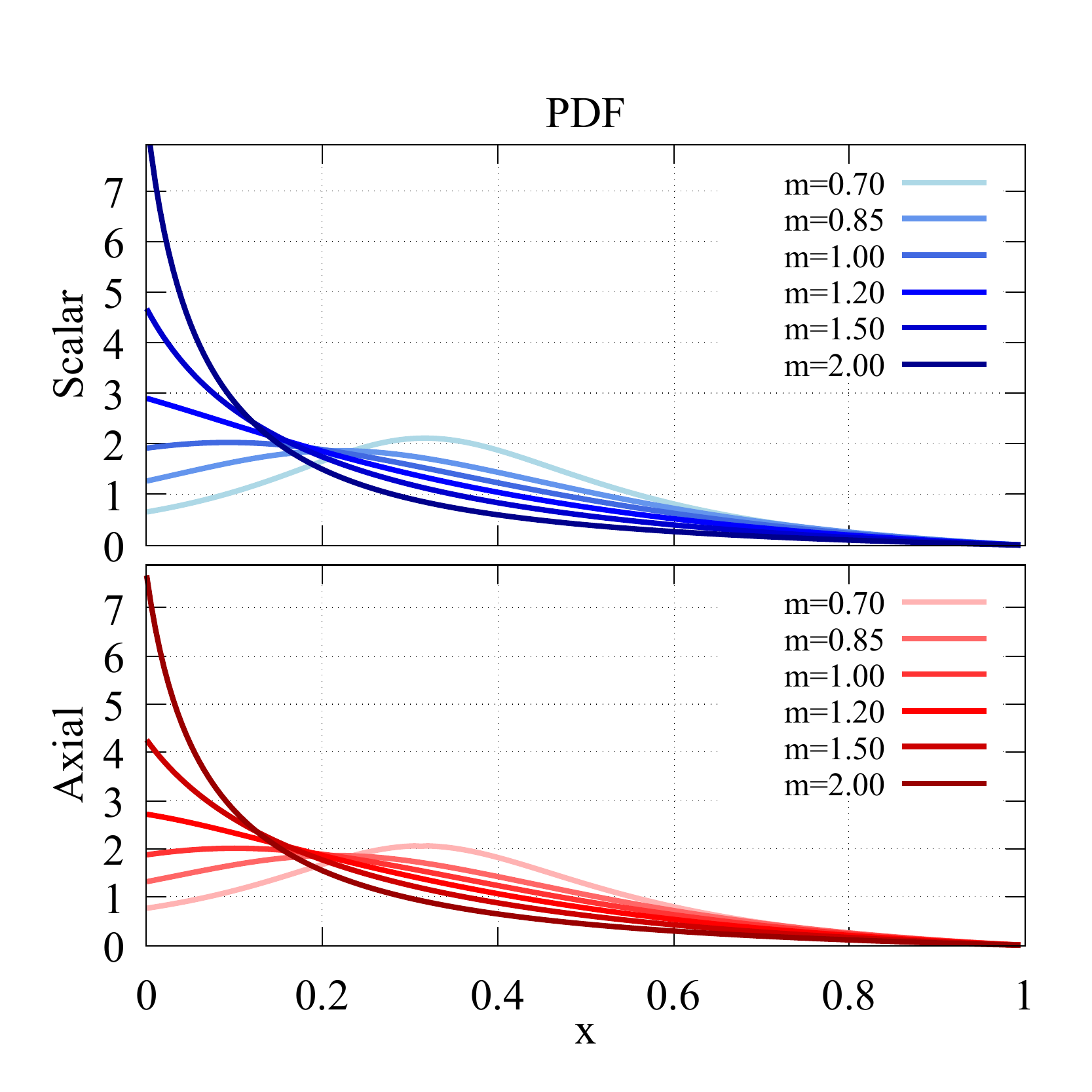}
    \end{subfigure}
    \vspace{-0.5cm}
    \caption{PDF as a function of $x$ for $\beta=1$ GeV and $M=0.939$ GeV. Left panel: results for different values of $m_q$ ($m_s=m_a=1$ GeV). Right panel: results for different values of $m_s=m_a=m$ ($m_q=0.35$ GeV). The mass values in the legend are given in GeV.}
    \label{fig:PDF_mq_m}
\end{figure}

\begin{figure}[htbp]
    \centering
    \begin{subfigure}[b]{0.51\linewidth}
        \centering
        \includegraphics[width=\linewidth]{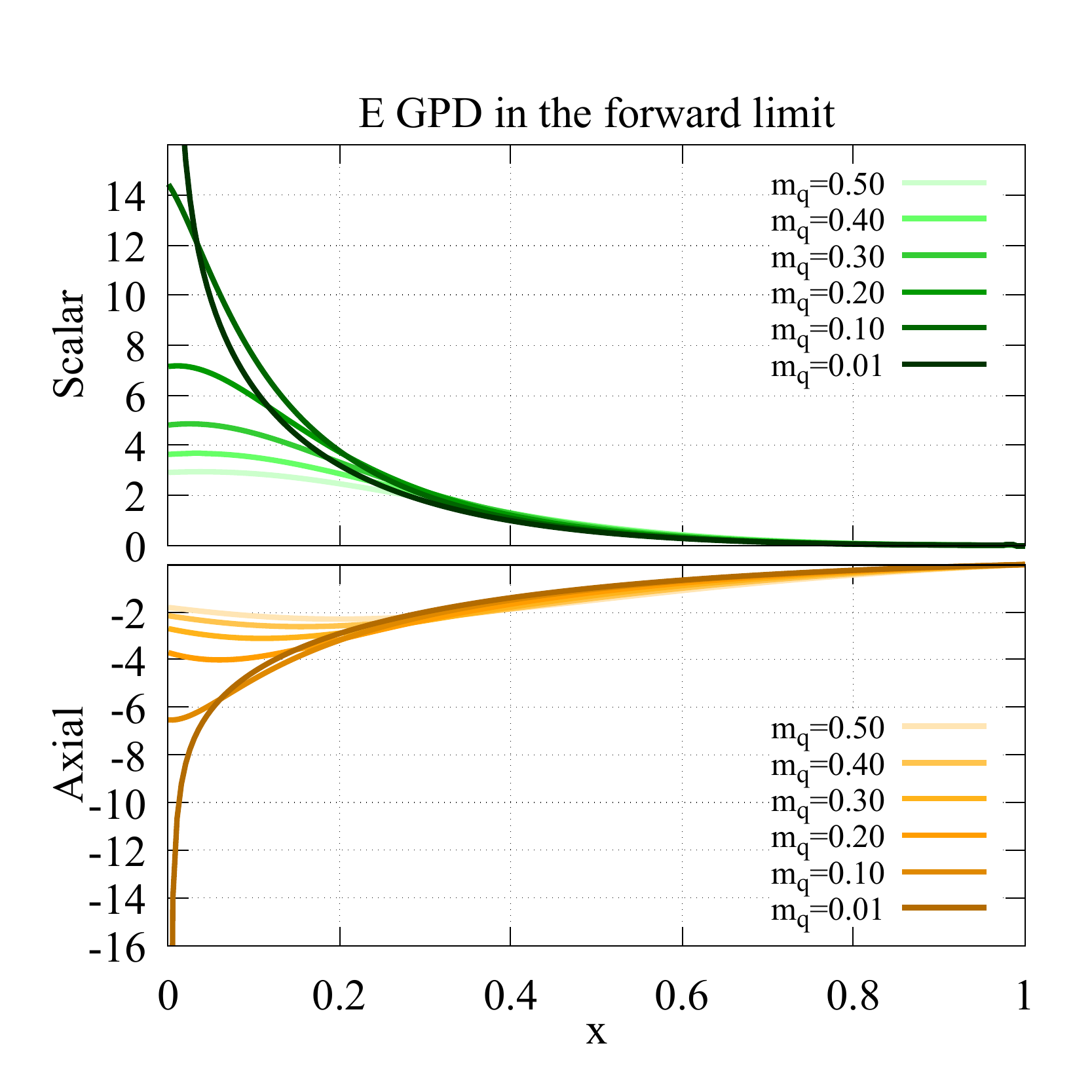}
    \end{subfigure}
\hspace{-0.7cm}
    \begin{subfigure}[b]{0.51\linewidth}
        \centering
        \includegraphics[width=\linewidth]{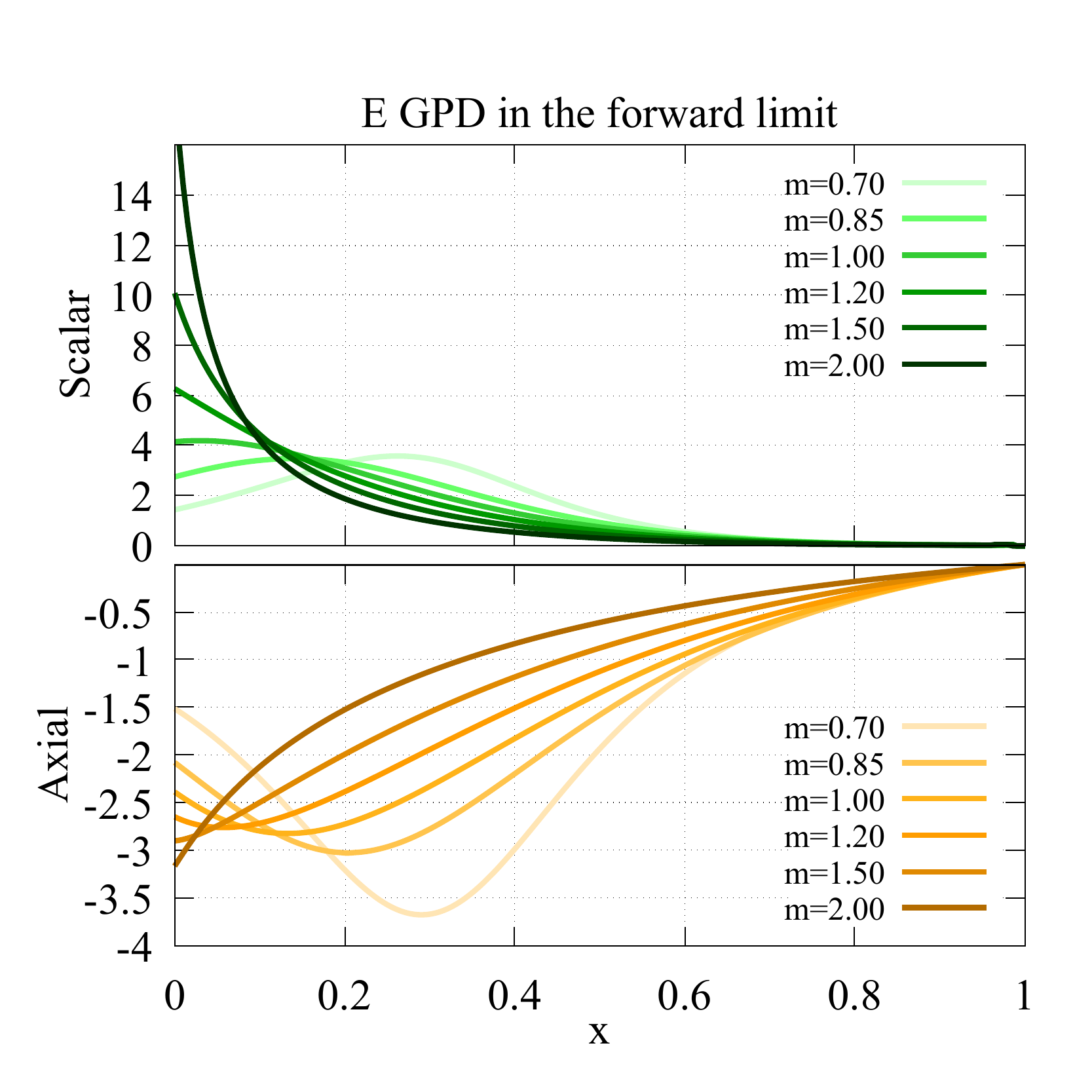}
    \end{subfigure}
    \vspace{-0.5cm}
    \caption{Zero momentum transfer $E$ GPD (forward limit) as a function of $x$ for $\beta=1$ GeV and $M=0.939$ GeV. Left panel: results for different values of $m_q$ ($m_s=m_a=1$ GeV). Right panel: results for different values of $m_s=m_a=m$ ($m_q=0.35$ GeV). The mass values in the legend are given in GeV.}
    \label{fig:E_FL_mqm}
\end{figure}

\begin{figure}[t]
    \centering
    \begin{subfigure}[b]{0.51\linewidth}
        \centering
        \includegraphics[width=\linewidth]{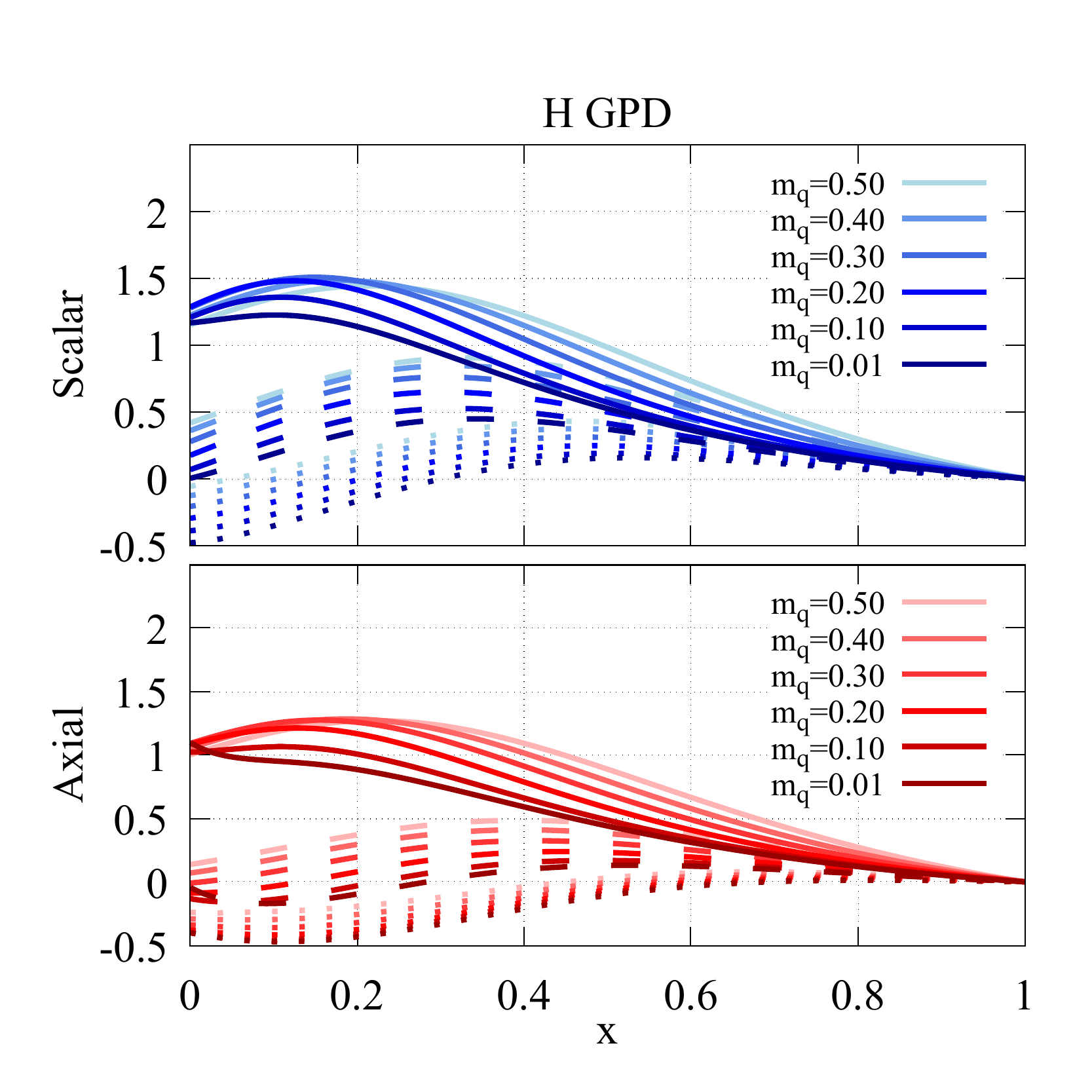}
    \end{subfigure}
\hspace{-0.7cm}
    \begin{subfigure}[b]{0.51\linewidth}
        \centering
        \includegraphics[width=\linewidth]{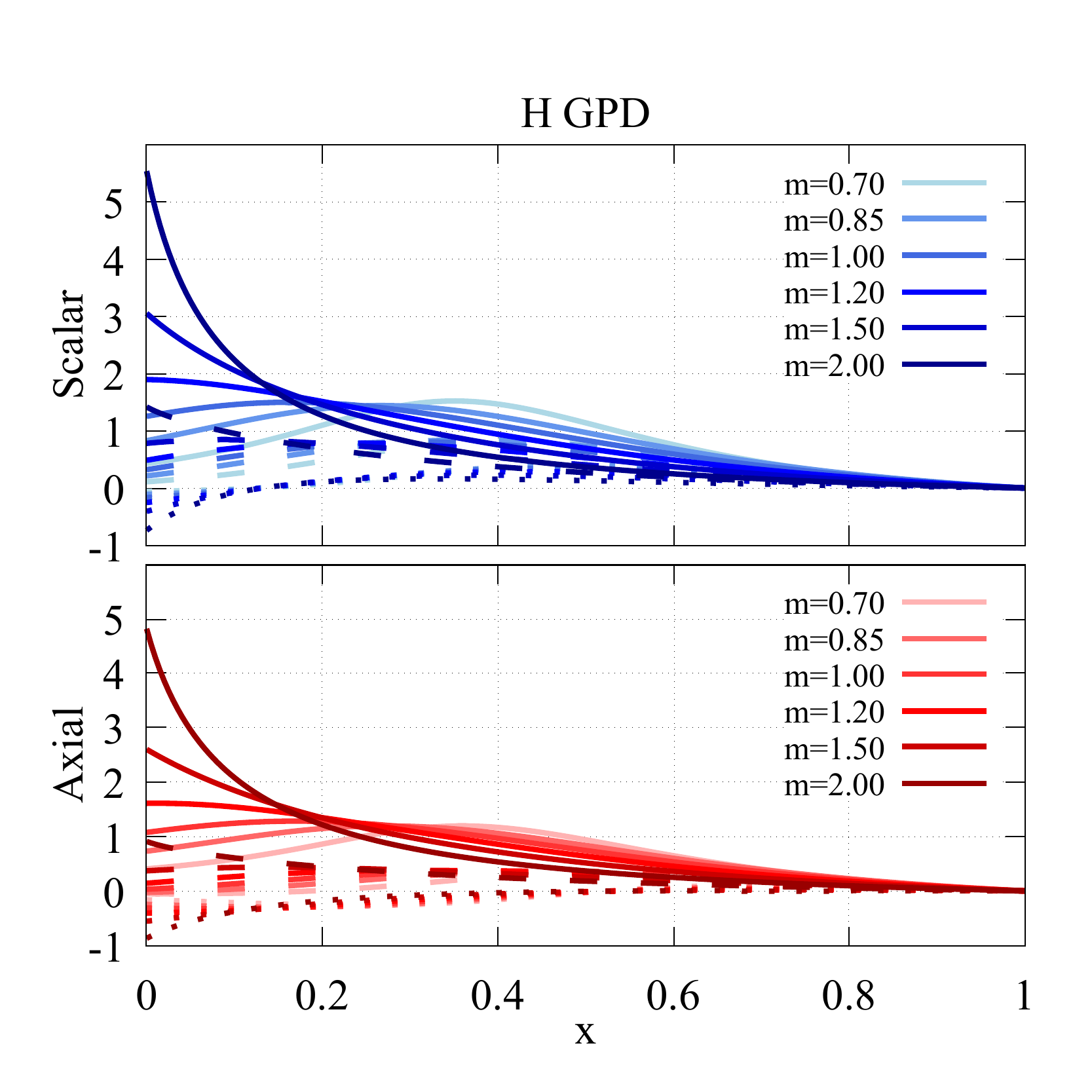}
    \end{subfigure}
    \vspace{-0.5cm}
    \caption{$H$ GPD as a function of $x$ for $\beta = 1$ GeV, $M = 0.939$ GeV, $|\vec\Delta_\bot|=0.5$ GeV (solid line), $|\vec\Delta_\bot|=1.0$ GeV (dashed line), $|\vec\Delta_\bot| = 1.5$ GeV (dotted line). Left panel: results for different values of $m_q$ ($m_s=m_a=1$ GeV). Right panel: results for different values of $m_s=m_a=m$ ($m_q=0.35$ GeV). The mass values in the legend are given in GeV.}
    \label{fig:H_Delta}
\end{figure}

We begin the numerical section by presenting the general characteristics of the GPDs. It is worth noting that the integral expressions given by \Cref{eq:H_GPD_scalar} describing GPDs (both scalar/axial and $H$/$E$) for zero skewness can be solved analytically. The resulting expressions are quite complex, so we define a few auxiliary quantities:
\begin{equation}
    \label{eq:common_definitions_alpha}
\alpha = (1-x)\sqrt{\vec\Delta_\bot^2},
\end{equation}
\begin{equation}
    \label{eq:common_definitions_D}
    D^d=\sqrt{(1-x)m_q^2 + \frac{\alpha^2}{4}  + x m_d^2 - (1-x) x M^2}\, ,
\end{equation}
\begin{equation}
    \label{eq:common_definitions_W}
W^d = \sqrt{\left(D^d + \Lambda^2\right)^2 -\alpha^2\Lambda^2}\, ,
\end{equation}
where $d$ stands for the scalar (S) or axial-vector (A) diquark.
Combining these expressions yields the basic functions used to generate analytical results for GPDs with zero skewness, given by
\begin{equation}
    \label{eq:ISA}
    I_1^{d} = \ln\left[ \frac{\left(D^{d}\right)^2 + \Lambda^2 - \alpha^2/2 +W^{d}}{2\left(D^{d}\right)^2 - \alpha^2/2} \right] ,\hspace{0.3cm}
    I_2^{d} = \frac{1}{\alpha D^{d}} \ln \left[ 
    \frac{(2 D^{d} + \alpha) (\left(D^{d}\right)^2 + \Lambda^2) }{\alpha (\left(D^{d}\right)^2 - \Lambda^2) + 2D^{d} W^{d}}
    \right].
\end{equation}
The results for the scalar $H$ GPD and $E$ GPD can then be written as follows:
\begin{equation}
    H^S(x,0,t) = \frac{g^2(1-x)}{16\pi^2}\left(I^S_1 + C_1^SI^S_2 \right), \hspace{1cm}
    E^S(x,0,t) =  \frac{g^2(1-x)}{16\pi^2} C_2^S I_2^S,
\end{equation}
where
\begin{equation}
    C_1^S = (m_q+Mx)^2 - \frac{\alpha^2}{4} - D^2_S, \hspace{1cm} C_2^S = 2(1-x)M(m_q+xM).
\end{equation}
In the forward limit, these expressions simplify considerably and can be written as:
\begin{equation}
\begin{split}
   H^S(x,0,0) &= q^S(x)=\frac{g^2(1-x)}{16\pi^2}\left( 
    \ln \left[\frac{(D^{S0})^2+\beta^2}{(D^{S0})^2} \right]
    +
    \frac{(m_q+xM)^2-(D^{S0})^2}{(D^{S0})^2((D^{S0})^2+\beta^2)} \beta^2
    \right), \\
    E^S(x,0,0)&=e_q^S(x) = \frac{g^2(1-x)}{16\pi^2}\frac{2(1-x)M(m_q+xM)\beta^2}{(D^{S0})^2((D^{S0})^2+\beta^2)},
\end{split}
\end{equation}
where $D^{S0} =\sqrt{ (1-x)m_q^2+xm_s^2-(1-x)xM^2}$.  In the same way, one can calculate the GPDs for the axial-vector diquark, including their forward limit form. Defining:
\begin{equation}
\begin{split}
    C^A_1 &= \frac{1}{2}\left(2+\frac{M^2}{m_a^2} - \frac{\vec\Delta^2_\bot}{4m_a^2} \right), \\
    C^A_2 & = \frac{1}{2}\Bigg[\left(2+\frac{M^2}{m_a^2} - \frac{\vec\Delta^2_\bot}{4m_a^2}\right)
    (m_q^2+x^2M^2)+2\left(4x -\frac{xM^2}{m_a^2} - \frac{(2+x)\vec\Delta^2_\bot}{4m_a^2}
    \right)m_qM \\
     &\hspace{2.8cm}-\frac{\Delta^2_\bot}{4}(1+x)
     \left(2(1+x)+(1-3x)\frac{M^2}{m^2_a} - (1+x)\frac{\vec\Delta^2_\bot}{4m_a^2} \right)\Bigg],\\
     C^A_3 &= -\frac{M^2}{m_a^2},\\
     C^A_4 &=\frac{1}{2}\Bigg[ -\frac{2m_q^2M^2 }{m_a^2}+2m_qM\left( 
     \frac{M^2(x-1)}{m_a^2} + \frac{\vec\Delta^2_\bot(1+x)}{4m_a^2} -4x
     \right)-4M^2x(x+1)\\
     &\hspace{7cm} + \frac{2xM^2}{m_a^2}\left(
     M^2+\frac{\vec\Delta^2_\bot}{4}
     \right) + \frac{M^2\vec\Delta^2_\bot}{2m_a^2}
     \Bigg],
\end{split}
\end{equation}
the $H$ and $E$ axial GPDs can then be written as:
\begin{equation}
\begin{split}
    H^A(x,0,t) = \frac{g^2(1-x)}{16\pi^2}\left[ C^A_1I^A_1+(C^A_2-C^A_1D^2_A)I^A_2\right],\\
    E^A(x,0,t) = \frac{g^2(1-x)}{16\pi^2}\left[
    C^
    A_3I^A_1+(C^A_4-C^A_3D^2_A)I^A_2
    \right].
    \end{split}
\end{equation}
The forward limit form is not as straightforward as in the case of scalar GPDs and still requires the use of auxiliary functions $C^A_{n}$; for $\vec\Delta^2_\bot=0$, these functions will be denoted as $C^A_{n0}$. Using this notation, the forward limit of axial-vector diquark GPDs can be written as:
\begin{equation}
\begin{split}
    H^A(x,0,0) = q^A(x) = \frac{g^2(1-x)}{16\pi^2}
    \left[ 
    C^A_{10} \ln\left[ \frac{(D^{A0})^2+\beta^2}{(D^{A0})^2}   \right]
    +
    \frac{C^A_{20} -C^A_{10} (D^{A0})^2 }{(D^{A0})^2 ((D^{A0})^2+\beta^2)}\beta^2
    \right], \\
        E^A(x,0,0) = e_q^A(x) = \frac{g^2(1-x)}{16\pi^2}
    \left[ C^A_{30}
     \ln\left[ \frac{(D^{A0})^2+\beta^2}{(D^{A0})^2}   \right]
    +
    \frac{C^A_{40} -C^A_{30}(D^{A0})^2 }{(D^{A0})^2 ((D^{A0})^2+\beta^2)}\beta^2
    \right] ,
\end{split}
\end{equation}
\begin{figure}[h!]
    \centering
    \begin{subfigure}[b]{0.49\linewidth}
        \centering
        \includegraphics[width=\linewidth]{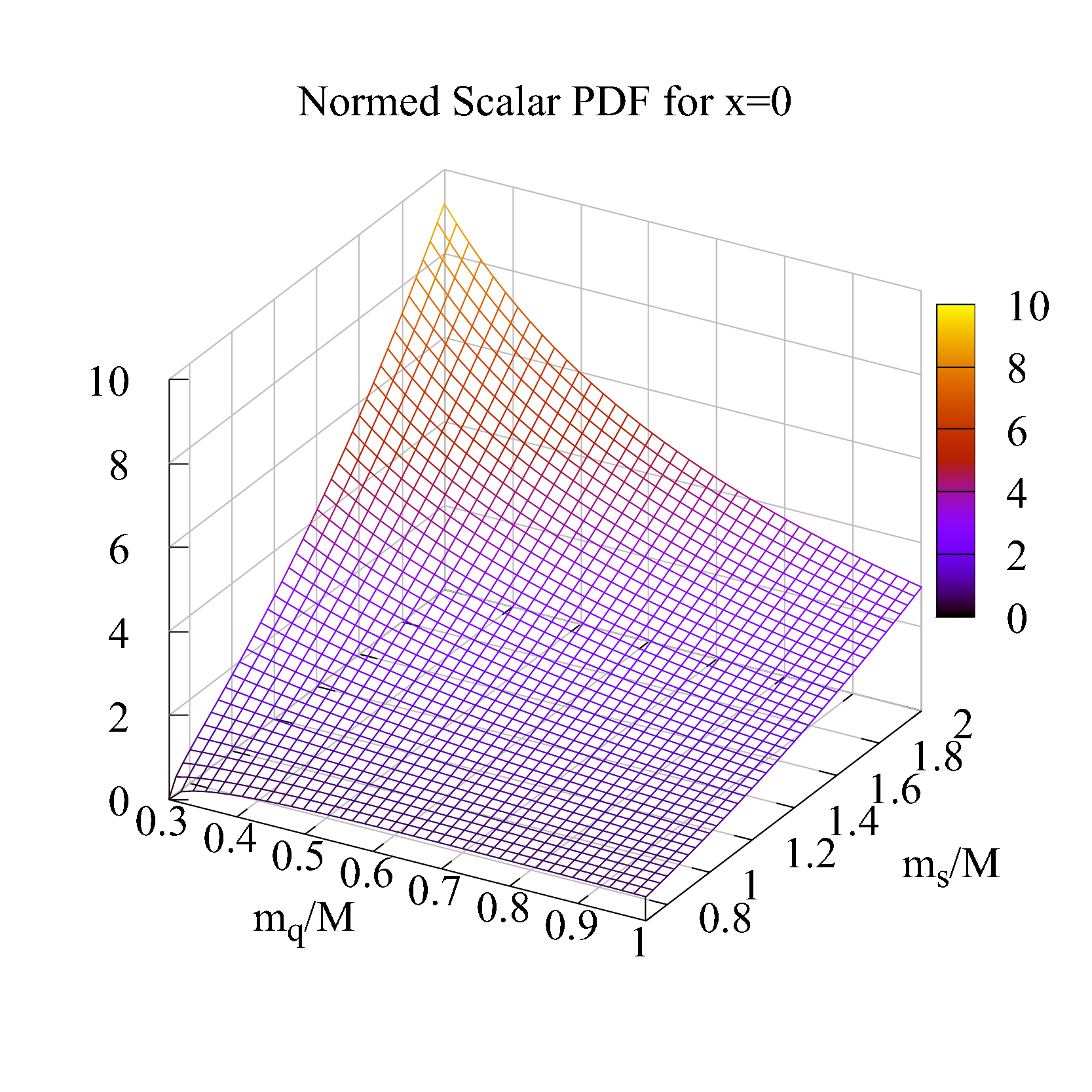}
    \end{subfigure}
\hspace{-0.5cm}
    \begin{subfigure}[b]{0.49\linewidth}
        \centering
        \includegraphics[width=\linewidth]{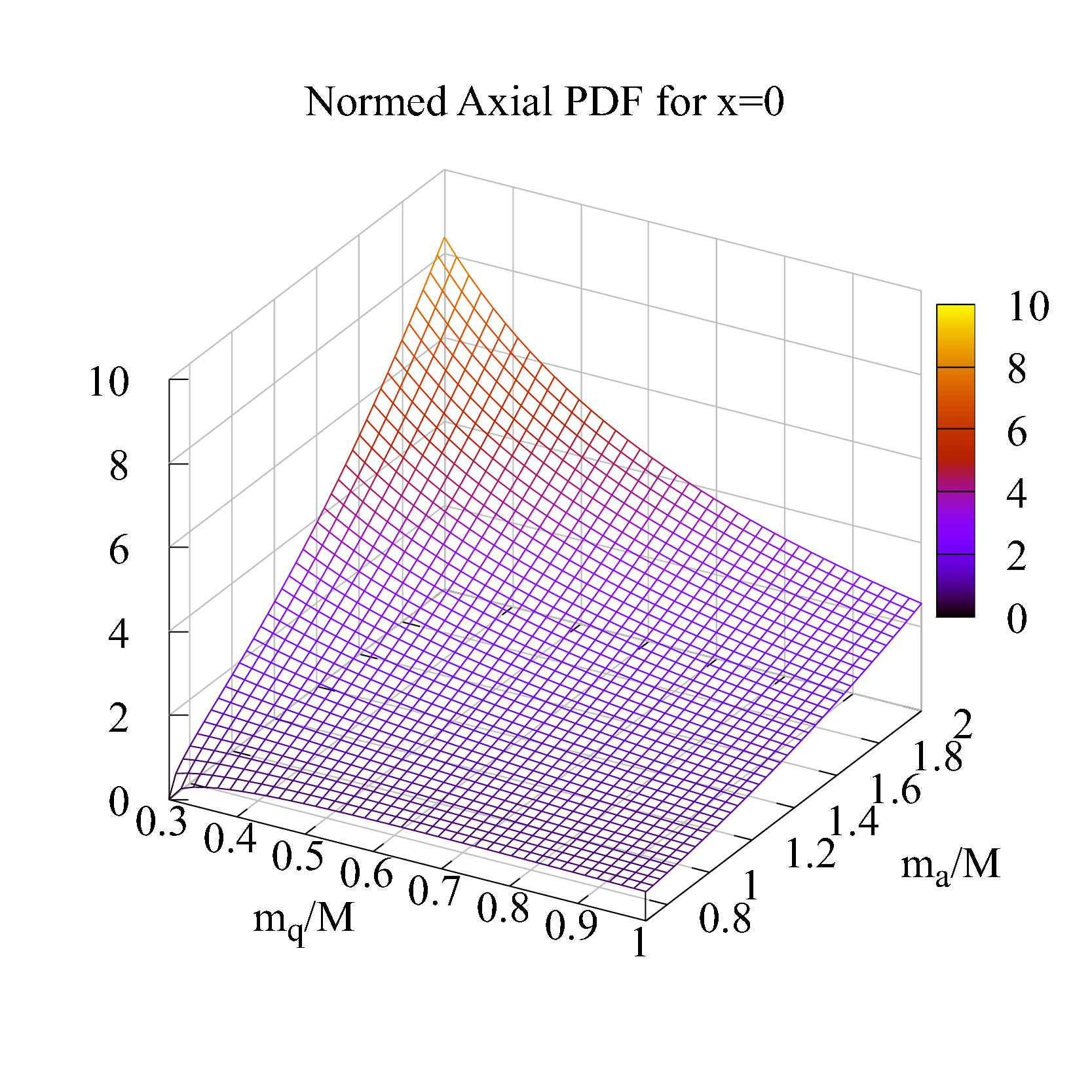}
    \end{subfigure}
    \vspace{-0.5cm}
    \caption{Normalized scalar PDF (left panel) and normalized axial PDF (right panel) for $x=0$, as a function of $m_q/M$ and $m_{s/a}/M$ for $M=1$ GeV, $\beta=1$ GeV.}
    \label{fig:norm}
\end{figure}
where $D^{A0} =\sqrt{ (1-x)m_q^2+xm_a^2-(1-x)xM^2}$. This formalism is useful not only from the point of view of numerical calculations, but also aids in the analysis of the behavior of zero-skewness GPDs. In particular, it allows us to study analytically the endpoint behavior
of the zero-skewness GPDs. The limit $x\to 0^+$ determines whether the
model can reproduce the steep rise of the valence-quark PDFs, whereas
the limit $x\to 1$ is essential for a consistent tomographic
interpretation, as it requires the $t$ dependence of the GPDs to vanish (see \Cref{eq:H_lim_1} and \Cref{eq:E_lim_1}). Later in this section,
we therefore examine the effect of the model parameters on the PDFs and
on the zero-skewness GPDs. This is particularly important for the $H$ GPDs in the forward limit (both axial and scalar) in the $x \to 0^+$ limit.
Based on experimental data and lattice calculations, we  expect a sharp rise in the PDFs for valence quarks as $x$ approaches zero (see, for instance \cite{Diehl:2013xca}). One might ask whether it is possible to reproduce such behavior in a DSM. For this purpose, one can calculate the forward limits of scalar and axial-vector diquark $H$ GPDs for $x\to0$; this yields the following expressions,
\begin{equation}
    \label{eq:limit_x_zero}
    q^S(x\to 0) = \frac{g^2}{16 \pi^2} \ln\left[\frac{m_q^2+\beta^2}{m_q^2}\right] ,
    \hspace{0.5cm}
    q^A(x\to 0 ) =  \frac{g^2}{16 \pi^2} \frac{1}{2}\left(2+\frac{M^2}{m_a^2}\right) \ln\left[\frac{m_q^2+\beta^2}{m_q^2}\right] .
\end{equation}
It might seem that, in the case of a scalar diquark, the result for $x$ close to zero is greater the smaller the value of $m_q$, whereas in the case of an axial-vector diquark, a smaller diquark mass $m_a$ also increases the values as $x$ approaches zero. However, we need to take normalization into account here, as the increase in value at a given point tells us nothing about the overall trend of the normalized distribution. Thus, if the integral from $x=0$ to $x=1$ of the forward limit scalar (or axial) GPD increases significantly upon changing a mass parameter, the increase in the normalization will reduce the value of the function at $x \to 0$, conversely, a decreased integral may lead to an increased value for the $x\to 0$ limit accordingly. \Cref{fig:norm} shows how the value at $x=0$ for the scalar and axial forward limits varies as a function of the ratio of the quark and the diquark masses to the proton mass. 
As can be seen, the PDF value at $x=0$ increases as $m_q$ decreases and $m_d$ increases, and the behaviors for scalar and axial PDFs are very similar. It is important to note that these observations suggest that, in the DSM, we cannot obtain a genuine divergence as $x$ approaches zero, except in cases where the valence quark (or the axial-vector diquark, for the axial case) has zero mass. This is because \Cref{eq:limit_x_zero} shows that a divergence in the unnormalized forward limit can occur only when the relevant mass parameter vanishes. Since the normalization integral remains finite, normalizing the distribution cannot introduce a divergence that is absent in the unnormalized function. Therefore, the normalized forward-limit GPDs also remains finite as $x\to 0$, provided that the relevant mass parameter is nonzero. 
So, with the right choice of parameters, we can obtain any value we like at $x=0$, but we can only mimic the true divergence, which we will not actually obtain. The PDFs for the scalar and axial-vector diquarks over the full range of $x$ are shown in \Cref{fig:PDF_mq_m}.   
As can be seen, the trends anticipated from \Cref{eq:limit_x_zero}: decreasing $m_q$ or increasing $m_{s/a}$ enhances the PDFs at small $x$. Obviously, the normalization condition implies that the small-$x$ enhancement is accompanied by PDF suppression at intermediate $x$. The curves corresponding to different masses cross at $x \approx 0.2$ and for lighter diquarks, the maximum of the PDF moves towards larger $x$. What is interesting, we observe a similar behavior for the $E$ GPD in the forward limit as shown in \Cref{fig:E_FL_mqm}, where it is important to notice that the $E$ axial GPD (in this limit) has a negative value over the entire $x$ range, regardless of the masses. In Appendix \ref{app:convergence}, we present analogous results for nonzero skewness GPDs.
 We analyzed only the forward limit GPDs for scalar and axial-vector diquarks, and not directly the forward-limit GPDs for up and down quarks. However, as stated above, the up quark GPDs can be written as a sum of scalar and axial GPDs, and the down quark functions are the same as for the axial case. So if scalar and axial PDFs increase significantly as $x$ approaches zero, then the up and the down quark PDFs will show the same behavior. The behavior of the $E$ GPDs is more complex because they are normalized not by themselves but by the forward limit $H$ GPDs. Moreover, because of the negative value of the forward limit axial $E$ GPD, one needs to be careful when transferring interpretations from the scalar and axial $E$ GPDs to the up quark $E$ GPD.
For the purposes of tomography, it is essential that the following equation is satisfied:
\begin{equation}
    \label{eq:H_lim_1}
    \lim_{x\to1}\frac{d}{dt}H(x,0,t)=0,
\end{equation}
which is derived from \Cref{eq:unpolarized_tomography} and describes a situation in which the proton’s entire momentum is carried by a single parton, so this parton is in the center of mass of the system. This, therefore, requires that the dependence on $t$ vanishes as $x \to 1$. For this reason, $q_{x=1}(\vec{b}_
\bot) = \delta(\vec{b}_\bot)$, which is ensured by \Cref{eq:H_lim_1}. The analytical form of the $H$ GPDs obtained for scalar and axial-vector diquark allows us to demonstrate analytically that this condition is always strictly satisfied.
For both axial-vector and scalar diquarks, there is a term $g^2(1-x)/(16\pi^2)$  that does not depend on $t$. It is therefore sufficient to show that, after taking the derivative with respect to $t$, the remaining terms do not contain a denominator factor that approaches zero faster.  When the calculations are carried out, the condition given by \Cref{eq:H_lim_1} is always satisfied within the physical region. This can also be seen from the calculation results presented in \Cref{fig:H_Delta}.
\begin{figure}[t]
    \centering
    \begin{subfigure}[b]{0.51\linewidth}
        \centering
        \includegraphics[width=\linewidth]{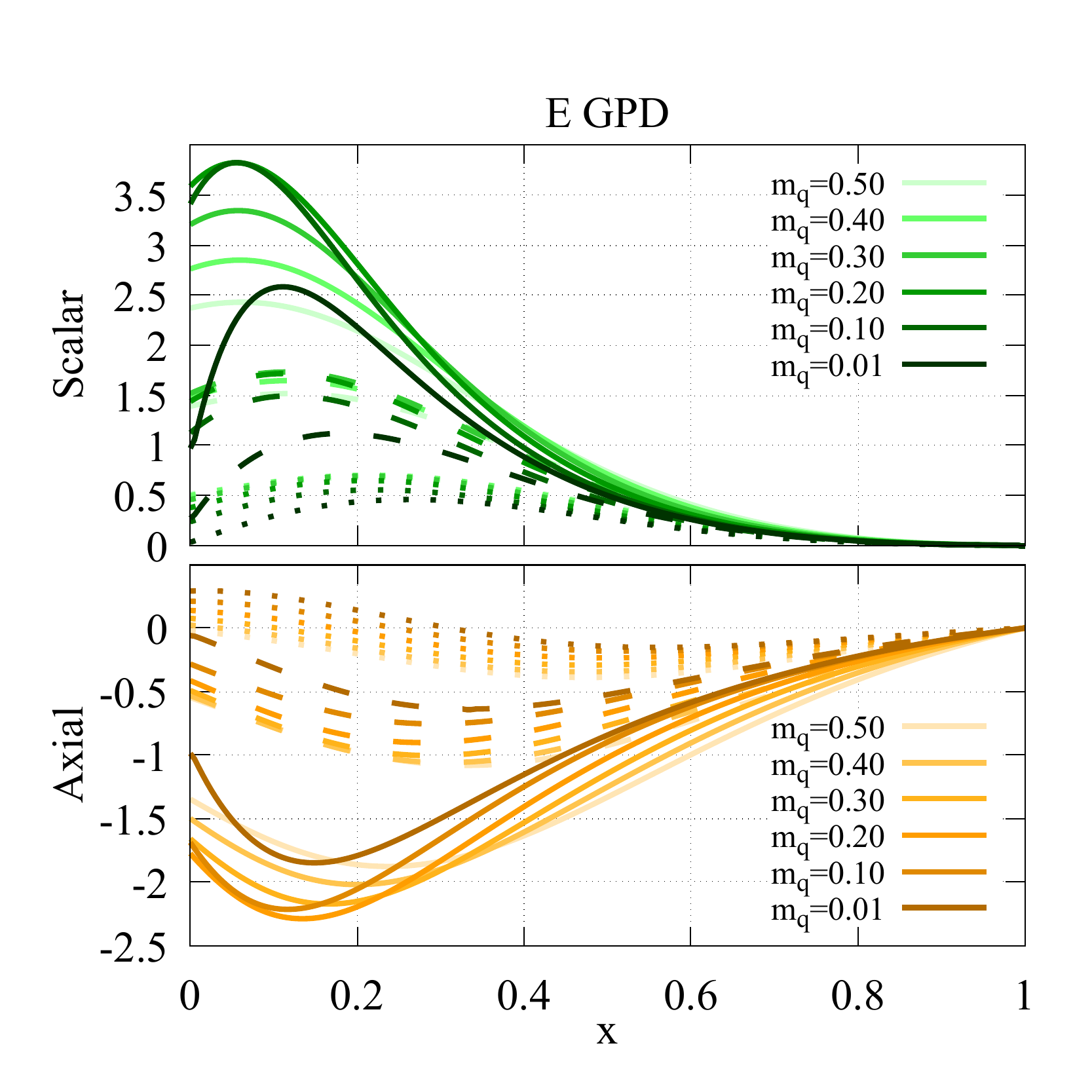}
    \end{subfigure}
\hspace{-0.7cm}
    \begin{subfigure}[b]{0.51\linewidth}
        \centering
        \includegraphics[width=\linewidth]{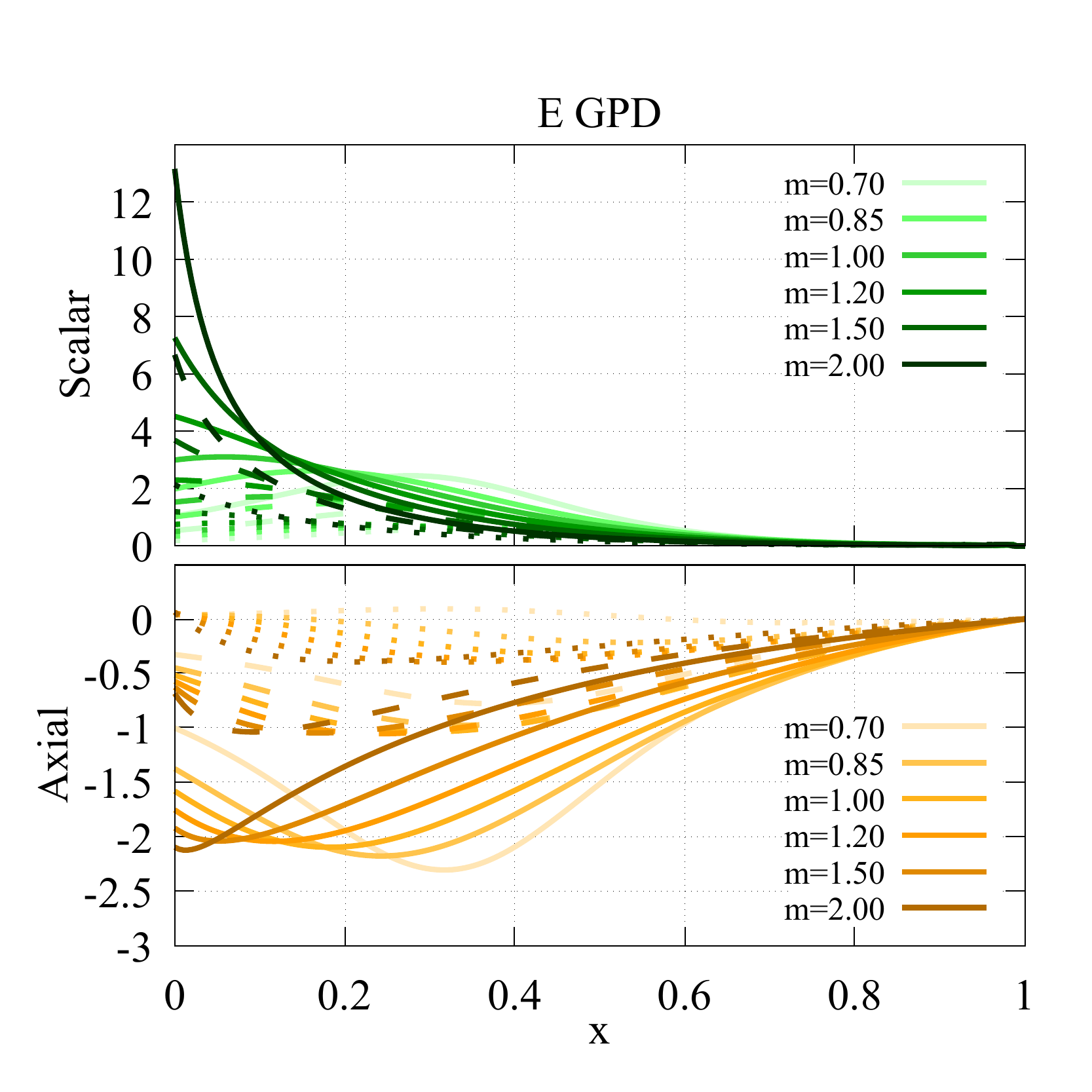}
    \end{subfigure}
    \vspace{-0.5cm}
    \caption{$E$ GPD as a function of $x$ for $\beta = 1$ GeV, $M = 0.939$ GeV, $|\vec\Delta_\bot|=0.5$ GeV (solid line), $|\vec\Delta_\bot|=1.0$ GeV (dashed line), $|\vec\Delta_\bot| = 1.5$ GeV (dotted line). Left panel: results for different values of $m_q$ ($m_s=m_a=1$ GeV). Right panel: results for different values of $m_s=m_a=m$ ($m_q=0.35$ GeV). The mass values in the legend are given in GeV.}
    \label{fig:E_delta}
\end{figure}
It is also interesting to note that the same behaviour as $x\to 1$ can be observed for the $E$ GPD, as shown in \Cref{fig:E_delta}. This means that, in the case of transverse polarization, there should be no shift in the distribution in $\vec{b}_\bot$ space when all the momentum is carried by a single parton, so the condition
\begin{equation}
\label{eq:E_lim_1}
    \lim_{x\to 1} \frac{d}{dt}E(x,0,t)=0,
\end{equation}
is also satisfied. As for the $H$ GPDs, this equality is proven analytically for both scalar and axial-vector diquarks.
\subsection{Calibration of the model}
In this work, we calibrate the DSM in two ways: using proton observables only, and using proton and neutron observables simultaneously. As we show below, the proton+neutron calibration can be regarded as a refinement of the proton-only one. The free parameters are:
\begin{align*}
    \text{quark mass} && \text{diquark masses} && \text{decomposition constant} && \sqrt{\text{average $k^2_\bot$}}\\
     m_q, \hspace{0.55cm} && m_d \equiv \{m_s, m_a\},\hspace{-0.15cm} && \lambda, \hspace{1.75cm} && \beta, \hspace{0.75cm}
\end{align*}
as free parameters of the model. For all the numerical calculations, we use $M=0.939$ GeV while $g_{\pm}$ is determined by the normalization condition. The model has been parametrized in terms of:
\begin{align}
 G_M^{p/n}(0), &&r_E^{p}, &&\left<r_E^2\right>^n,  &&r_M^{p/n}, && A(0), && J(0).
\end{align}
Radii without a square denote $r = \sqrt{\langle r^2 \rangle}$.
The values $G_E^p(0)=1$ and $G_E^n(0)=0$ follow from the considerations
above and hold regardless of the model parameters, so these quantities
are not included in the fitting procedure. In the fitting procedure, we minimize the maximum deviation from the reference values rather than the mean deviation. This choice ensures a satisfactory fit for every parameter, rather than fitting some parameters very well at the expense of others. We estimate the model uncertainty as follows: instead of fitting to a single set of values, we fit to values drawn from a Gaussian distribution centered on the reference value, with the uncertainty treated as one standard deviation. Thus, for a quantity $\chi_0$ with uncertainty $\Delta\chi$, written as $\chi_0 \pm \Delta\chi$ (or equivalently $\chi_0(\Delta\chi)$), the fitting procedure draws a value randomly from a Gaussian distribution around $\chi_0$, with the probability of choosing a value $\chi$ given by:
\begin{equation}
    \text{probability} = \frac{1}{\Delta\chi\sqrt{2\pi}}\, \exp{\left[-\frac{1}{2} \left(\frac{\chi - \chi_0}{\Delta\chi} \right)^2 \right]}\, .
\end{equation}
Values closer to the central value $\chi_0$ are therefore more likely to be generated during the fitting procedure. This procedure leads to potentially asymmetric uncertainties, which we denote as ${\chi_0}^{(+\Delta\chi)}_{(-\Delta \chi)}$ relative to the central value, where the central value is the value obtained from the fit to $\chi_0$, and $(\pm\Delta\chi)$ are the uncertainties corresponding to values greater than or less than the central value. The reference values of the nucleon electromagnetic observables used for comparison with the calculated results are taken from the Particle Data Group (PDG) compilation \cite{PDG2024}. The adopted values are:
\[
r_E^p = 0.8409(4)\,\mathrm{fm}, \quad
r_M^p = 0.851(26)\,\mathrm{fm}, \quad
r_M^n = 0.864(9)\,\mathrm{fm},
\]
\[
\left<r_E^2\right>^n = -0.1155(17)\,\mathrm{fm}^2,
\quad
G_M^p(0) = 2.79284734463(82),
\quad
G_M^n(0) = -1.91304276(45)\, .
\] 
For the GFFs, we adopt reference values approximately corresponding to
the lattice results of Hackett et al. \cite{PhysRevLett.132.251904}, namely
$A(0)=0.50(5)$ and $J(0)=0.250(25)$. Since these are rounded values
rather than the exact lattice numbers, we assign them a 10\% uncertainty. 
\[
A(0) = 0.50(5), \qquad J(0) = 0.250(25)\, .
\]
The uncertainty bands shown in the subsequent figures are constructed as follows. We perform the fitting procedure described above, drawing the reference values from Gaussian distributions centered on the adopted central values, and compute for each fit the value of the minimax criterion introduced above. We then sort the fits by increasing value of this criterion and retain the first $68\%$ of them, corresponding to approximately one standard deviation. The retained fits are plotted together, and the uncertainty band is defined as the area enclosed between the extremal curves of this set at each value of the relevant variable. The central curve corresponds to the fit to the exact reference values, i.e., the fit in which the reference values were not sampled from Gaussian distributions.

\subsubsection{Proton fit}
\begin{figure}
    \centering
    \includegraphics[width=1\linewidth]{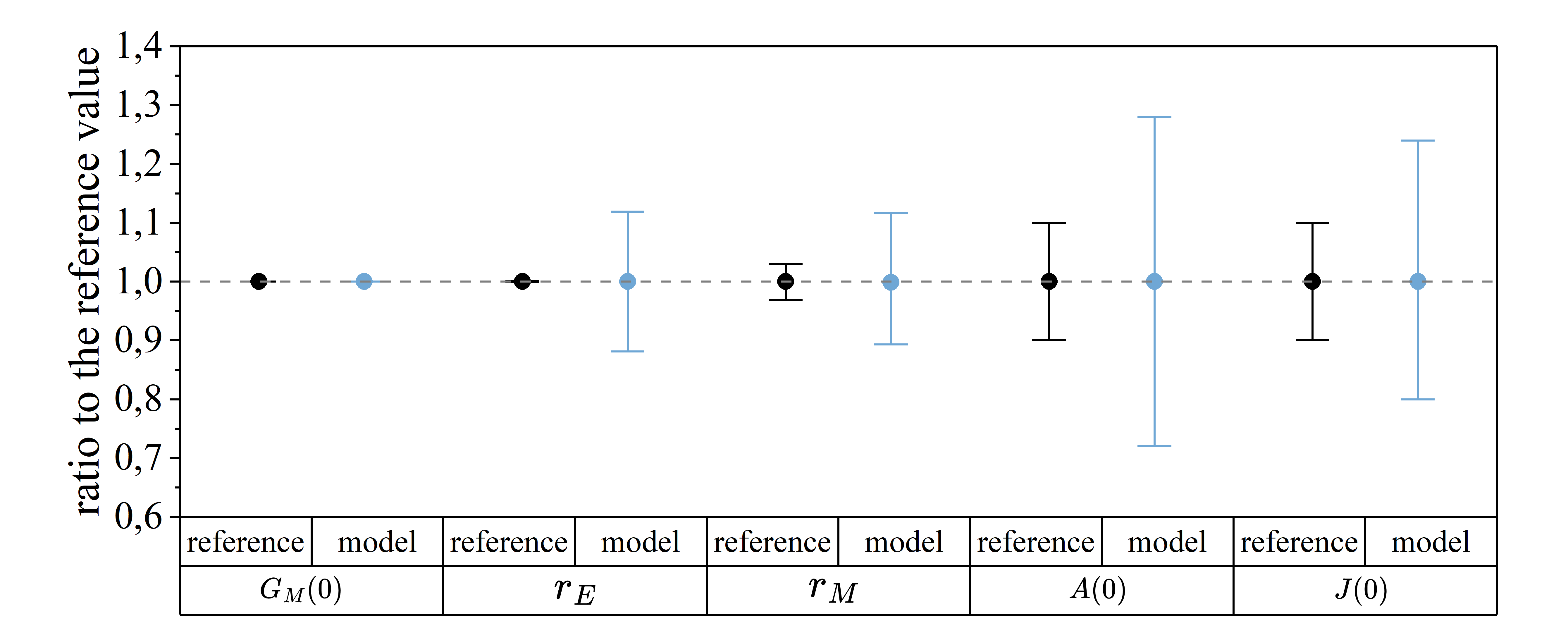}
    \caption{Proton magnetic moment, electromagnetic radii, and GFFs reference values with uncertainties (black) and model results (blue) relative to the reference value. For a fit including only quantities describing the proton.}
    \label{fig:fit_results_proton_proton}
\end{figure}

\begin{figure}
    \centering
    \includegraphics[width=1\linewidth]{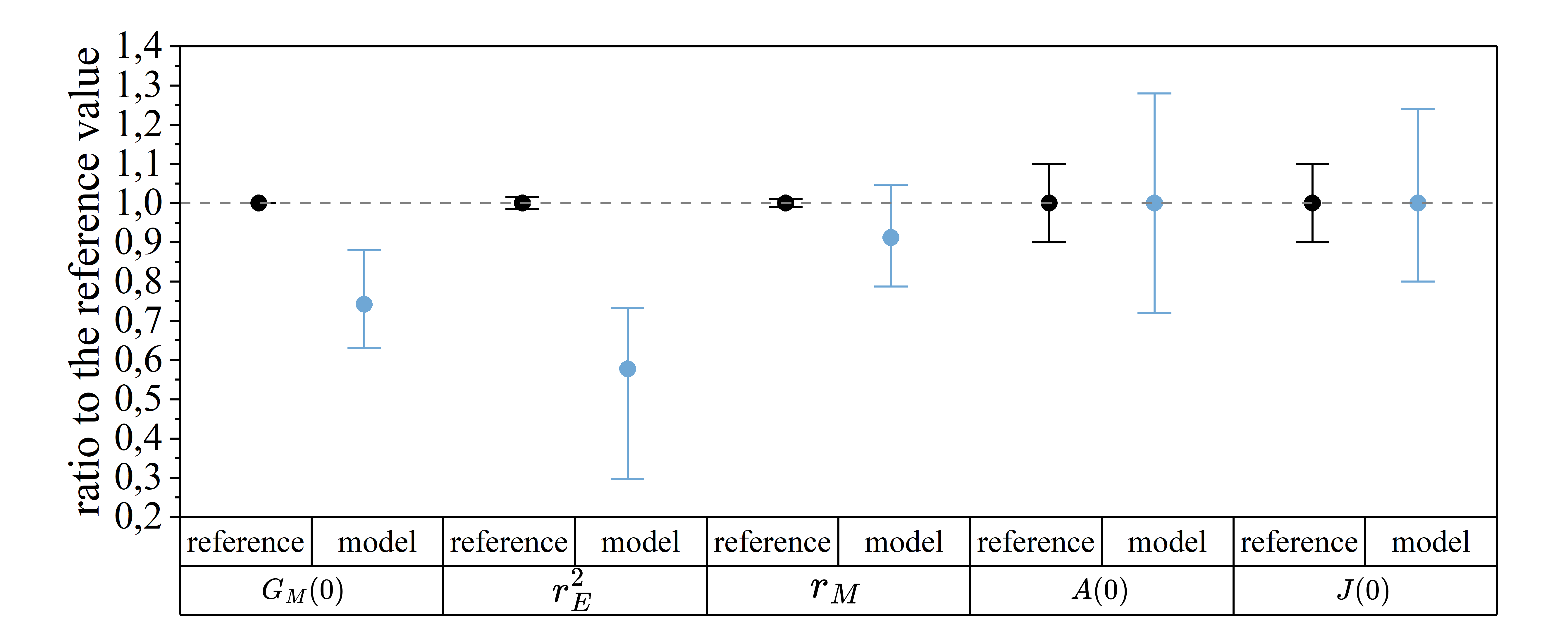}
    \caption{Neutron magnetic moment, electromagnetic radii, and GFFs reference values with uncertainties (black) and model results (blue) relative to the reference value. For a fit including only quantities describing the proton.}
    \label{fig:fit_results_proton_neutron}
\end{figure}

In this section, we present results for the DSM parametrized to the proton only.
In \Cref{tab:results}, we collect all fitted parameters and fit results and compare them with the reference data. 
\begin{table}[h!]
\centering
\caption{Model results compared to reference parameters, for the model parametrized only for the proton.}
\label{tab:results}
\begin{tblr}{
  width = \linewidth,
  colspec = {Q[130]Q[223]Q[190]Q[130]Q[221]},
  cell{2}{1} = {c=3}{0.552\linewidth,c},
  cell{2}{4} = {r=2}{},
  cell{2}{5} = {r=2}{},
  cell{4}{4} = {r=2}{},
  cell{4}{5} = {r=2}{},
  cell{6}{1} = {c=3}{0.552\linewidth,c},
  cell{6}{4} = {r=2}{},
  cell{6}{5} = {r=2}{},
  cell{8}{4} = {r=2}{},
  cell{8}{5} = {r=2}{},
  cell{10}{4} = {r=2}{},
  cell{10}{5} = {r=2}{},
  vline{4} = {1-11}{},
  vline{2} = {2,6}{},
  hline{1-2,12} = {-}{},
  hline{3,6-7,10} = {1-3}{dashed},
}
Observable & Reference value & Model value & Parameter & Calibrated value \\
Proton     &                 &             &\Large $m_q$        &\Large $0.118^{(+62)}_{(-36)} \, \text{GeV}$                 \\
$G_M^p(0)$ & $2.79284734463(82)$  & $2.79279^{(+35)}_{(-23)}$ &           &                  \\
$r_E^p$ & $0.8409(4)\, \text{fm}$ & $0.8409^{(+10)}_{(-10)}\,\text{fm}$ & \Large$m_s$ & \Large$1.30^{(+37)}_{(-32)} \, \text{GeV}$                 \\
$r_M^p$  & $0.851(26)\, \text{fm}$ & $0.85^{(+10)}_{(-9)}\,\text{fm}$ &           &                  \\
Neutron    &                 &             & \Large $
m_a$        &\Large $1.9^{(+1.6)}_{(-0.6)}\, \text{GeV}$                  \\
$G_M^n(0)$ &$-1.91304276(45)$ & $-1.42^{(+26)}_{(-21)}$&           &                  \\
$\left<r_E^2\right>^n$ & $-0.1155(17)\,\text{fm}^2$ & $-0.067^{(+18)}_{(-32)}\,\text{fm}^2$             & \Large$\lambda$    & \Large $1.41^{(+22)}_{(-13)}$                \\
$r_M^n$  & $0.864(9)\, \text{fm}$& $0.79^{(+12)}_{(-11)}\, \text{fm}$ &           &                  \\
$A(0)$       &  $0.50(5)$  & $0.50^{(+13)}_{(-14)}$ & \Large$\beta$      &\Large  $0.80^{(+10)}_{(-14)}\, \text{GeV}$                \\
$J(0)$      &   $0.250(25)$  & $0.250^{(+62)}_{(-47)}$  &           &                  
\end{tblr}
\end{table}
 For easier comparison, we also present the data as a ratio to the reference value in \Cref{fig:fit_results_proton_proton} for the proton and in \Cref{fig:fit_results_proton_neutron} for the neutron. Note that we present results for both hadrons while performing a parametrization that only includes the proton. As can be seen, the quantities describing the proton can be fitted nearly perfectly, but even though the central values are close to ideal, the uncertainties are significantly larger than those of the reference values. This is particularly noticeable for GFFs, where the uncertainty reaches about $30\%$ of the central value. However, it is just a reflection of the fact that we chose $10\%$ uncertainty for the reference value. Despite performing fits only for the proton, we get results for the neutron that are qualitatively consistent with reference values. The reference value for the magnetic radius lies within the model uncertainty. The maximum value (inside uncertainty) for the neutron magnetic moment is about $10\%$ lower than the reference value. The worst result is for the mean-square charge radius of the neutron, which is almost two times smaller than the reference value. However, this value is very small, so the percentage difference does not translate directly into a difference in absolute value; even so, the reference value lies somewhat outside our uncertainty range. It is also worth noting that the uncertainties of our model parameters do not change their values significantly, despite the non-negligible percentage differences, particularly for the quark mass $m_q$ and the scalar-diquark mass $m_s$. The only exception to this observation is the axial diquark mass $m_a$, which has a large upper uncertainty limit. This may suggest that, in an overall fit, the ratio of the axial-diquark mass to the quark mass is important, but that fitting a single quantity is highly sensitive to $m_a$. In this scenario, a single quantity can strongly affect the axial-diquark mass without changing the fit quality for the other parameters. Note that the calibrated diquark masses are larger than the nucleon mass, $M < m_d$, and the range $M > m_d$ lies outside the uncertainty range of $m_s$ and $m_a$ obtained from the fit. One can also note that the calibrated $\lambda$ value is very close to the commonly chosen $\lambda=1.5$, and that its uncertainty range includes this value. The calibrated model parameters appear consistent with the discussion in the theoretical section, where we argued that, for good PDFs, the diquark masses should be much larger than the quark mass. This leads to the conclusion that this fit, despite containing no direct information about PDFs, should be consistent with experimental results (at least qualitatively).  
\subsubsection{Proton + Neutron fit}
\begin{figure}
    \centering
    \includegraphics[width=1\linewidth]{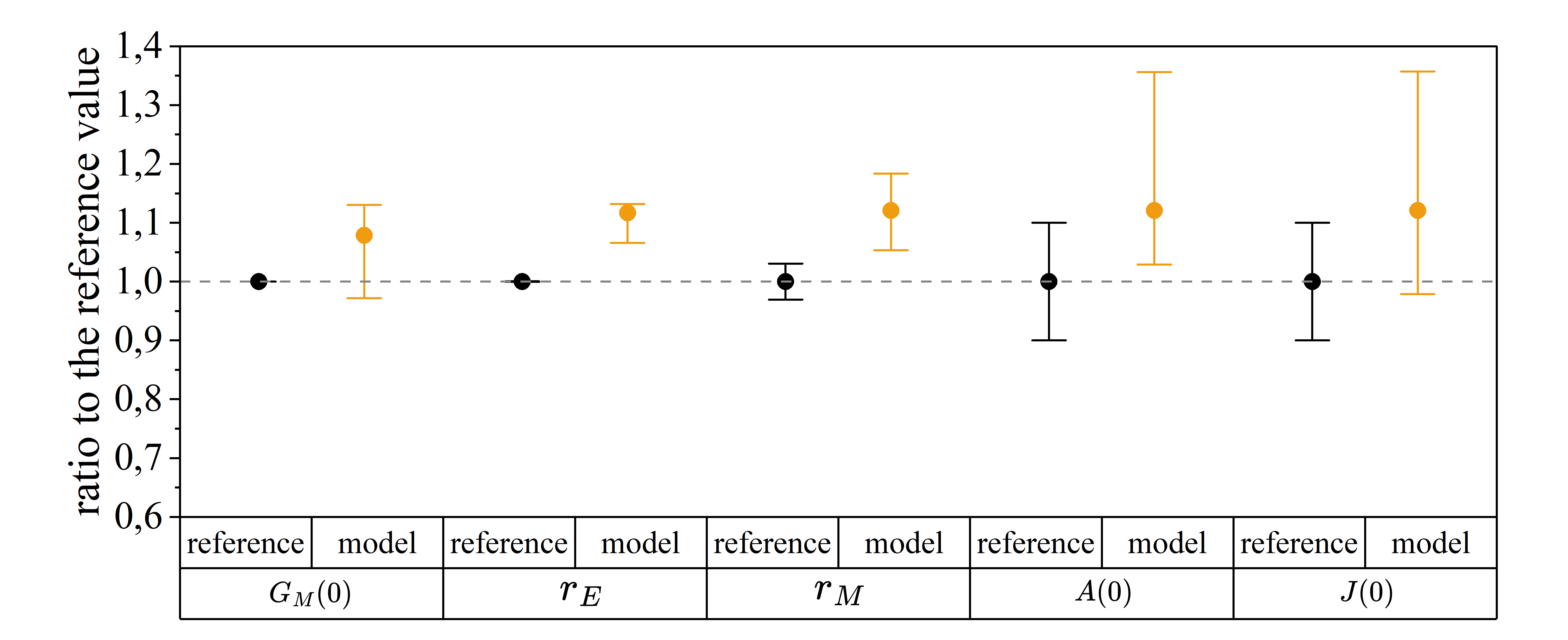}
    \caption{Proton magnetic moment, electromagnetic radii and GFFs reference values with uncertainties (black) and model results (orange) relative to the reference value. For a fit including only quantities describing 
simultaneously the proton and the neutron.}
    \label{fig:fit_results_proton+neutron_proton_d}
\end{figure}

\begin{figure}
    \centering
    \includegraphics[width=1\linewidth]{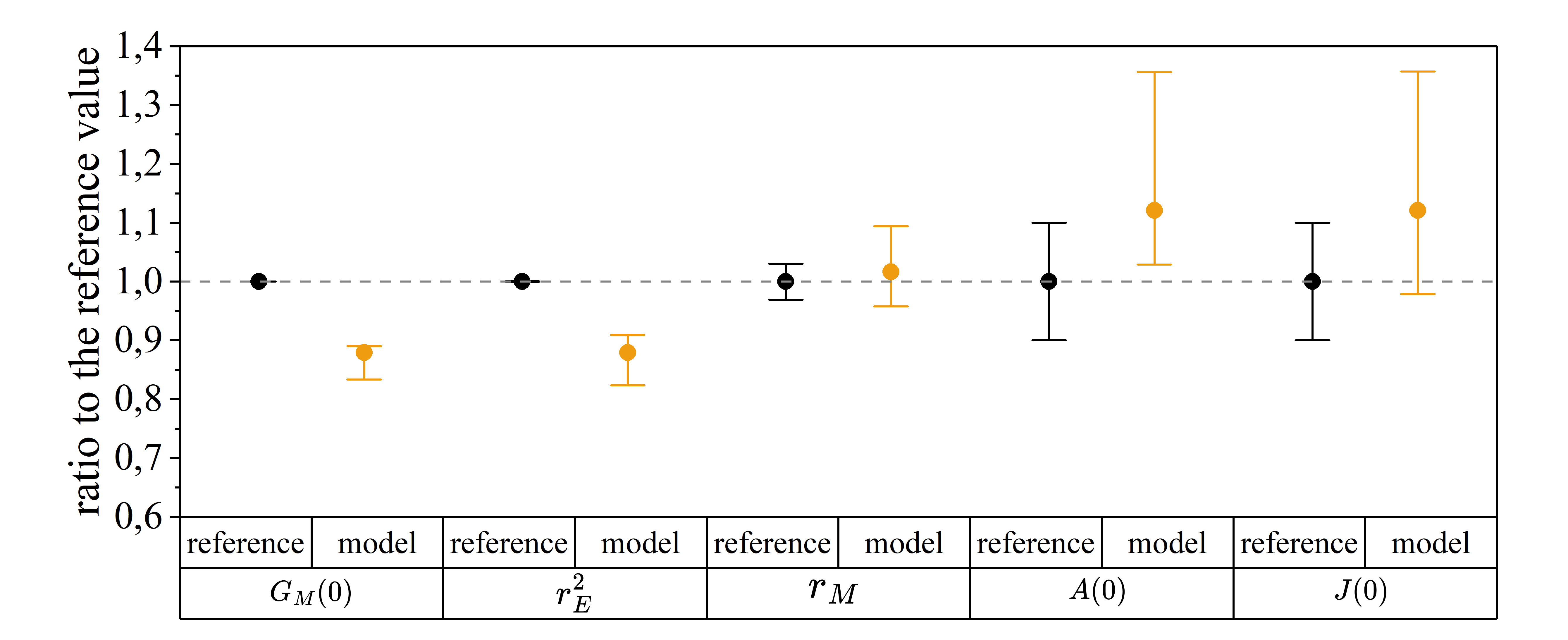}
    \caption{Neutron magnetic moment, electromagnetic radii and GFFs reference values with uncertainties (black) and model results (orange) relative to the reference value. For a fit including only quantities 
simultaneously describing the proton and the neutron.}
    \label{fig:fit_results_proton+neutron_neutron_d}
\end{figure}

\begin{figure}
    \centering
    \includegraphics[width=1\linewidth]{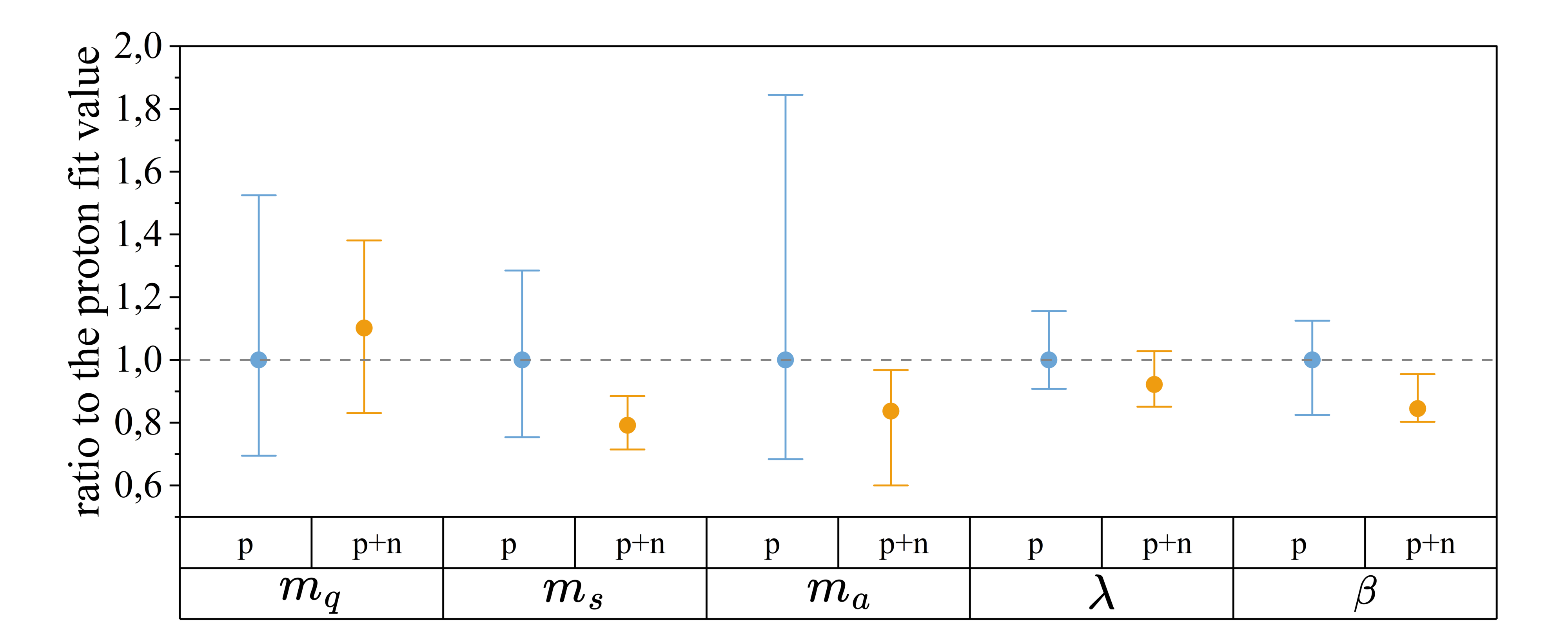}
    \caption{Comparison of obtained model parameters for the fit including only the proton (p) and the fit simultaneously including the proton and the neutron (p+n). The results are shown as a ratio to the central value obtained for the fit including only the proton.}
    \label{fig:fit_results_comparission}
\end{figure}

Our main interest is fitting the model to the proton, but a natural extension is to check whether the model can simultaneously describe both the proton and the neutron. In this section, we calibrate the model including not only quantities describing the proton but also the neutron. The fitting procedure is the same as above; the only difference being that more observables are included. We collect the results of the fit in \Cref{tab:results_p+n_d}. 
\begin{table}[h!]
\centering
\caption{Model results compared to reference parameter, for a model parametrized to the proton and the neutron at the same time.}
\label{tab:results_p+n_d}
\begin{tblr}{
  width = \linewidth,
  colspec = {Q[130]Q[223]Q[200]Q[130]Q[221]},
  cell{2}{1} = {c=3}{0.552\linewidth,c},
  cell{2}{4} = {r=2}{},
  cell{2}{5} = {r=2}{},
  cell{4}{4} = {r=2}{},
  cell{4}{5} = {r=2}{},
  cell{6}{1} = {c=3}{0.552\linewidth,c},
  cell{6}{4} = {r=2}{},
  cell{6}{5} = {r=2}{},
  cell{8}{4} = {r=2}{},
  cell{8}{5} = {r=2}{},
  cell{10}{4} = {r=2}{},
  cell{10}{5} = {r=2}{},
  vline{4} = {1-11}{},
  vline{2} = {2,6}{},
  hline{1-2,12} = {-}{},
  hline{3,6-7,10} = {1-3}{dashed},
}
Observable & Reference value & Model value & Parameter & Calibrated value \\
Proton     &                 &             &\Large $m_q$        &\Large $0.130^{(+33)}_{(-32)} \, \text{GeV}$                 \\
$G_M^p(0)$ & $2.79284734463(82)$  & $3.01^{(+14)}_{(-30)}$ &           &                  \\
$r_E^p$ & $0.8409(4)\, \text{fm}$ & $0.939^{(+12)}_{(-43)}\,\text{fm}$ & \Large$m_s$ & \Large$1.03^{(+12)}_{(-10)} \, \text{GeV}$                 \\
$r_M^p$  & $0.851(26)\, \text{fm}$ & $0.954^{(+5)}_{(-5)}\,\text{fm}$ &           &                  \\
Neutron    &                 &             & \Large $
m_a$        &\Large $1.59^{(+25)}_{(-45)}\, \text{GeV}$                  \\
$G_M^n(0)$ &$-1.91304276(45)$ & $-1.682^{(+21)}_{(-87)}$&           &                  \\
$\left<r_E^2\right>^n$ & $-0.1155(17)\,[\text{fm}^2]$ & $-0.1015^{(+35)}_{(-64)}\,\text{fm}^2$             & \Large$\lambda$    & \Large $1.30^{(+15)}_{(-10)}$                \\
$r_M^n$  & $0.864(9)\, \text{fm}$& $0.878^{(+67)}_{(-51)}\, \text{fm}$ &           &                  \\
$A(0)$       &  $0.50(5)$  & $0.56^{(+12)}_{(-5)}$ & \Large$\beta$      &\Large  $0.676^{(+88)}_{(-34)}\, \text{GeV}$                \\
$J(0)$      &   $0.250(25)$  & $0.280^{(+59)}_{(-35)}$  &           &                  
\end{tblr}
\end{table}
Following the same logic as in the previous subsection, we also present the results as a ratio to the reference value in \Cref{fig:fit_results_proton+neutron_proton_d} for the proton, and in \Cref{fig:fit_results_proton+neutron_neutron_d} for the neutron.
As can be seen, despite the more complex fit involving more quantities, the fit results (including uncertainties) differ from the reference values by at most about $10\%$. However, in most cases, the reference value does not fall within the uncertainty range of the obtained results. It is important to note that the parameter values obtained from the proton-only and proton+neutron fits agree with each other, as shown in \Cref{fig:fit_results_comparission}. This means that the proton+neutron fit can be treated as an improvement of the proton-only fit rather than as an entirely different result. As can be seen in this figure, the results for both approaches always share a common range of values, ensuring a satisfactory result for both fits. Furthermore, only a small portion of the proton+neutron uncertainty does not overlap with the proton-only uncertainty. In addition, the uncertainties of the proton+neutron fit are always smaller than those of the proton-only fit. This reinforces the earlier conclusion that adding the neutron does not produce a new fit but rather improves the proton-only fit. The larger uncertainties of the proton-only calibration reflect the fact that fewer observables constrain the parameters, which leaves room for correlated compensation among them. In the remainder of this paper, we will present both the results for a fit that includes only the proton (represented in blue) and those for a fit that also takes the neutron into account (represented in orange). One might also ask whether a neutron-only fit would be consistent with the two approaches presented here, but we do not discuss this possibility.
\subsection{Form factors results}
\begin{figure}[htbp]
    \centering
    \begin{subfigure}[b]{0.51\linewidth}
        \centering
        \includegraphics[width=\linewidth]{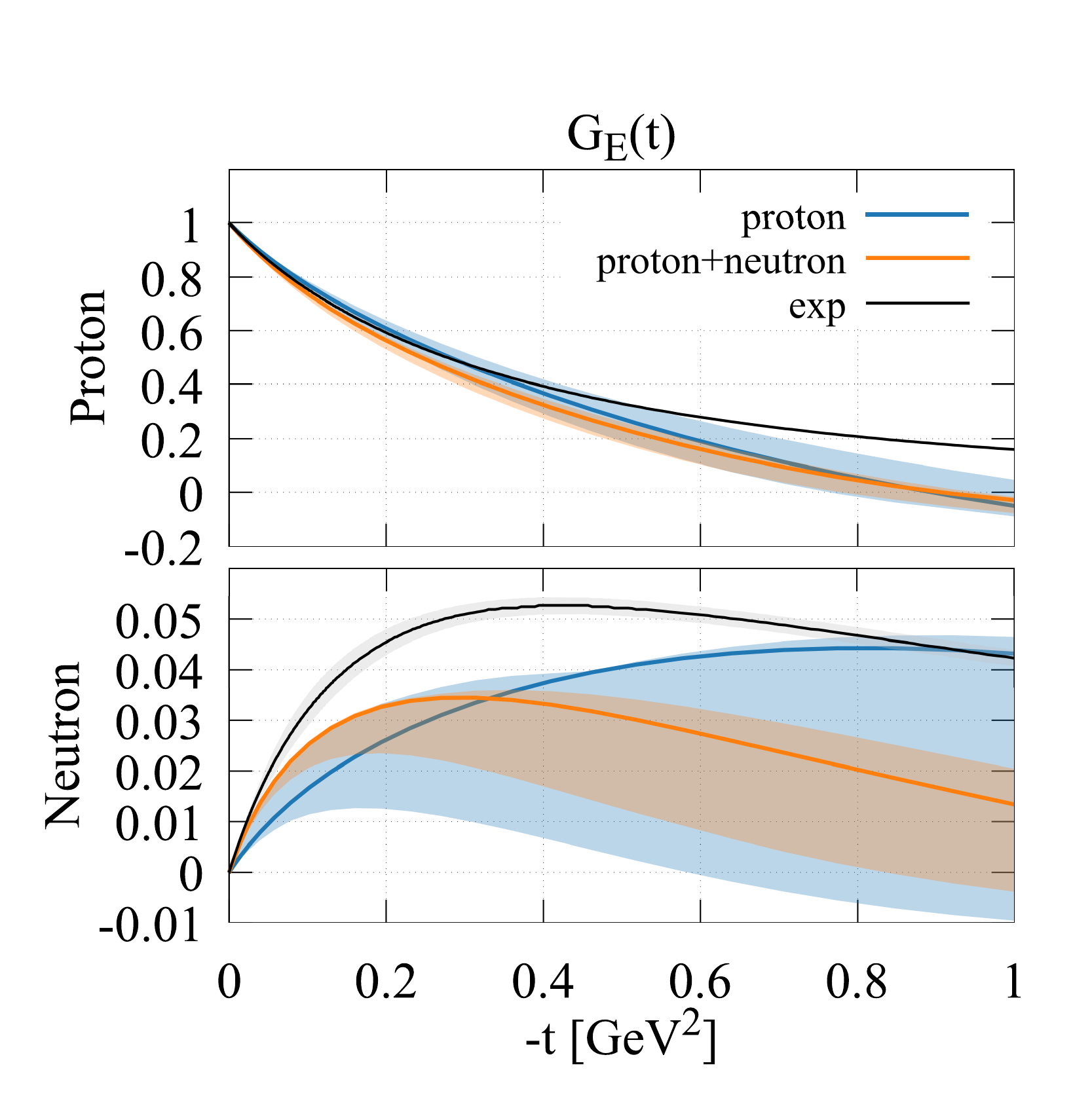}
    \end{subfigure}
\hspace{-0.7cm}
    \begin{subfigure}[b]{0.51\linewidth}
        \centering
        \includegraphics[width=\linewidth]{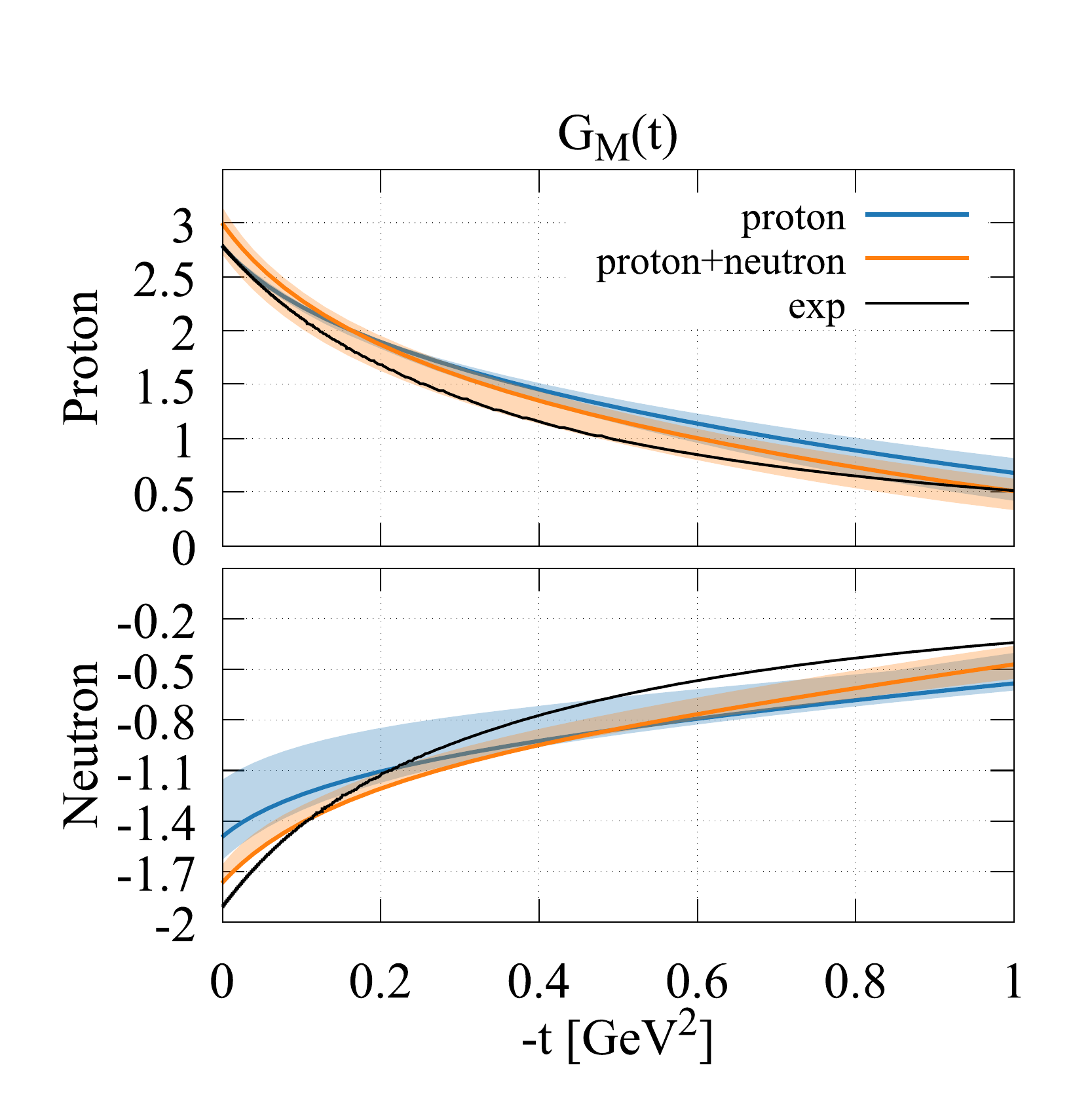}
    \end{subfigure}
    \vspace{-0.5cm}
    \caption{EMFFs as a function of $-t$ obtained for the proton and the neutron, calculated from results for a fit including only the proton (blue), and a fit that included both the proton and the neutron (orange), compared to experimental data (black) \cite{YE20188}.}
    \label{fig:EMFF}
\end{figure}

\begin{figure}
        \centering
        \includegraphics[width=\linewidth]{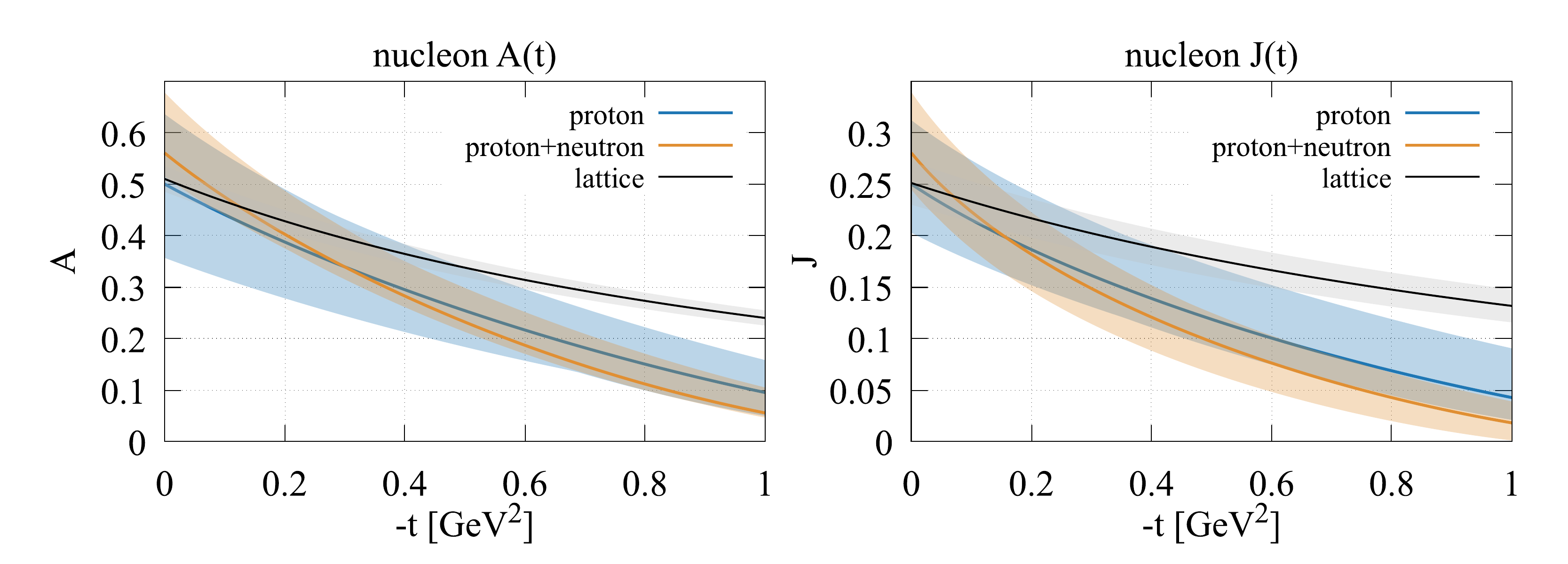}
    \caption{GFFs as a function of $-t$ obtained for the proton and the neutron, calculated from results for a fit including only the proton (blue) and a fit that includes both the proton and the neutron (orange), compared to lattice data (black) \cite{PhysRevLett.132.251904}.}
    \label{fig:GFF nucleon}
\end{figure}

In this section, we present our results for EMFFs and GFFs compared to reference data (experimental or lattice). For all the plots, we follow the same color code as in the previous section, to facilitate comparisons. For the reference value, we use a black color; for results of fitting only to the proton, we use a blue color, and for results we get from fit that also includes the neutron, we use an orange color. Results for EMFFs for both hadrons are shown in \Cref{fig:EMFF}. Despite calibrating the model only for $-t \to 0$, the results we obtained have the same characteristics as the reference data for the range $-t \in [0,1]$ GeV$^2$. Form factors values for $-t$ larger than $1$ GeV$^2$ are not examined here, because of the simplified nature of the model, the main limitation being the absence of gluons. As mentioned in the previous section, the value of $G_E(-t=0)$ for both proton and neutron is correctly normalized by construction and is independent of the parameters. It is important to note that the fit including only the proton (blue) is in almost perfect agreement for proton EMFFs for both the value of $G_{E/M}$ and for the value of $\frac{\text{d}G_{E/M}}{\text{d}t}$  at $-t=0$. However, this does not mean that the proton-only fit agrees much better with the proton reference data at higher $-t$ than the proton+neutron fit (orange), as can be seen in \Cref{fig:EMFF}. Results obtained from the proton-only fit are still in reasonably good agreement with neutron EMFFs even though neutron was not included during the parametrization. The fit including both the neutron and the proton (orange) gives results that are within the uncertainty range consistent with the fit that takes only the proton into consideration. It is also worth recalling the interesting fact discussed above that the proton+neutron fit yields smaller uncertainty bands than the proton-only fit. The largest discrepancies with the reference data are observed for $G_E^{n}$; this may be due to the fact that these values are very small, so any simplifications in the model may result in larger percentage discrepancies.
The results for GFFs, presented in \Cref{fig:GFF nucleon} for both nucleons, show greater discrepancies from the reference data \cite{PhysRevLett.132.251904} (in this case, lattice data rather than experimental data) than those for EMFFs. In addition, all features related to the characteristics of the charts and their relationships with each other for both kinds of fit for EMFFs apply to GFFs. It can be seen here that both $A(t)$ and $J(t)$ decay too quickly as $-t$ increases compared to the reference data. The larger discrepancies observed for GFFs compared to EMFFs may be due to the absence of gluons; since gluons carry no electric charge, GFFs may be more sensitive to their absence than EMFFs. 
\begin{figure}[htbp]
    \centering
    \begin{subfigure}[b]{0.51\linewidth}
        \centering
        \includegraphics[width=\linewidth]{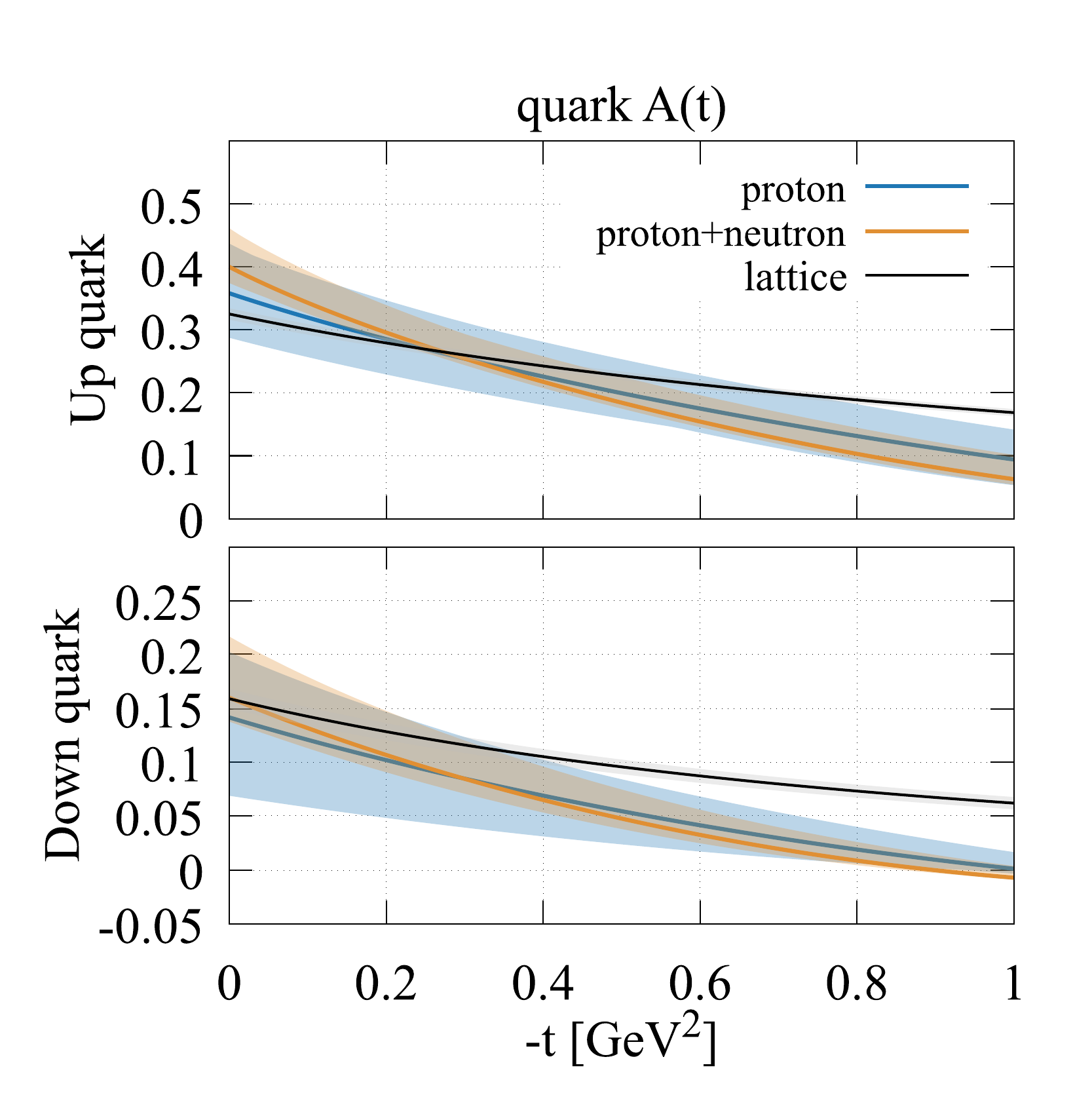}
    \end{subfigure}
\hspace{-0.7cm}
    \begin{subfigure}[b]{0.51\linewidth}
        \centering
        \includegraphics[width=\linewidth]{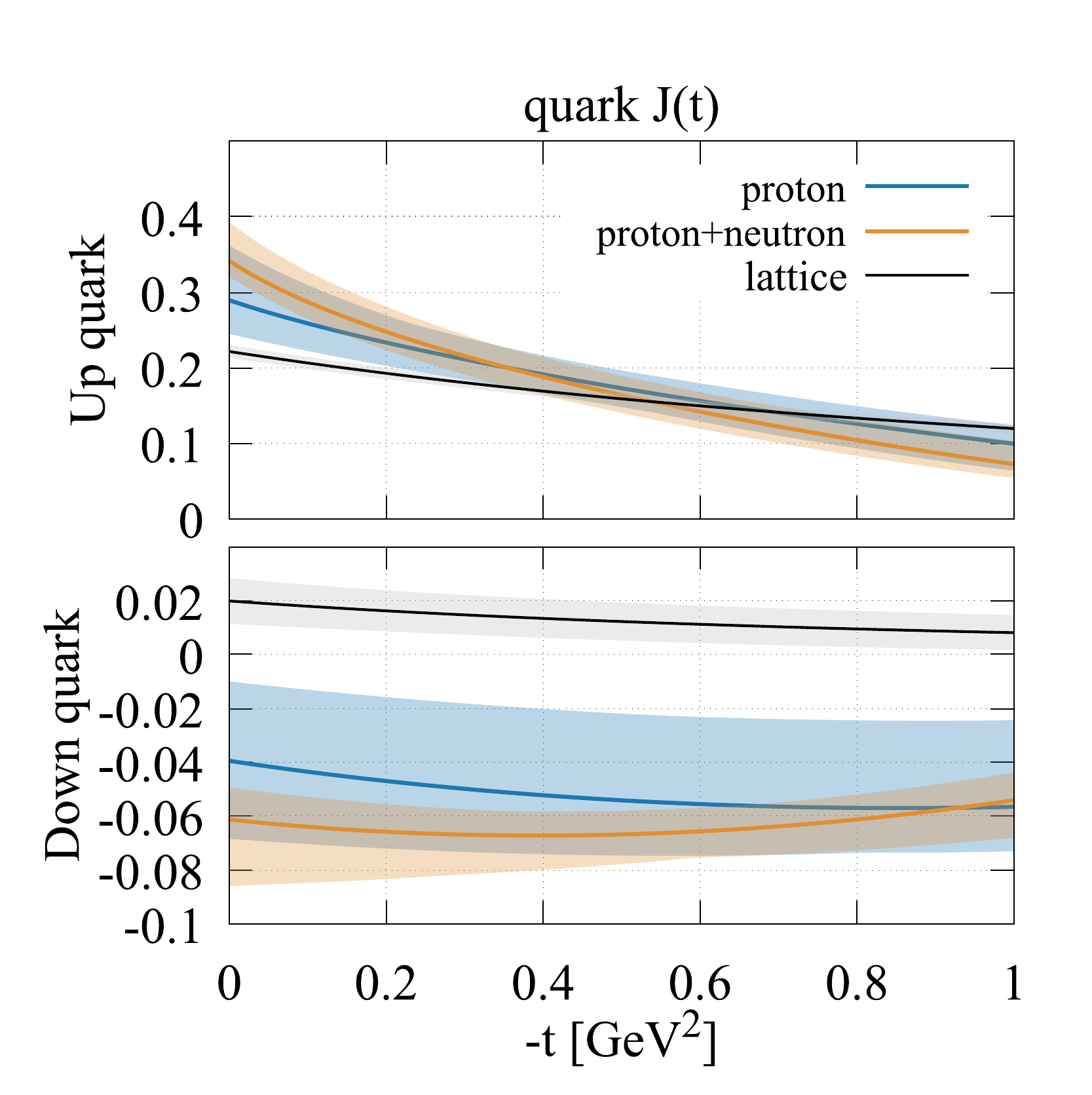}
    \end{subfigure}
    \vspace{-0.5cm}
    \caption{GFFs as a function of $-t$ obtained for up and down quarks in the proton, calculated from results for a fit including only the proton (blue), and a fit that included both the proton and the neutron (orange), compared to lattice data (black) \cite{PhysRevLett.132.251904}.}
    \label{fig:GFF_quarks}
\end{figure}

The results presented in \Cref{fig:GFF_quarks}, for flavor-decomposed GFFs, are very important; they demonstrate the model's qualitative agreement with the reference data not only for whole nucleons but also for the quarks building the nucleon. Again, the results obtained do not perfectly match the lattice data, but despite the lack of a direct fit to the quantities describing the quarks, the results obtained are rather close to the lattice results. This suggests an internal consistency of the model. As above, the characteristics of the graphs are similar to those discussed above. The most interesting difference is the opposite sign of $J_{\text{down}}(t)$ compared to the reference results. Although the sign is different, the results obtained behave similarly to the reference results, changing very little as $t$ varies, and the value itself is near zero. However, a negative value is not out of the question, and there are earlier works suggesting that $J^{\text{down}}(0)$ could be negative, for example \cite{Diehl:2013xca}. The value determined in the aforementioned study is: $J^{\text{down}} = -0.004^{+0.010}_{-0.016}$. It is therefore consistent (in terms of sign) with our results and falls within the range of values determined for the fit that considers only the proton.
\subsection{PDFs, GPDs, and tomography results}
In this section, we present the main results of this paper: the PDFs, the GPDs, and the proton tomography for both up and down quarks. It should be emphasized that the model was tuned with respect to the FF-related quantities at $t=0$, so all the results in the previous section for $t \neq 0$ are model predictions. Thus, the above results only have a single point, $t=0$, connecting them to the reference data. In contrast, the results shown in this section are pure model predictions, made without any $t$-dependent input during the parameter fitting process. 

\subsubsection{PDFs and GPDs}
\begin{figure}[htbp]
    \centering
    \begin{subfigure}[b]{0.51\linewidth}
        \centering
        \includegraphics[width=\linewidth]{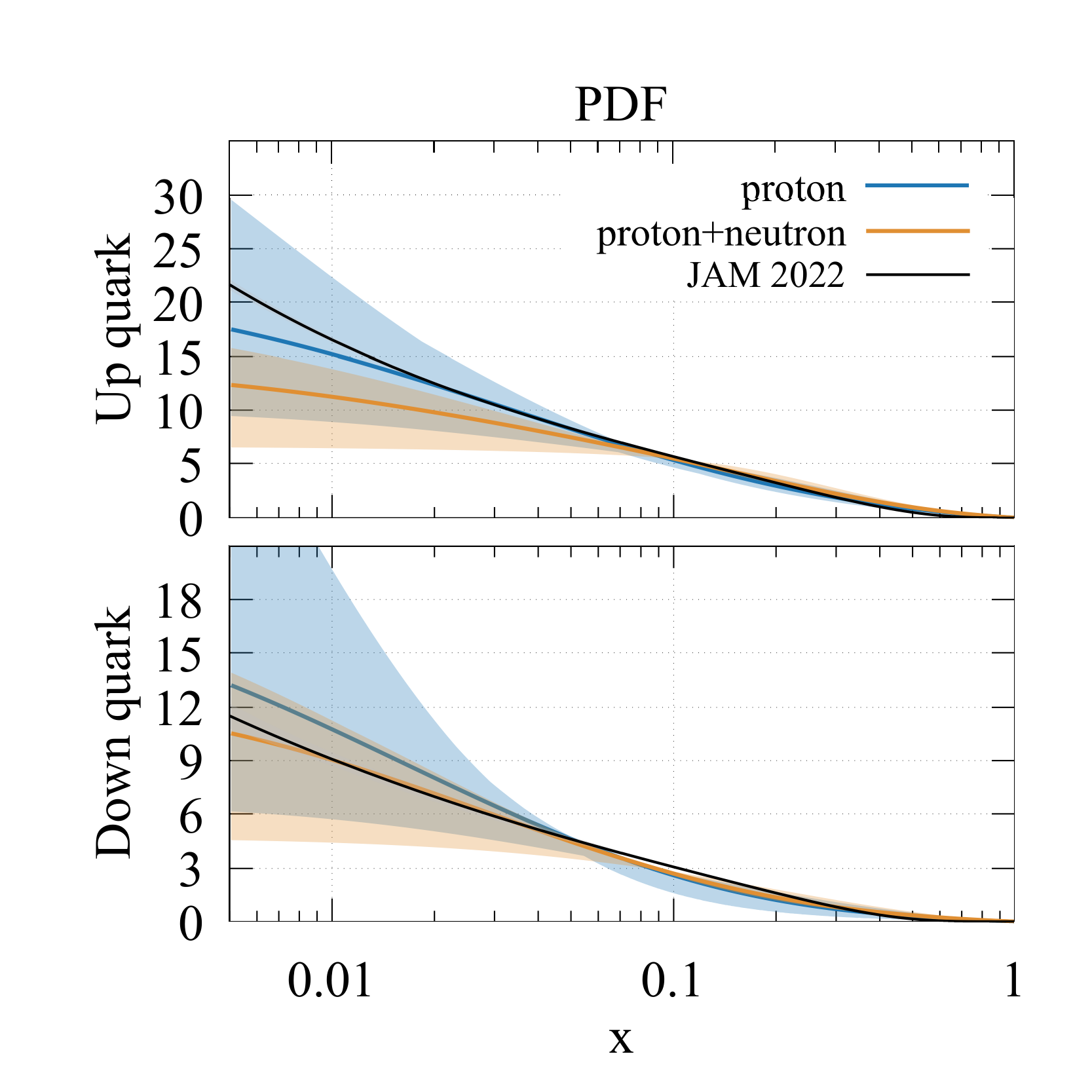}
    \end{subfigure}
\hspace{-0.7cm}
    \begin{subfigure}[b]{0.51\linewidth}
        \centering
        \includegraphics[width=\linewidth]{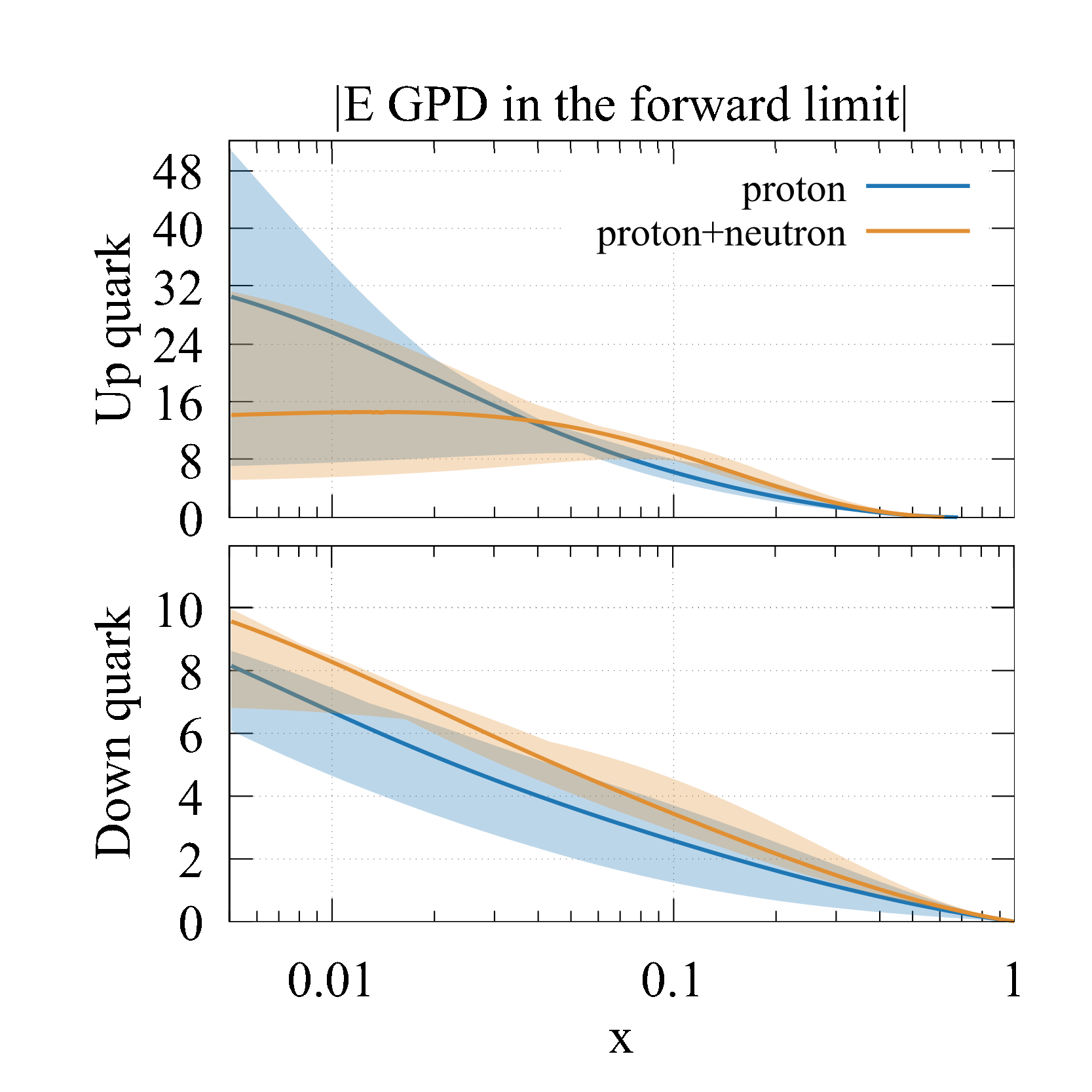}
    \end{subfigure}
    \vspace{-0.5cm}
    \caption{The PDF and the absolute value of the $E$ GPD in the forward limit as a function of $x$, obtained for up and down quarks in the proton, calculated from results for the fit including only the proton (blue), and fit that includes both the proton and the neutron (orange), compared to JAM 2022 for the PDF (black) \cite{Moffat_2021}. The absolute value for the $E$ GPD only affects the down quark plot, because the values for the up quark $E$ GPD are positive over the whole $x$ range, in contrast to the down quark GPD, which is negative over the entire range.}
    \label{fig:PDF}
\end{figure}

In \Cref{fig:PDF}, we show our PDFs and the forward limit $E$ GPDs results for both kinds of fits. The PDF results are compared to data from the Jefferson Lab Angular Momentum Collaboration  (JAM 2022) \cite{Moffat_2021}. For both the PDF and the forward limit $E$ GPD, we present results using a logarithmic scale for the axis with Bjorken $x$. We use a logarithmic scale here to make the graph easier to read, because the resulting curves show a sharp increase in values for small $x$ (within the studied range). 

We present results for the down quark forward limit $E$ GPDs as an absolute value because, as follows from previous results, the forward limit $E$ GPD for the down quark should be negative over the whole range (see \Cref{fig:E_FL_mqm})\footnote{It is worth recalling that GPDs for the down quark in the proton have only axial-vector diquark contribution.}, and we find it easier to compare to the forward limit $E$ GPD for the up quark if the values on both graphs are positive. The results we obtained are consistent with JAM 2022 reference data for both proton-only and proton+neutron fit. The largest differences are observed in the PDF plot for the down quark at $x\approx 0.1$; this is the only part where the reference values lie outside the uncertainty band for both fits. For the up quark PDF, we obtained better agreement with the reference data for the proton-only fit. However, for the down quark PDF, better agreement is observed for the proton+neutron fit. The greatest differences between the two approaches are evident for very small values of $x$, less than $0.05$ in the case of the central value. Still, the uncertainty bands overlap across the entire range. For the proton+neutron fit, the reference value for the up quark falls outside the uncertainty band at small $x$. The situation is different for the forward limit $E$ GPD, where the values for both the up and down quarks differ significantly from one another across the entire $x$ range, depending on the fitting procedure chosen. Nevertheless, there are still overlapping points in the majority of the range. In all of the plots, we observe the characteristics discussed in \Cref{sec:General characteristics of GPDs}, where, for $m_q \ll m_d$, the absolute values of the forward-limit GPDs increase sharply as $x$ decreases. Note that because the forward limit $E$ GPD for the axial-vector diquark is negative over the entire range, and because GPDs for the up quark are calculated as ${E^{\text{up}}} = \lambda E^S+(2-\lambda)E^A$, the value of the forward limit $|E|$ GPD for the up quark does not have to increase rapidly for small $x$. For the up quark, the proton+neutron calibration predicts a plateau in the
central value of $|E|$ GPD for small $x$, while the lower edge of the uncertainty
band decreases towards smaller $x$. A similar, though much weaker, decrease
of the lower band is also present in the proton-only fit. Since the band represents alternative
model curves, this implies that some allowed parameter sets yield a falling
$|E|$ for the up quark as $x \to 0$.
\begin{figure}[htbp] 
    \centering
    \begin{subfigure}[b]{0.51\linewidth}
        \centering
        \includegraphics[width=\linewidth]{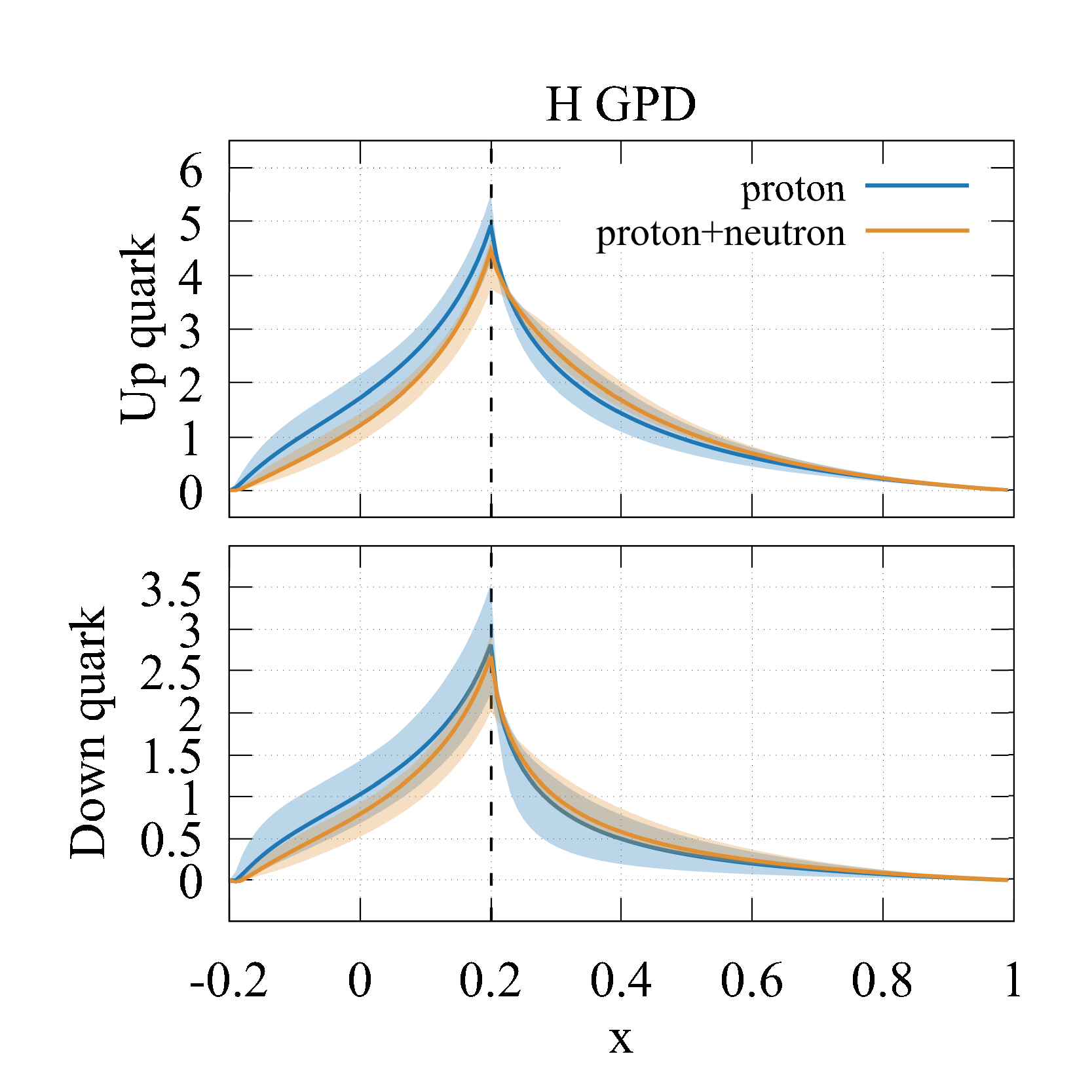}
    \end{subfigure}
\hspace{-0.7cm}
    \begin{subfigure}[b]{0.51\linewidth}
        \centering
        \includegraphics[width=\linewidth]{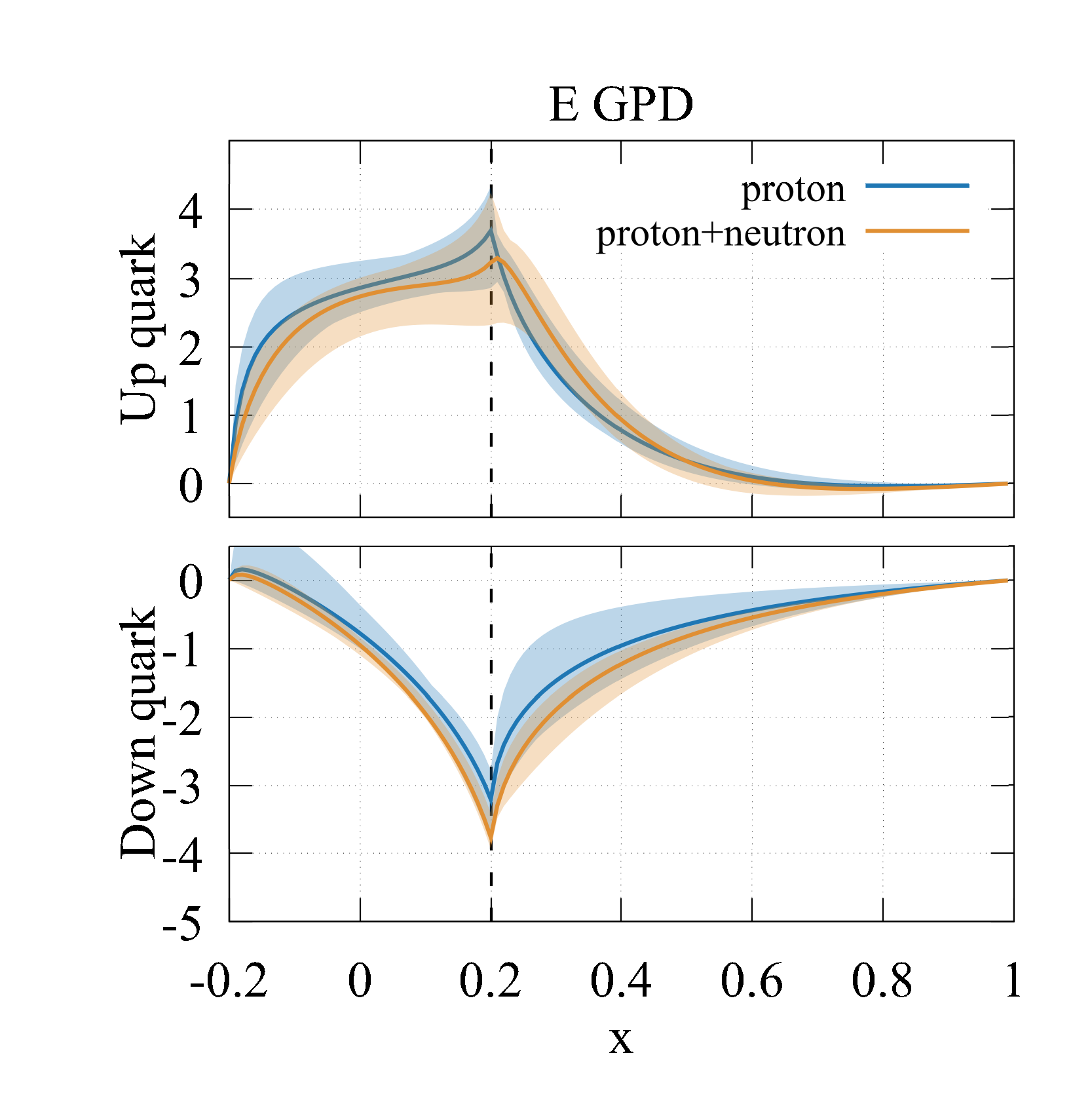}
    \end{subfigure}
    \vspace{-0.5cm}
    \caption{Zero transverse momentum transfer ($|\vec\Delta_\perp|=0$) $H$ and $E$ GPDs with $\xi =0.2$ as a function of $x$ obtained for up and down quarks in a proton, calculated from results for the fit including only the proton (blue) and the fit that includes both the proton and the neutron (orange). The boundary of the ERBL region is indicated by a vertical dashed line.}
    \label{fig:H_E_GPDs}
\end{figure}

A natural extension of the calculations involving the forward limit GPDs is the calculation of nonzero momentum transfer GPDs in this model for our calibrated parameters. We show results for both kinds of fits, for $\xi=0.2$ and $|\vec\Delta_\perp| = 0$ in \Cref{fig:H_E_GPDs}. Unlike in lattice calculations based on the quasi-distribution approach \cite{Alexandrou:2020zbe,Chu:2025kew}, using the DSM, we can obtain values for $x \approx \pm \xi$. Thus, a natural continuation of this work is to check how our results compare to other calculations, determine values for nonzero momentum transfer GPDs, and to see whether this model can fill the gaps in lattice calculations. We tested the model for several different values of $\xi$. For any value of $\xi$, the scalar and axial GPDs always exhibit a transition from the ERBL to the DGLAP region, in which the function increases in the ERBL region and decreases in the DGLAP region. Thus, the GPDs are continuous at $x=\xi$, with discontinuous first derivatives
\footnote{For the up quark $E$ GPD, it can be seen that the peak of the GPD does not coincide perfectly with the line $x=\xi$. This is because the $E$ GPD for an up quark is the sum of the scalar and the axial $E$ GPDs, where the axial $E$ GPD has a negative sign. This may result in a slight shift of the peak, depending on the parameters. Whereas for the other GPDs for the up and down quarks, the peak is always at $x=\xi$, regardless of the parameters.}.

\subsubsection{Tomography}
\label{Sec:Tomogrpahy}
\begin{figure}
        \centering
        \includegraphics[width=1.0\linewidth]{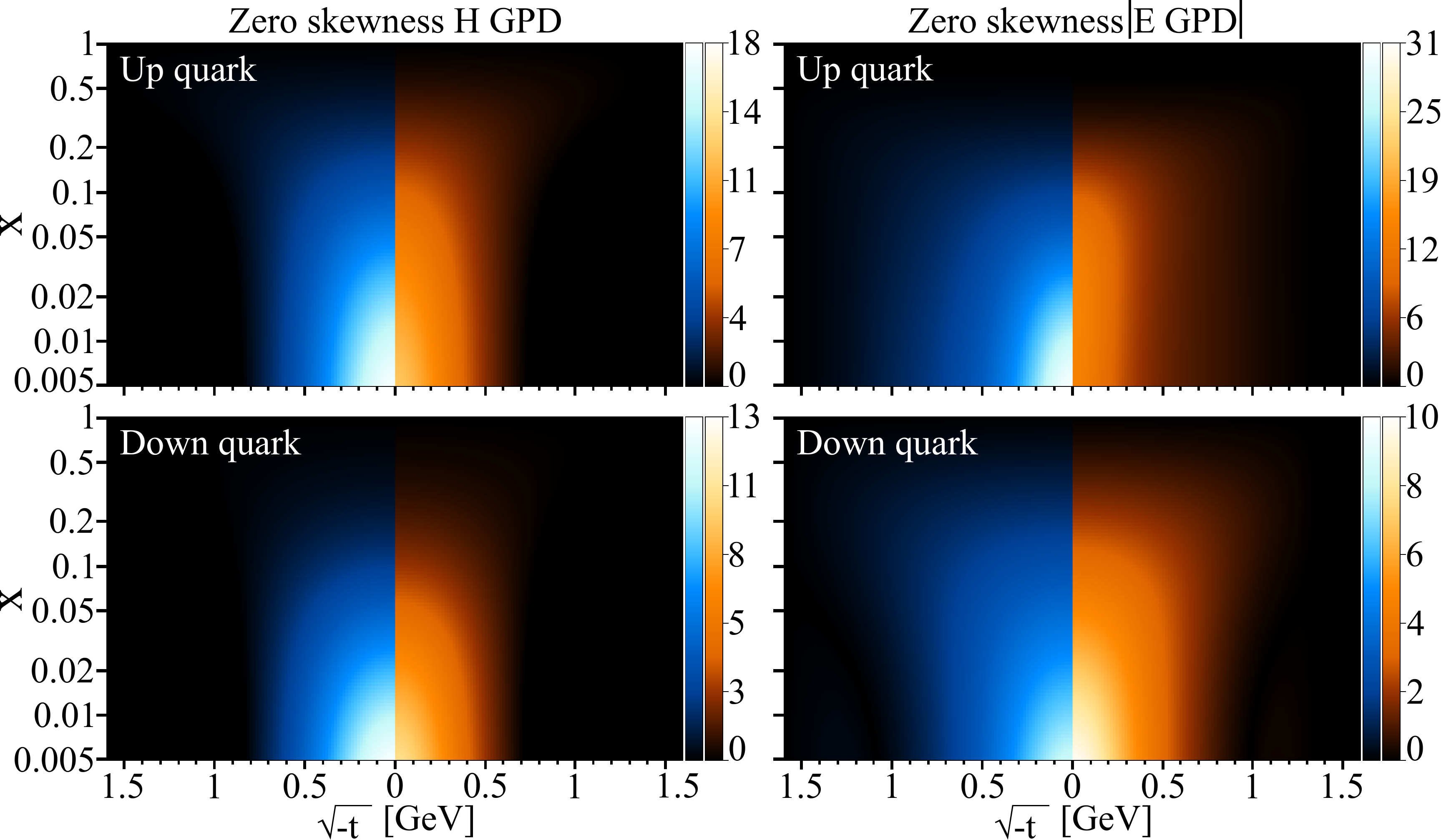}
    \caption{Zero skewness GPDs as function of $x$ and $\sqrt{-t}$. For the fit including the proton (blue) and both the proton and the neutron (orange). The results are presented only for the central value without uncertainty bands. Note that the vertical axis uses a logarithmic scale.}
    \label{fig:impact_GPD}
\end{figure}

In this section, we present our proton tomography results for both the up and the down quarks, using both fitting procedures. The tomography results depend on both the forward limit GPDs and the cutoff form. For this reason, we firstly present our results for zero-skewness GPDs as a function of $x$ and $\sqrt{-t}$   \footnote{For zero-skewness GPDs: $\sqrt{-t}=|\vec\Delta_\perp|$.} in \Cref{fig:impact_GPD}.

Again, we use a logarithmic scale for the axis containing $x$ (vertical) and show results for the $E$ GPDs as an absolute value, because values for zero-skewness down quark $E$ GPD are negative over the whole range. For $H$ GPDs for both quarks and the $E$ GPD for the up quark, we can see that the results for the proton-only fit have higher values at small $x$, regardless of the value of $\sqrt{-t}$. The situation is different only for the down quark $E$ GPD, where the values for the proton+neutron fit are higher at small $x$. But as $-t$ increases, the results for the proton-only fit become larger in magnitude than those for the proton+neutron fit. This is a consequence of the smaller $\beta$ obtained with the fit that takes both nucleons into account, compared to the fit that only takes the proton into account. In all of the plots, one can see that the results for the proton+neutron fit decay more slowly with increasing $x$ than those for the proton-only fit. The exception is large $-t$, for which the smaller $\beta$ value causes the results for the proton+neutron fit to vanish due to the cutoff. From this, we can conclude that, in general, the tomography for the proton-only fit should have greater magnitude at small $x$ and a greater polarization effect for the up quark, but a lower polarization effect for the down quark (compared to the effect on the up quark), in terms of magnitude. The magnitude of the proton+neutron fit should approach that of the proton-only fit, or even exceed it, as $x$ increases. This is because the PDF values for the proton+neutron fit are larger than for the proton-only fit at high $x$.  However, one needs to remember that the proton+neutron fit has a lower $\beta$, so values may have larger magnitude but decay faster as the $\sqrt{-t}$ increases. 
\begin{figure}[!h]
        \centering
        \includegraphics[width=1.0\linewidth]{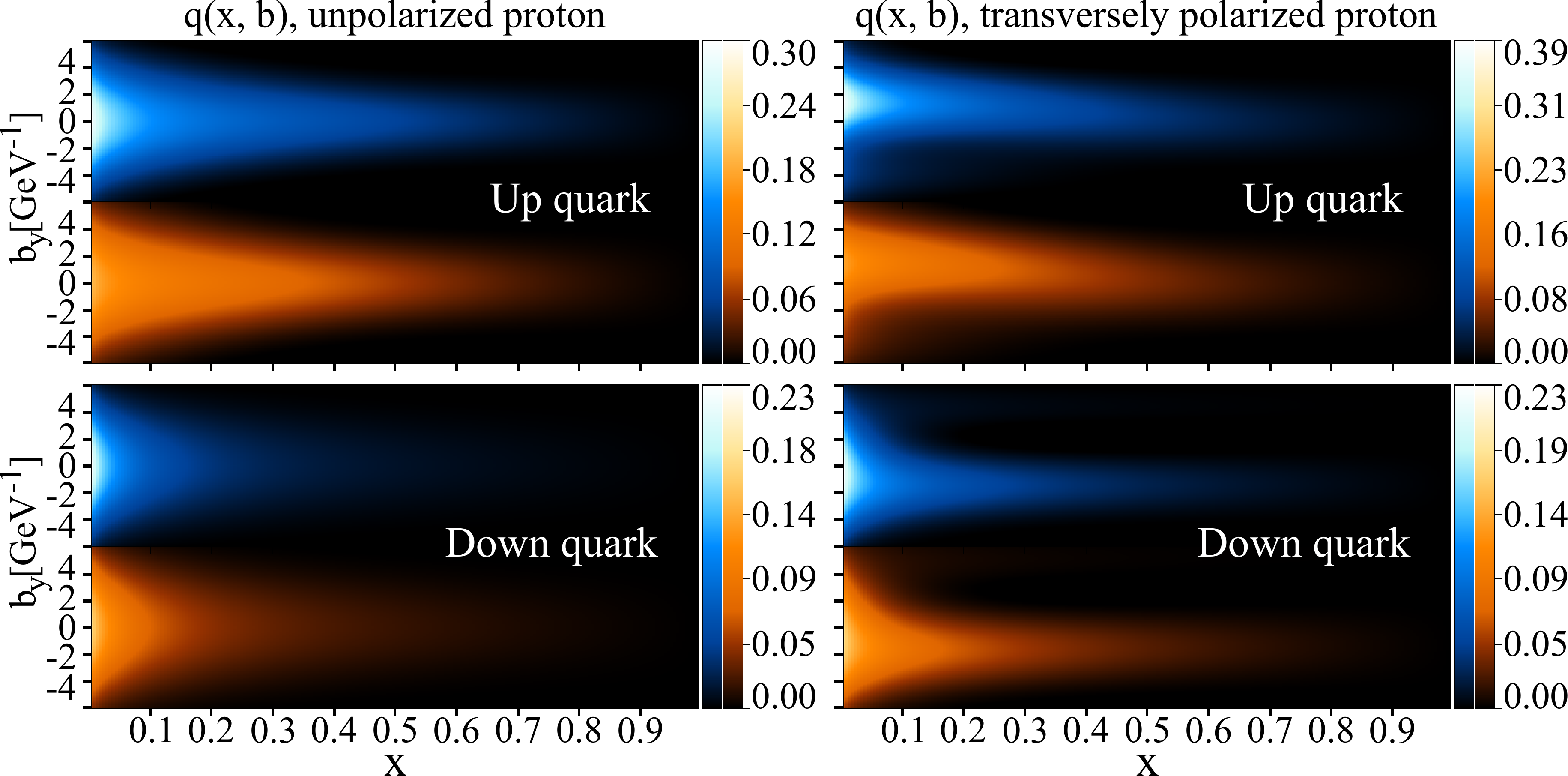}
    \caption{Proton GPDs in the impact parameter space for both the up and the down quark. Results for the central value obtained for the fit taking only the proton into consideration are shown in blue; results for the fit that take both the proton and the neutron into account are shown in orange. The proton is polarized along the $x$ axis.}
    \label{fig:impact_b}
\end{figure}

Because the results go to zero more slowly for the proton-only fit, we expect them to be narrower in tomography after the Fourier transform than the results for the proton+neutron fit. For the latter, we expect a slower decay as $b_\bot$ increases. For the same reason, we expect that polarized tomography for small $x$ will have a more shifted maximum for the proton+neutron fit than for the proton-only fit, especially for the down quark. However, it is difficult to make a prediction for larger values of $x$ because the values themselves are so small that the cutoff values should not have such a significant impact. In \Cref{fig:impact_b}, we show proton GPDs in the impact parameter space for both the up and the down quark. 

A look at the GPDs in impact parameter space provides a good qualitative insight into the obtained results. We can see that the above predictions based on the zero-skewness GPDs plots are consistent with our results. The distributions for the proton+neutron fit are broader (in the sense of a slower rate of decline with increasing $x$) than those for the proton-only fit. It can also be seen that, in terms of magnitude, the results at small $x$ are larger for the proton-only fit. Also, as $x$ increases, the differences in the resulting values of $q(x,b)$ disappear. It is difficult to verify, based on these plots, the above predictions regarding the polarization effect for both fitting methods. However, it is clearly visible that as $x$ increases, the effect of polarization becomes smaller, regardless of the quark flavor or the fitting procedure used. However, the above plots do not provide a complete picture, as they focus solely on $b_y$ while completely ignoring $b_x$. This is not a problem for unpolarized distributions, which are radially symmetric. For a transversely polarized proton, however, this does not provide a complete picture. For this reason, to provide a more quantitative description, we use complete two-dimensional heatmaps along with cross-sections for $b_x=0$, for given values of $x$. 

We present the results for $x=0.1$ in \Cref{fig:tomography_x01} and for $x=0.3$ in \Cref{fig:tomography_x03}.  
We also examined the results for $x=0.5$, $x=0.7$, and $x=0.9$, and show them in Appendix \ref{app:additional}. In these plots, we can quantitatively compare the distributions for different fits and partons. We can see that the distributions for the proton+neutron fit are broader and have smaller magnitude for both values of $x$, as expected above. It is also easy to see that the above prediction that, for the proton+neutron fit, the shift in the maximum will be greater than for a proton-only fit, is correct. From the higher values of $x$ examined, we also know that, for sufficiently large $x$, the maximum value for the proton-only fit can become smaller than that for the proton+neutron fit in the case of the down quark. In contrast, for the up quark, the maximum will always be greater for the proton-only fit, with the difference diminishing as $x$ increases. We have also observed that polarization becomes less significant as the value of $x$ increases; this is particularly noticeable for the up quark. In general, polarization has greater impact on the down quark distributions. The results we obtained are qualitatively comparable to other calculations related to proton tomography (see, for example, \cite{Cichy:2024afd}). Additionally, for both the tomography and the earlier results, the fitting procedure does not have a dramatic effect on the results. The negative values appearing in the cross section, especially for larger $x$, are numerical artifacts resulting from the cutoff imposed on the GPDs at $|\vec\Delta_\bot|>2\beta$, which induces oscillations upon Fourier transform. The model appears to be internally consistent, and adding a neutron to the fit, despite changing the obtained parameters, does not result in completely different behavior of subsequent results, which are only indirectly derived from the fitting. This suggests that such a parameterization procedure could also work for other hadrons to which this model could be applied. 
\FloatBarrier

\begin{figure}[h]
    \centering
    \begin{subfigure}[b]{\linewidth}
        \centering
        \includegraphics[width=\linewidth]{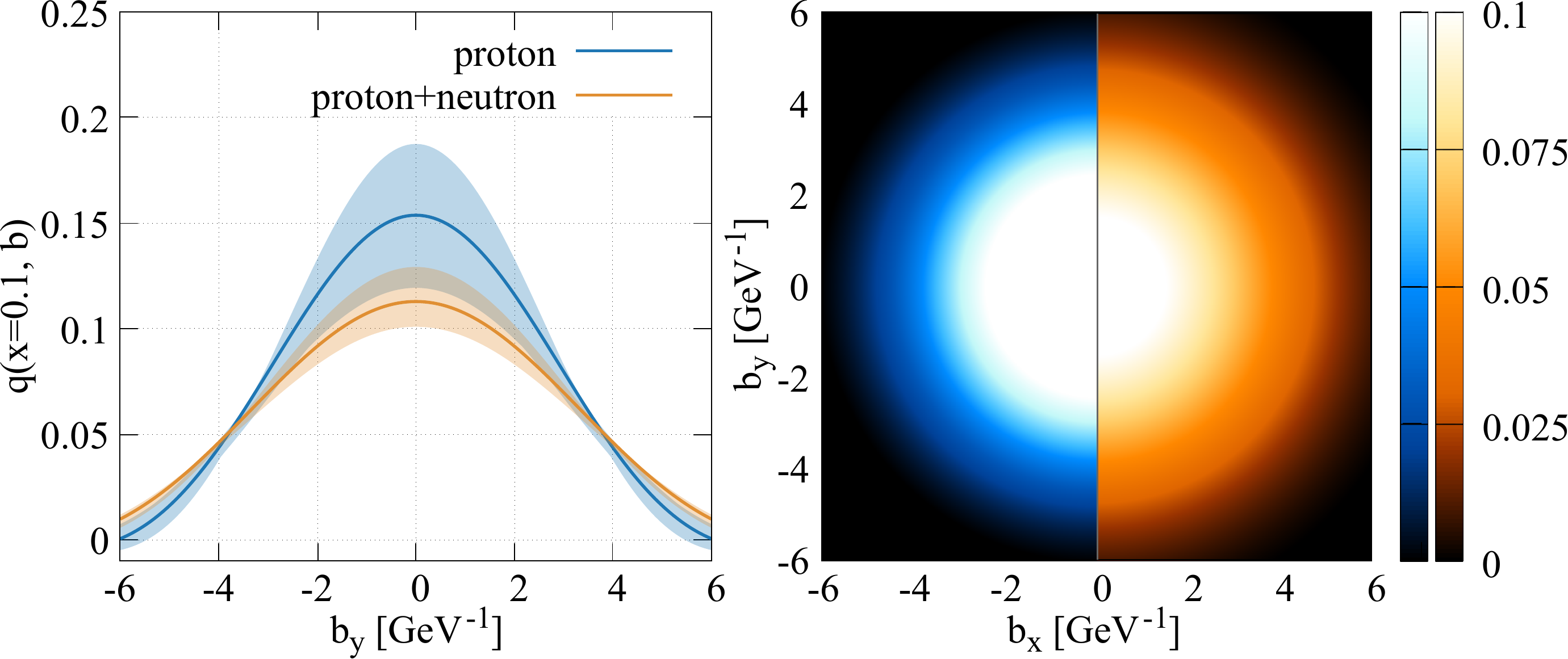}
        \caption{$x=0.1$, unpolarized up quark in proton.}
        \label{subfig:tom_up}
    \end{subfigure}
\end{figure}
\begin{figure}[p]
    \ContinuedFloat
    \centering
    \begin{subfigure}[b]{\linewidth}
        \centering
        \includegraphics[width=\linewidth]{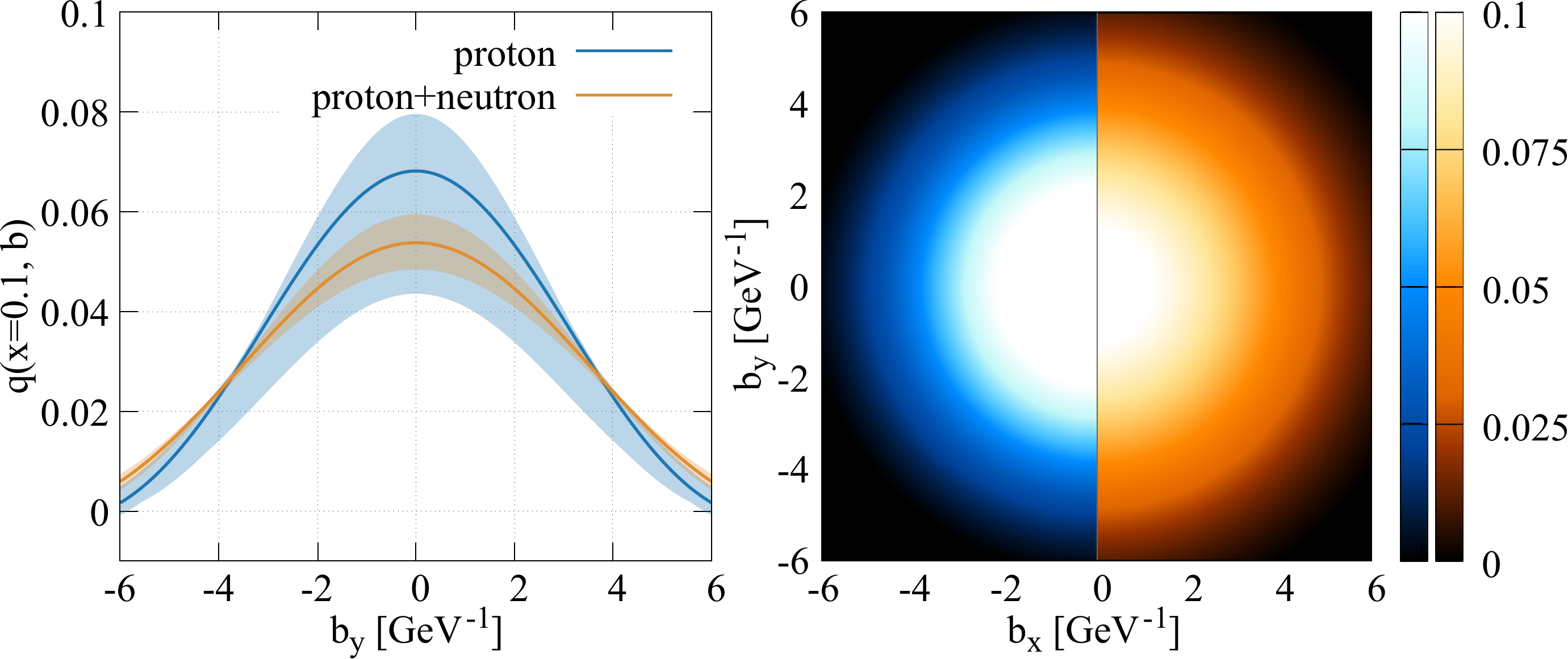}
        \caption{$x=0.1$, unpolarized down quark in proton (heatmap is multiplied by two).}
        \label{subfig:tom_down}
    \end{subfigure}
\end{figure}
\begin{figure}[p]
    \ContinuedFloat
    \centering
    \begin{subfigure}[b]{\linewidth}
        \centering
        \includegraphics[width=\linewidth]{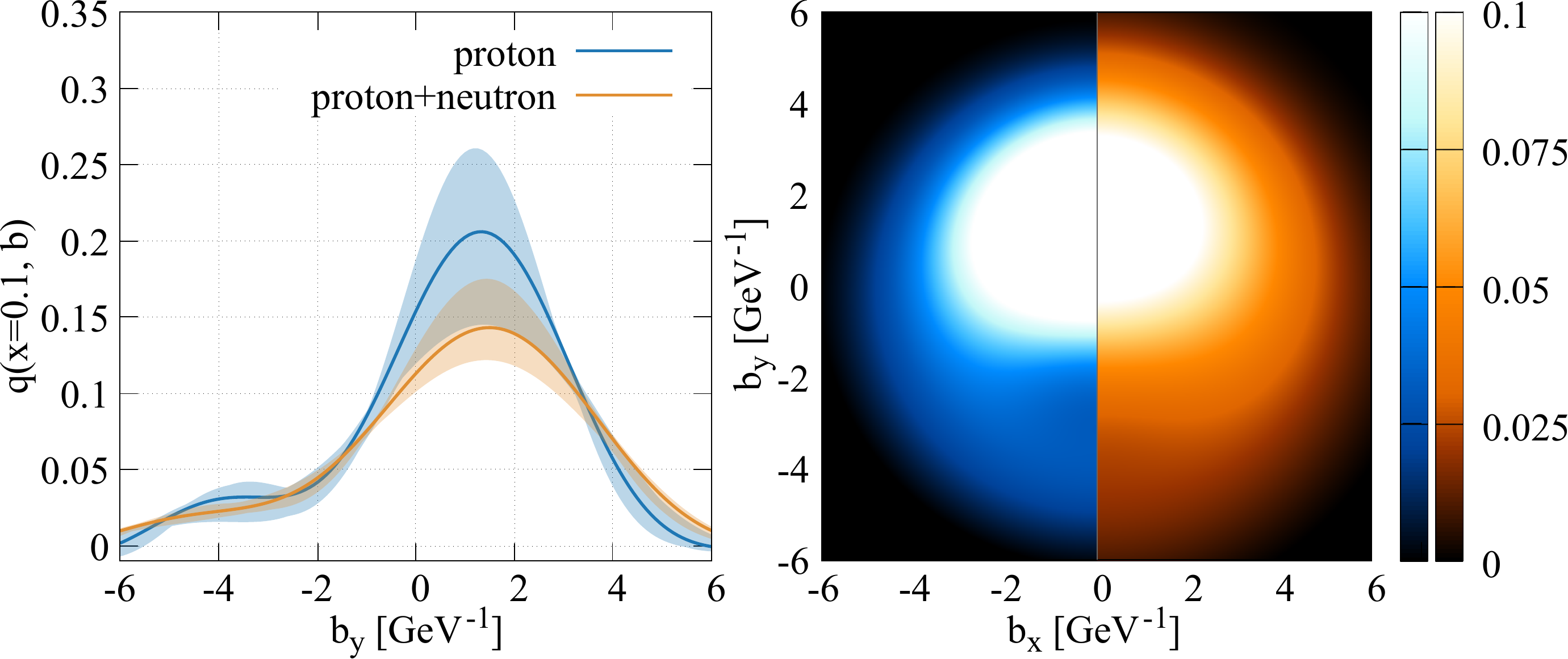}
        \caption{$x=0.1$, transversely polarized up quark in proton.}
        \label{subfig:tom_up_pol}
    \end{subfigure}
\end{figure}
\begin{figure}[p]
    \ContinuedFloat
    \centering
    \begin{subfigure}[b]{\linewidth}
        \centering
        \includegraphics[width=\linewidth]{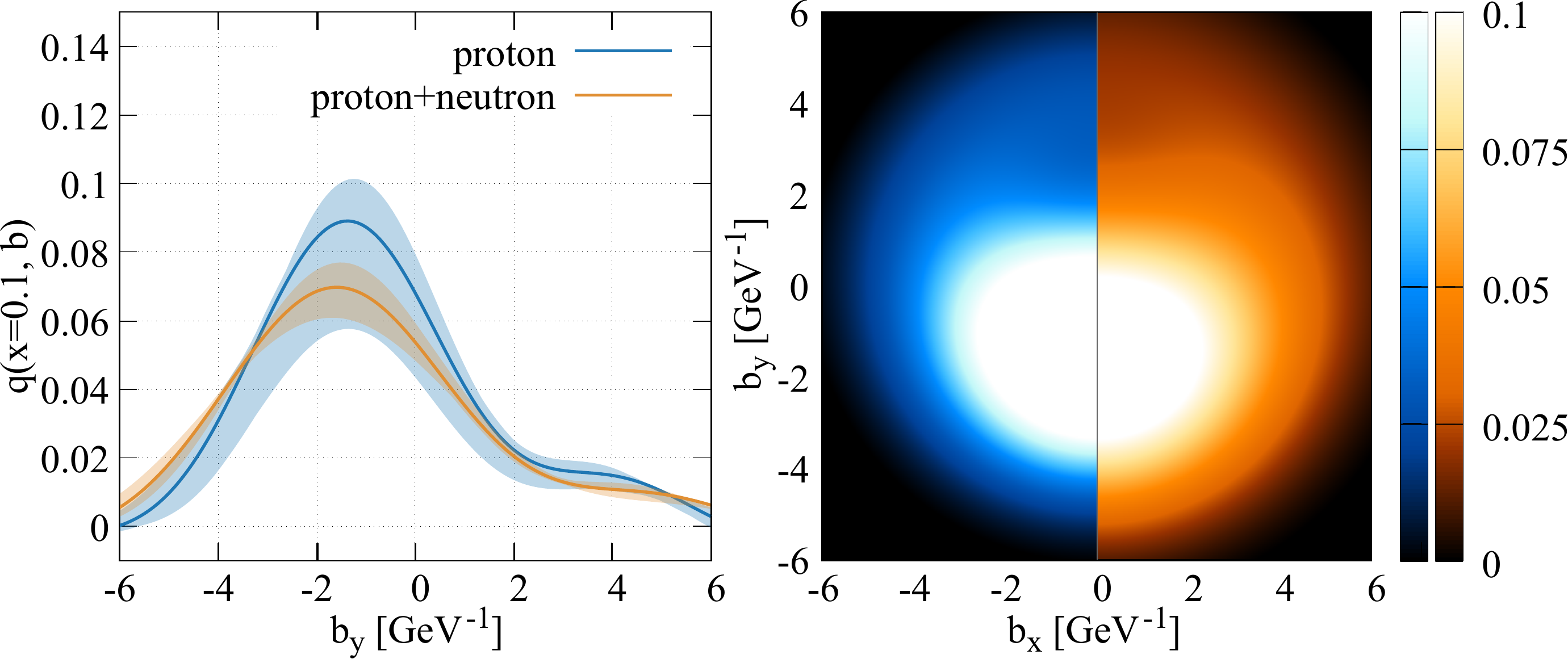}
        \caption{$x=0.1$, transversely polarized down quark in proton (heatmap is multiplied by two).}
        \label{subfig:tom_down_pol}
    \end{subfigure}
        \caption{Nucleon tomography for unpolarized and transversely polarized proton for $x=0.1$. The left column shows the cross-section for $b_x=0$, of the heat map shown in the right column. We present both the result for the fit including only protons (blue) and both nucleons (orange). We show the central result on the heatmap.}
        \label{fig:tomography_x01}
\end{figure}
\begin{figure}[p]
    \centering
    \begin{subfigure}[b]{\linewidth}
        \centering
        \includegraphics[width=\linewidth]{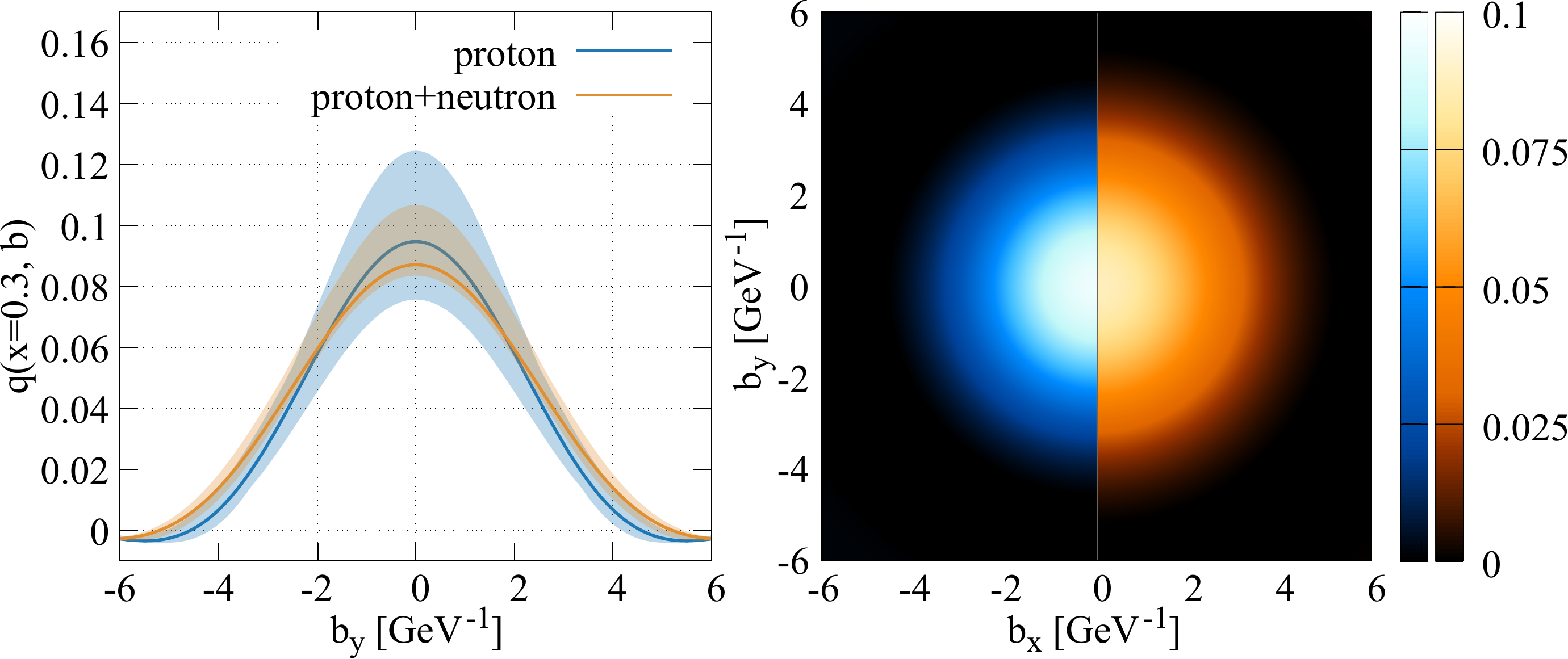}
        \caption{$x=0.3$, unpolarized up quark in proton.}
        \label{subfig:tom_up_03}
    \end{subfigure}
\end{figure}
\begin{figure}[p]
    \ContinuedFloat
    \centering
    \begin{subfigure}[b]{\linewidth}
        \centering
        \includegraphics[width=\linewidth]{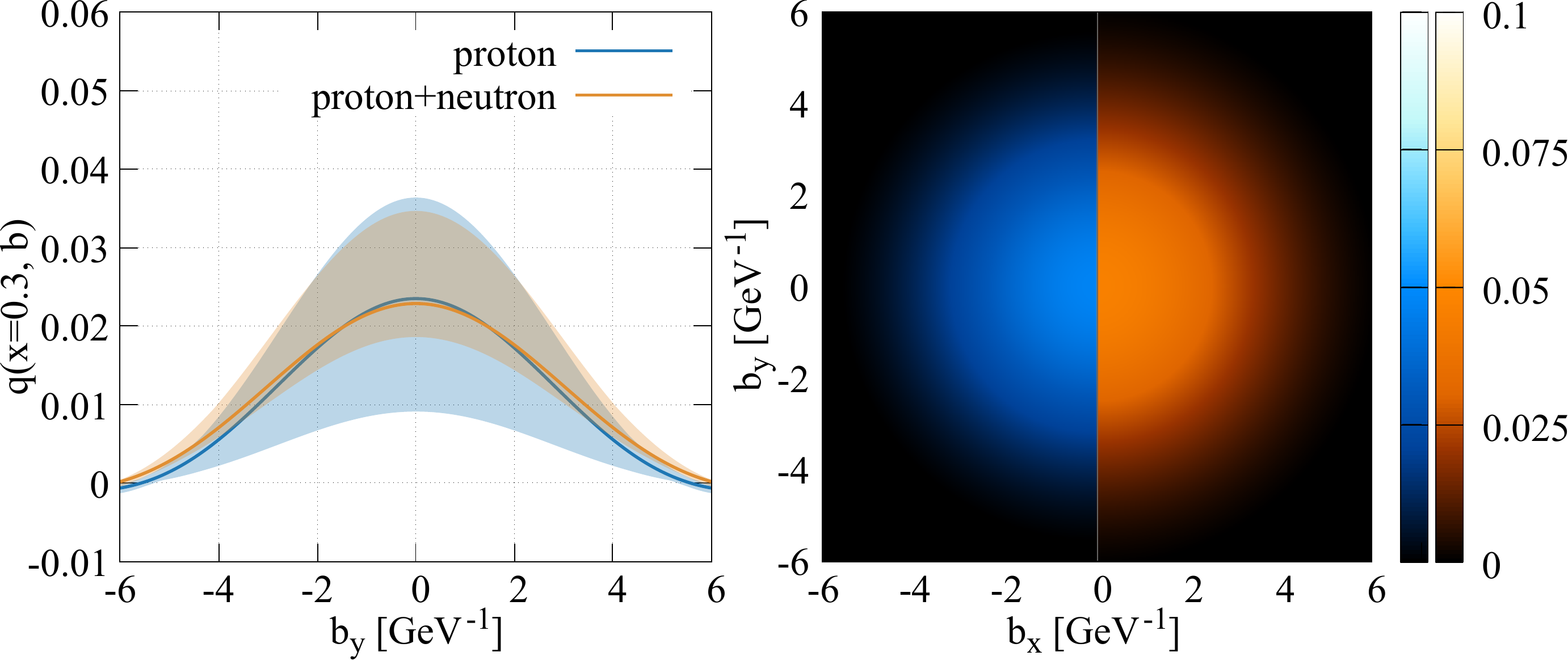}
        \caption{$x=0.3$, unpolarized down quark in proton (heatmap is multiplied by two).}
        \label{subfig:tom_down_03}
    \end{subfigure}
    \centering
    \begin{subfigure}[b]{\linewidth}
        \centering
        \includegraphics[width=\linewidth]{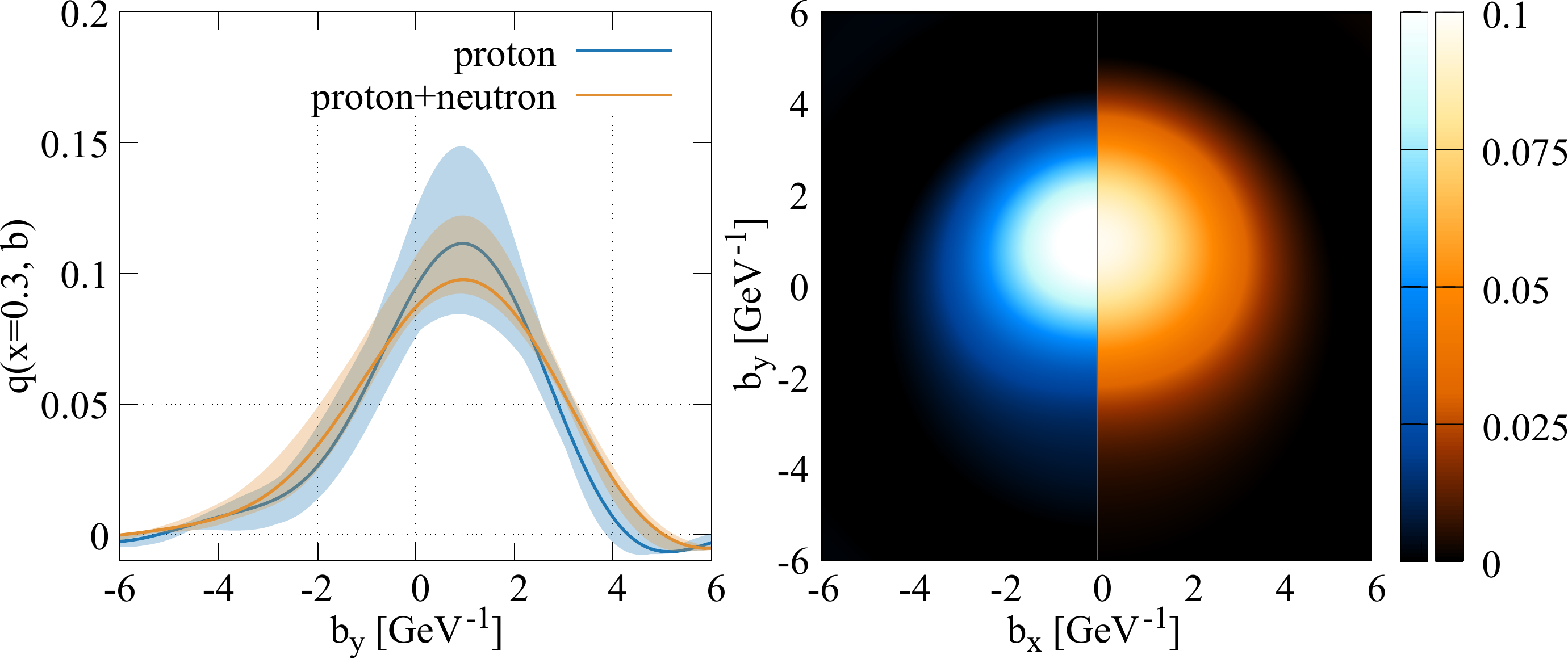}
        \caption{$x=0.3$, transversely polarized up quark in a proton.}
        \label{subfig:tom_up_pol_03}
    \end{subfigure}
    \centering
    \begin{subfigure}[b]{\linewidth}
        \centering
        \includegraphics[width=\linewidth]{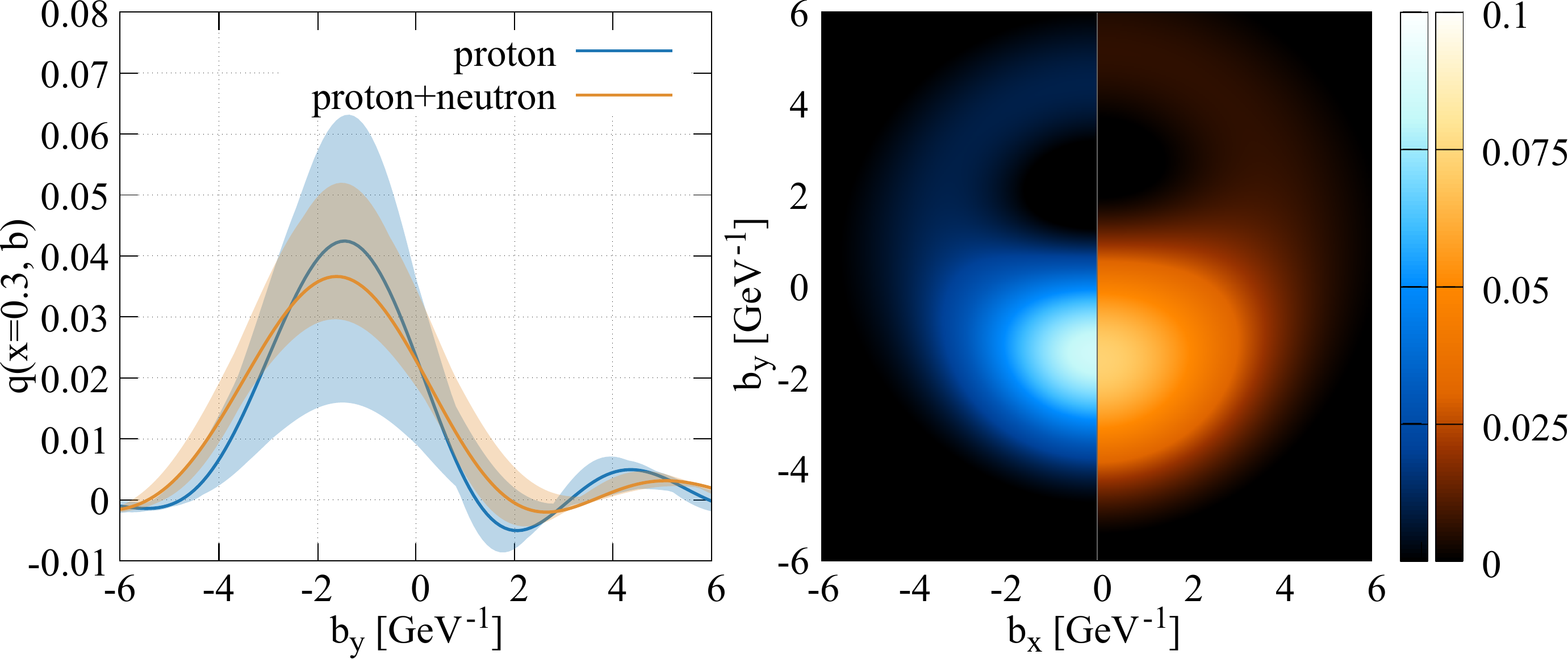}
        \caption{$x=0.3$, transversely polarized down quark in proton (heatmap is multiplied by two).}
        \label{subfig:tom_down_pol_03}
    \end{subfigure}
        \caption{Nucleon tomography for unpolarized and transversely polarized proton for $x=0.3$. The left column shows the cross-section for $b_x=0$, of the heatmap shown in the right column. We present both the result for the fit including only the proton (blue) and both nucleons (orange). We show the central result on the heatmap.}
        \label{fig:tomography_x03}
\end{figure}

\FloatBarrier

\FloatBarrier

\section{Summary and prospects}
\label{Sec:summary and prospects}
In this work, we studied valence-quark nucleon tomography in a spectator model containing both scalar and axial-vector diquark configurations. 
We calculated the twist-2 vector GPDs,
$H^{q}(x,\xi,t)$ and $E^{q}(x,\xi,t)$ and derived analytic expressions for their zero-skewness limits. The scalar and axial-vector contributions were combined to obtain the flavor-separated GPDs, $H^{\text{up},\text{down}}$ and $E^{\text{up},\text{down}}$. 
A central new result is the explicit analytic expression for the axial-vector-diquark contribution to the helicity-flip GPD $E^{q}$. 
As an independent check, we derived the corresponding
quasi-GPD and verified its reduction to the light-cone result in the infinite-momentum limit.

We also investigated the regularization of the transverse-momentum integration. 
A fixed hard cutoff, as is typically used in the literature, was found to produce an unphysical
growth of the axial-vector contribution at sufficiently large $\lvert\vec{\Delta}_{\perp}\rvert$. 
Therefore, we introduced the momentum-transfer-dependent quasi-hard cutoff (see Eq.~(\ref{eq:quasihardcutoff})).
This prescription treats the transverse momenta entering the initial and final nucleon--quark--diquark vertices symmetrically, reduces to the usual hard cutoff in the forward limit, and prevents the
unphysical large-$\lvert\vec{\Delta}_{\perp}\rvert$
behavior encountered with a fixed integration boundary.

The analytic results allowed us to examine the endpoint behavior of the distributions. For finite quark and diquark masses, the valence-quark PDFs remain finite as $x\to0$. 
Decreasing the active-quark mass or increasing the spectator mass enhances the small-$x$ distributions, but does not produce a genuine small-$x$ divergence. We also demonstrated that the derivatives of GPDs with respect to $t$ vanish in the $x\to1$ limit.
This result has a direct tomographic interpretation: a quark carrying the full longitudinal momentum of the nucleon becomes localized at the transverse center of momentum.

The Mellin moments of the calculated GPDs were used to obtain the
electromagnetic and gravitational form factors entering the
calibration of the model. We considered two complementary
calibrations:
\begin{itemize}
\item The first used proton observables only, leaving neutron quantities as predictions. It accurately reproduced the proton observables included as input and gave a qualitatively consistent description of the neutron, although the magnitude of the neutron mean-square charge radius was underestimated. 
\item  In the second calibration, proton and neutron observables were included simultaneously. The resulting observables generally differed from their reference values by approximately ten percent or less.
\end{itemize}
The model parameters obtained from the two calibrations have overlapping uncertainty ranges. Thus, including neutron information does not require a qualitatively different parameter set. Instead, it reduces the parameter uncertainties and more strongly constrains the
axial-vector sector, which controls an essential part of the flavor dependence. This consistency supports the use of a common scalar-plus-axial-vector spectator description for the proton and neutron at the present level of accuracy.

Using the two calibrated parameter sets, we calculated the valence-quark PDFs, $u_{v}(x)=H^{\text{up}}(x,0,0),\;
d_{v}(x)=H^{\text{down}}(x,0,0)$, as well as zero-skewness GPDs,
$H^{\text{up},\text{down}}(x,0,t)$ and $E^{\text{up},\text{down}}(x,0,t)$.
We then Fourier transformed the zero-skewness GPDs with respect to the transverse momentum transfer to obtain the impact-parameter distributions of unpolarized up and down quarks in unpolarized and transversely polarized nucleons. For an unpolarized nucleon, the resulting
transverse distributions are azimuthally symmetric. 
As $x$ increases, they become more localized near the transverse center of momentum, consistent with the analytically established $x\to1$ behavior. 
For a transversely polarized nucleon, the contribution from $E^{q}$ generates a sideways deformation of the
quark distribution. The results obtained from the proton-only and proton--neutron calibrations are qualitatively similar, although their differences reflect the additional constraints imposed by the neutron observables.

Several directions for future work naturally follow from the present analysis. For quantitative comparisons with
lattice-QCD determinations of PDFs and GPDs, the distributions obtained here should be associated with a well-defined low model scale and subsequently evolved to the scale at which the comparison is performed. The calculated electromagnetic and gravitational form
factors, as well as the transverse spatial distributions, can also be confronted more systematically with 
lattice-QCD results. Further studies of the regulator dependence and of more general nucleon-quark-diquark vertex structures will be important for assessing the model dependence of the predictions.
Finally, extending the framework to nonzero skewness, helicity and transversity GPDs, and explicit sea-quark and gluon degrees of freedom would provide a more complete description of nucleon structure beyond the present valence-quark approximation.

\section{Acknowledgements}
J.~T was supported by the Adam Mickiewicz University in Poznań under the Initiative for Excellence – Research University (ID-UB) grant No. 212/64/UAM/0015. S.~B. acknowledges support from the U.S. National Science Foundation under Grant No.~PHY-2609761. K.~C. is supported by the National Science Centre (Poland) grant OPUS No. 2021/43/B/ST2/00497.

\bibliographystyle{JHEP}
\bibliography{bibliography}

\appendix
\section{Cutoff discussion}
\label{NR}
\begin{figure}[htbp]
    \centering
        \centering
        \includegraphics[width=\linewidth]{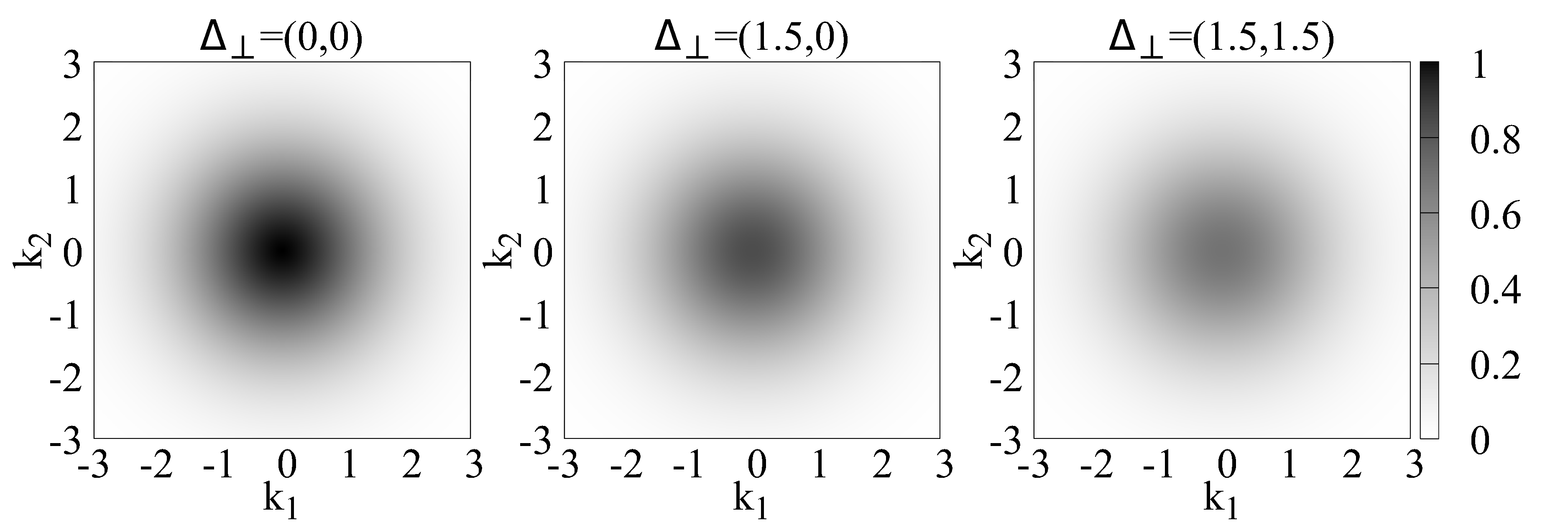}
    \vspace{-0.5cm}
    \caption{$g_+\left(\vec{k}_\bot+ \vec{\Delta}_\bot/2 \right)\cdot g_-\left(\vec{k}_\bot- \vec{\Delta}_\bot/2 \right)$ value for form factor given by \Cref{eq:cutoff_soft}, for different values of $\vec{\Delta}_\bot$, and $\Lambda = 2.5\,[\text{GeV}]$. All values in plots are given in GeV.}
    \label{fig:soft_cutoff}
\end{figure}

\begin{figure}[htbp]
    \centering
        \centering
        \includegraphics[width=\linewidth]{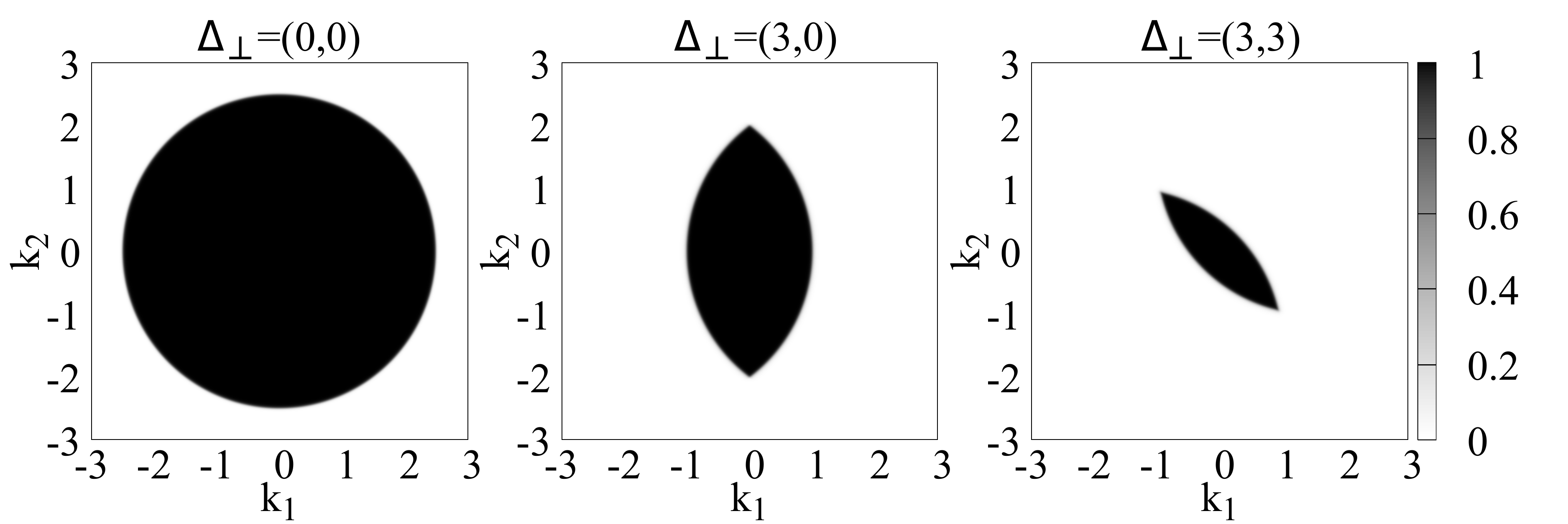}
    \vspace{-0.5cm}
    \caption{$g_+\left(\vec{k}_\bot+ \vec{\Delta}_\bot/2 \right)\cdot g_-\left(\vec{k}_\bot- \vec{\Delta}_\bot/2 \right)$ value for form factor given by \Cref{eq:cutoff_quasisoft}, for different values of $\vec{\Delta}_\bot$, $\Lambda = 2.5\,[\text{GeV}]$ and $\alpha = 0.05$. All values in plots are given in GeV.}
    \label{fig:quasisoft_cutoff}
\end{figure}

\begin{figure}[htbp]
    \centering
        \centering
        \includegraphics[width=\linewidth]{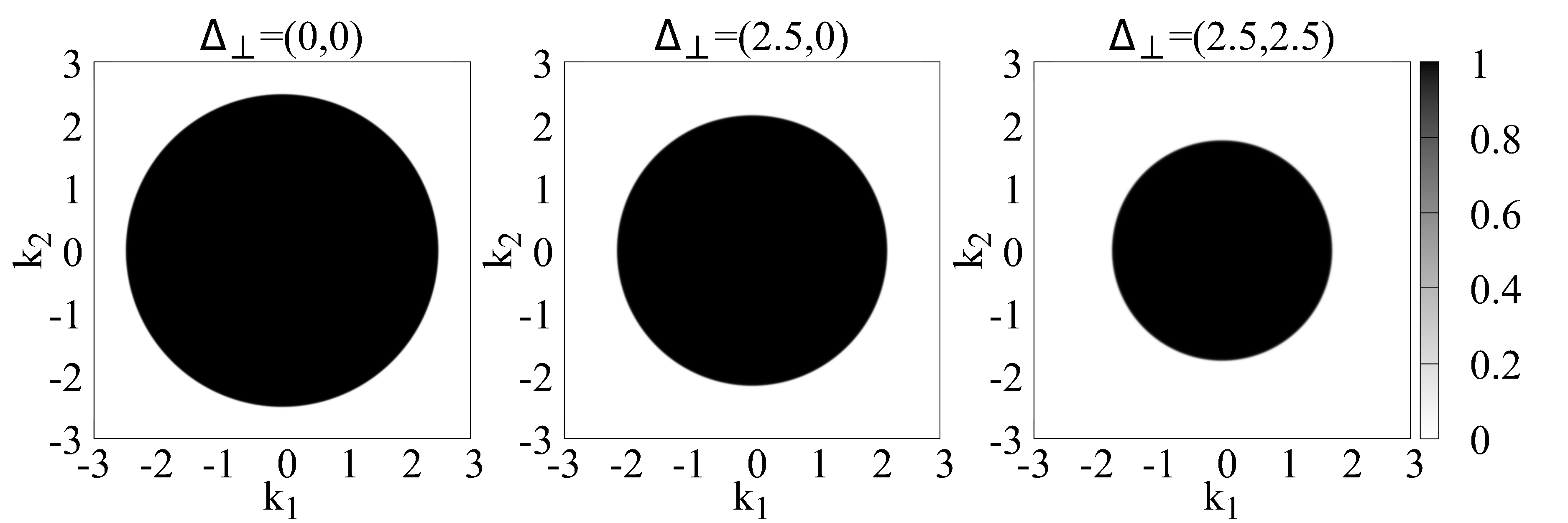}
    \vspace{-0.5cm}
    \caption{Soft cutoff value given by \Cref{eq:cutoff_quasisoft}, for different values of $\vec{\Delta}_\bot$, $\Lambda = 2.5\sqrt{2}\,[\text{GeV}]$ and $\alpha = 0.05$. All values on plots are given in GeV.}
    \label{fig:quasisoft_real_cutoff}
\end{figure}

 We tested several different approaches to determining the cutoff. These included a hard cutoff with a fixed integration limit (like in \cite{Bhattacharya:2019cme}), and a soft cutoff employing a Gaussian function\footnote{ One can also think about dipolar cutoff \cite{GAMBERG2015112}, but this kind of cutoff is $\vec{\Delta}_\bot$ independent.}\cite{PhysRevD.101.114005} where the coupling constant $g$ is multiplied by an exponential form factor and becomes a function of $\vec{k}_\bot\pm \vec{\Delta}_\bot/2$ (for more information, see \cite{tan2024chiraleventwist3gpdsproton}),
  \begin{equation}
  \label{eq:cutoff_soft}
     g \to g\left(\vec{k}_\bot\pm \vec{\Delta}_\bot/2 \right) = g_\pm \exp{\left[-\frac{\left(\vec{k}_\bot \pm \vec{\Delta}_\bot/2\right)^2}{\Lambda^2}\right]}\, .
 \end{equation}
  We did not intend to modulate the values of the resulting functions using a cutoff; for this reason, we rejected the Gaussian form, because, as one can see in \Cref{fig:soft_cutoff}, the value of $g_+\left(\vec{k}_\bot+ \vec{\Delta}_\bot/2 \right)\cdot g_-\left(\vec{k}_\bot- \vec{\Delta}_\bot/2 \right)$  is not a sudden drop to zero at the cutoff limit, but rather its value decreases gradually to zero, which results in significant modulation of the function’s value.
For this reason, we do not use a Gaussian form, as well as all other functional forms exhibiting significant variation in values within the range from $\vec{k}_\bot \pm \vec{\Delta}_\bot/2=0$ to the specified cutoff value, where the function’s value should fade away. 
Instead, we are looking for form factors such that the resulting value is an almost constant function, up to a certain limit beyond which it begins to decay rapidly. We tested several forms of exponential functions $g_\pm(\vec k_\bot\pm\vec\Delta_\bot/2)$  such that they do not significantly affect the value of the integral, apart from reducing the function’s value to zero above a specified cutoff value. However, it was not possible to find a suitable functional such that it would retain its radial symmetry as the value of $\vec\Delta_\bot$ increases (i.e., regardless of the components). For example, we tested the following form factor:
 \begin{equation}
 \label{eq:cutoff_quasisoft}
    g \to g\left(\vec{k}_\bot\pm \vec{\Delta}_\bot/2 \right)  = g_\pm\cdot \frac{\sqrt{1+\exp{(\Lambda^2/\alpha)}}}{\sqrt{ \exp{(k_T\pm \Delta_T/2)^2/\alpha +\exp{(\Lambda^2/\alpha)} } } } \, .
\end{equation}
At a first glance, such a term satisfies all the requirements, but a problem arises, because $g_+\left(\vec{k}_\bot+ \vec{\Delta}_\bot/2 \right)\cdot g_-\left(\vec{k}_\bot- \vec{\Delta}_\bot/2 \right)$   is not a function of $\vec\Delta_\bot^2$ but of $\vec{\Delta}_\bot$, so it depends not only on the magnitude of the transverse momentum transfer but also on its components. For this reason, it is not shrinking in a radially symmetrical way, as can be seen in \Cref{fig:quasisoft_cutoff}.
It is possible to obtain GPD-modulating functions in the intended manner, for example:
 \begin{equation}
     \,\frac{1+e^{\frac{-\Lambda^2}{\alpha}}}{1+e^{\frac{2k_\bot^2+\frac{1}{2} \vec\Delta_\bot^2-\Lambda^2 }{\alpha}}} \, ,
 \end{equation}
 then this kind of cutoff shrinks radially with respect to increasing magnitude of $\vec{\Delta}_\bot$, as shown in \Cref{fig:quasisoft_real_cutoff}.
However, it is not possible to obtain this form as a product of the functions $g_+(k_\bot+\vec\Delta_\bot/2)\cdot g_-(k_\bot-\vec\Delta_\bot/2)$. Leaving the hard cutoff in place, however, the hard cutoff causes a problem with the axial-vector GPD for large magnitudes of $|\vec\Delta_\bot|$. This kind of cutoff does not depend on $\vec{\Delta}_\bot$, and only limits the $\vec{k}_\bot$ value. But we observed behavior where, for larger values of $|\vec\Delta_\bot|$, axial-vector diquark GPDs, after an initial drop, begin to rise to infinity as $|\vec\Delta_\bot|$ increases.
For this reason, we propose a hard cutoff with integration limits dependent on the value of $|\vec\Delta_\bot|$ as an alternative to the standard hard cutoff. A cutoff defined in this way is referred to as a quasi-hard cutoff. In the GPD structure, $\vec k_\bot \pm \vec\Delta_\bot/2$ naturally appears as the transverse momentum of the hadron at each of the interaction vertices. Hence (as in the case of a soft cutoff), we define a quasi-hard cutoff using precisely these quantities. Since we are working in the symmetric frame, we need not concern ourselves with the components of the transverse momentum, but only with its magnitude; therefore, we calculate the sum of the squares of the transverse momenta of the hadrons entering and leaving the scattering event. This quantity is related to the average perpendicular momentum; we then restrict this value to a certain quantity $2\beta^2$, which gives us the following equation,
\begin{equation}
    \label{eq:cutoff_nierownosc}
    2k^2_\bot + \vec\Delta^2_\bot/2 \leq 2\beta^2.
\end{equation}
Solving for $k_\bot^2$ gives the maximum transverse momentum $\Lambda$:
\begin{equation}
    \label{eq:cutoff}
    \Lambda = \sqrt{\beta^2 - \vec\Delta^2_\bot/4}.
\end{equation}
In the forward limit ($|\vec\Delta_\bot|=0$), the maximum average parton transverse momentum is $\Lambda=\beta$. So $\beta^2$ can be interpreted as the average squared transverse parton momentum. For $\Lambda=0$, the value $\beta = |\vec\Delta_\bot|/2 \xrightarrow{} 2\beta = |\vec\Delta_\bot|$. This means, therefore, that two times the root of the average squared momentum is equal to the maximum value of the momentum transfer; this describes a situation in which a parton changes its momentum from $\beta$ to $-\beta$. Interestingly, \Cref{eq:cutoff} is a transformed version of the equation describing the uncertainty in the parton’s transverse momentum $ \left<k_\bot^2\right> - \left< k_\bot\right>^2  = \beta^2 - \Lambda^2 = \vec\Delta^2_\bot/4$.

\FloatBarrier
\section{Convergence of axial-vector diquark QGPD}
\label{app:convergence}

\begin{figure}[htbp]
    \centering
    \begin{subfigure}[b]{0.51\linewidth}
        \centering
        \includegraphics[width=\linewidth]{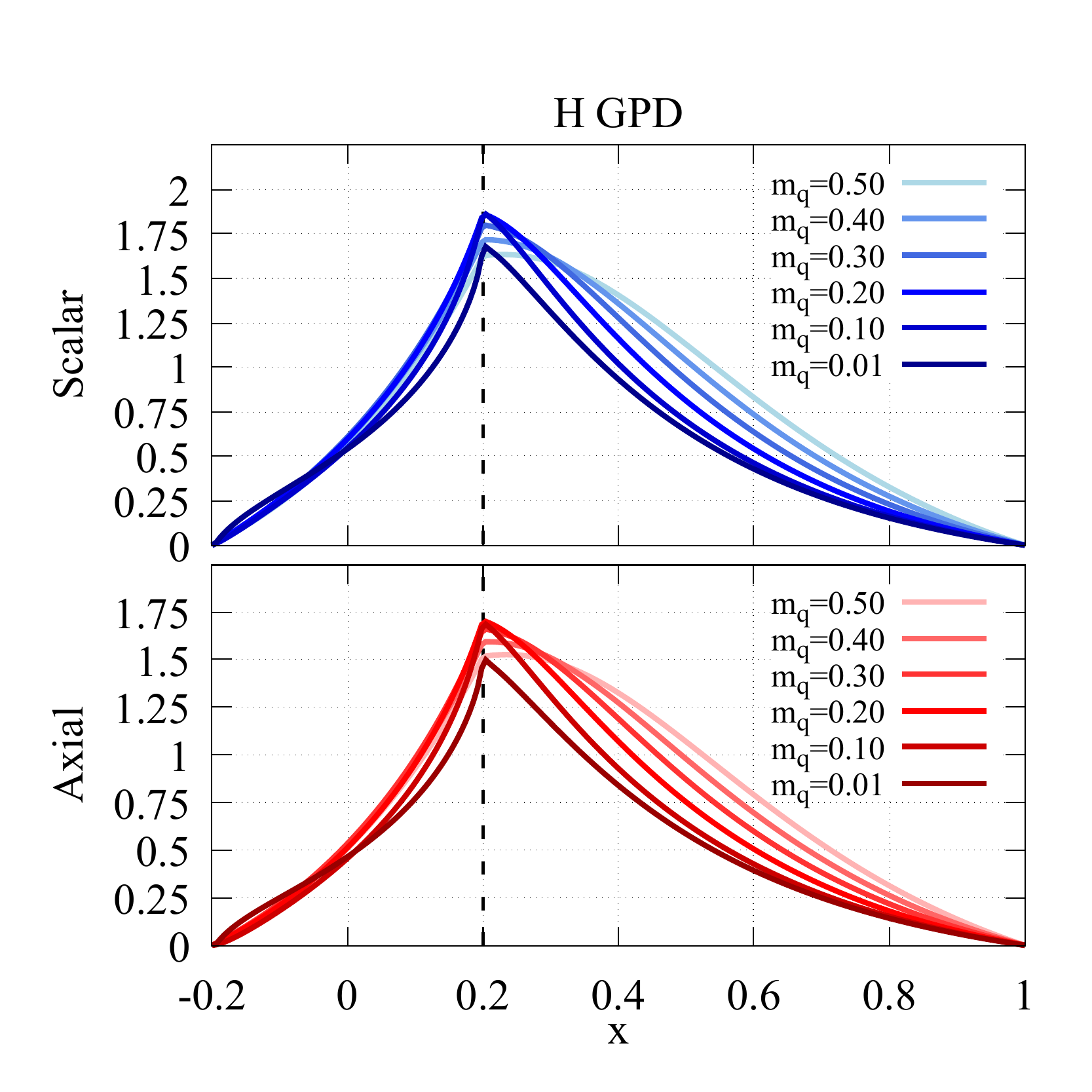}
    \end{subfigure}
\hspace{-0.7cm}
    \begin{subfigure}[b]{0.51\linewidth}
        \centering
        \includegraphics[width=\linewidth]{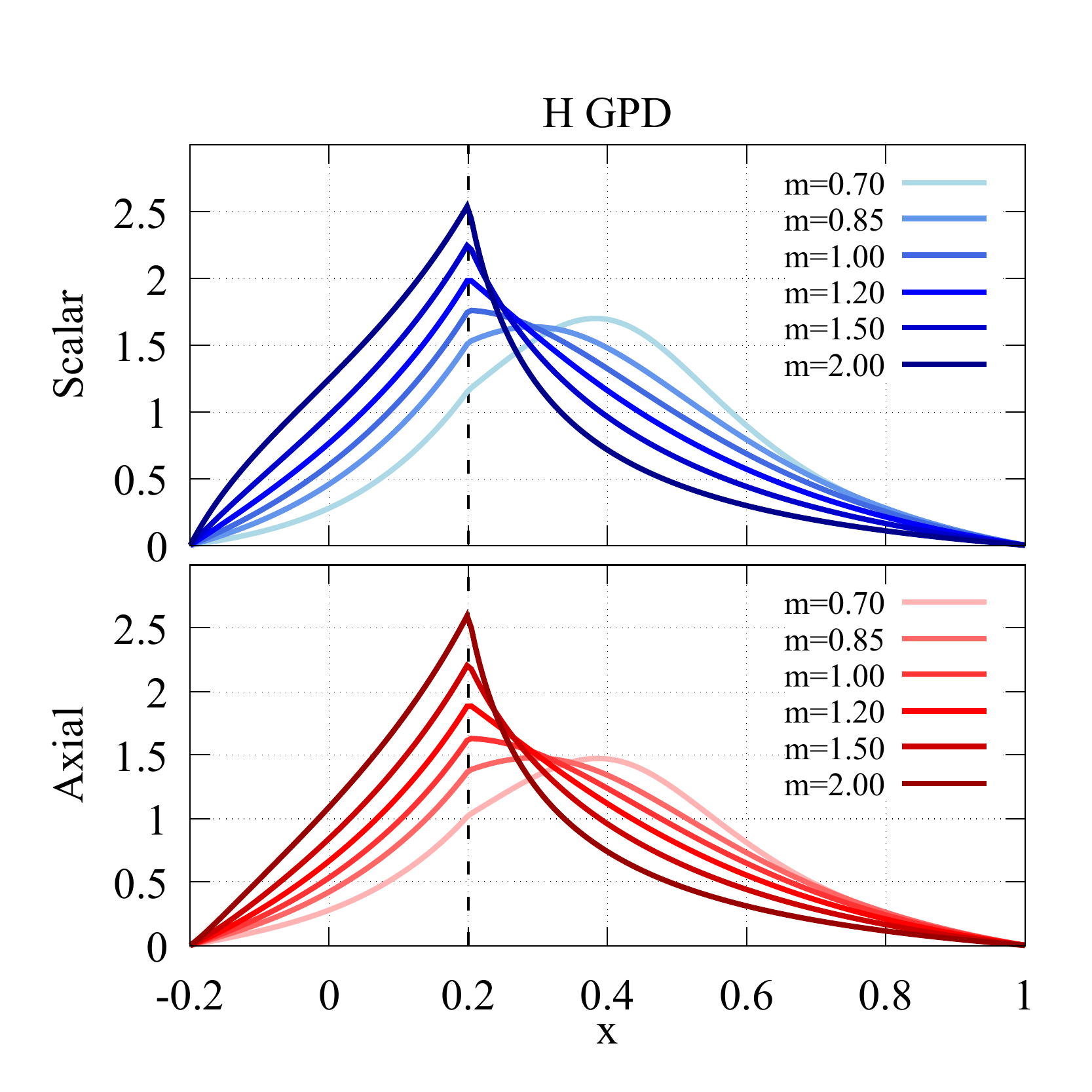}
    \end{subfigure}
    \vspace{-0.5cm}
    \caption{Zero transverse momentum transfer $H$ GPD as a function of $x$ for $\beta = 1$ GeV, $M = 0.939$ GeV and $\xi=0.2$. Left panel: results for different values of $m_q$ ($m_s=m_a=1$ GeV). Right panel: results for different values of $m_s=m_a=m$ ($m_q=0.35$ GeV). The mass values in the legend are given in GeV.}
    \label{fig:H_xi_FL_mqm}
\end{figure}

\begin{figure}[htbp]
    \centering
    \begin{subfigure}[b]{0.51\linewidth}
        \centering
        \includegraphics[width=\linewidth]{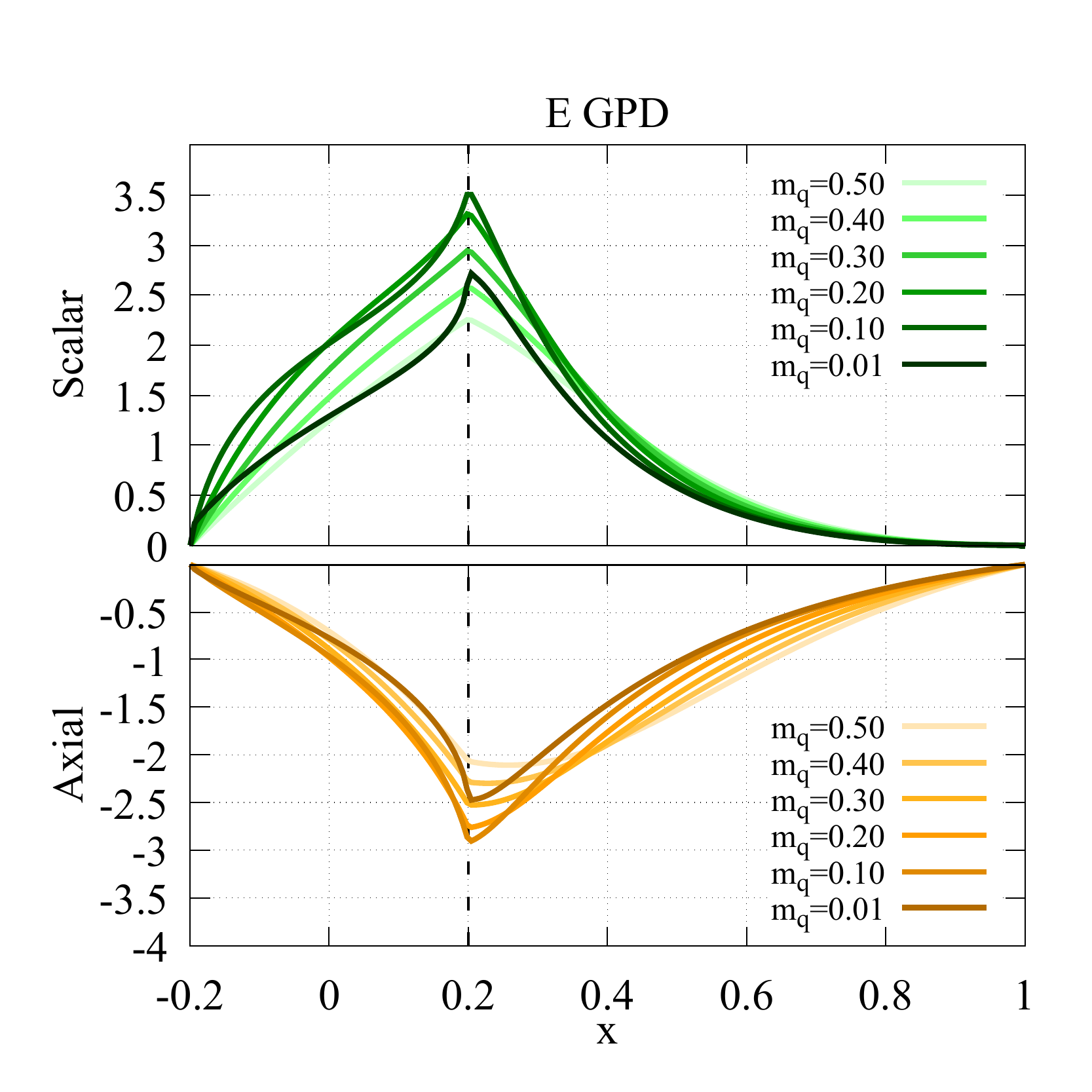}
    \end{subfigure}
\hspace{-0.7cm}
    \begin{subfigure}[b]{0.51\linewidth}
        \centering
        \includegraphics[width=\linewidth]{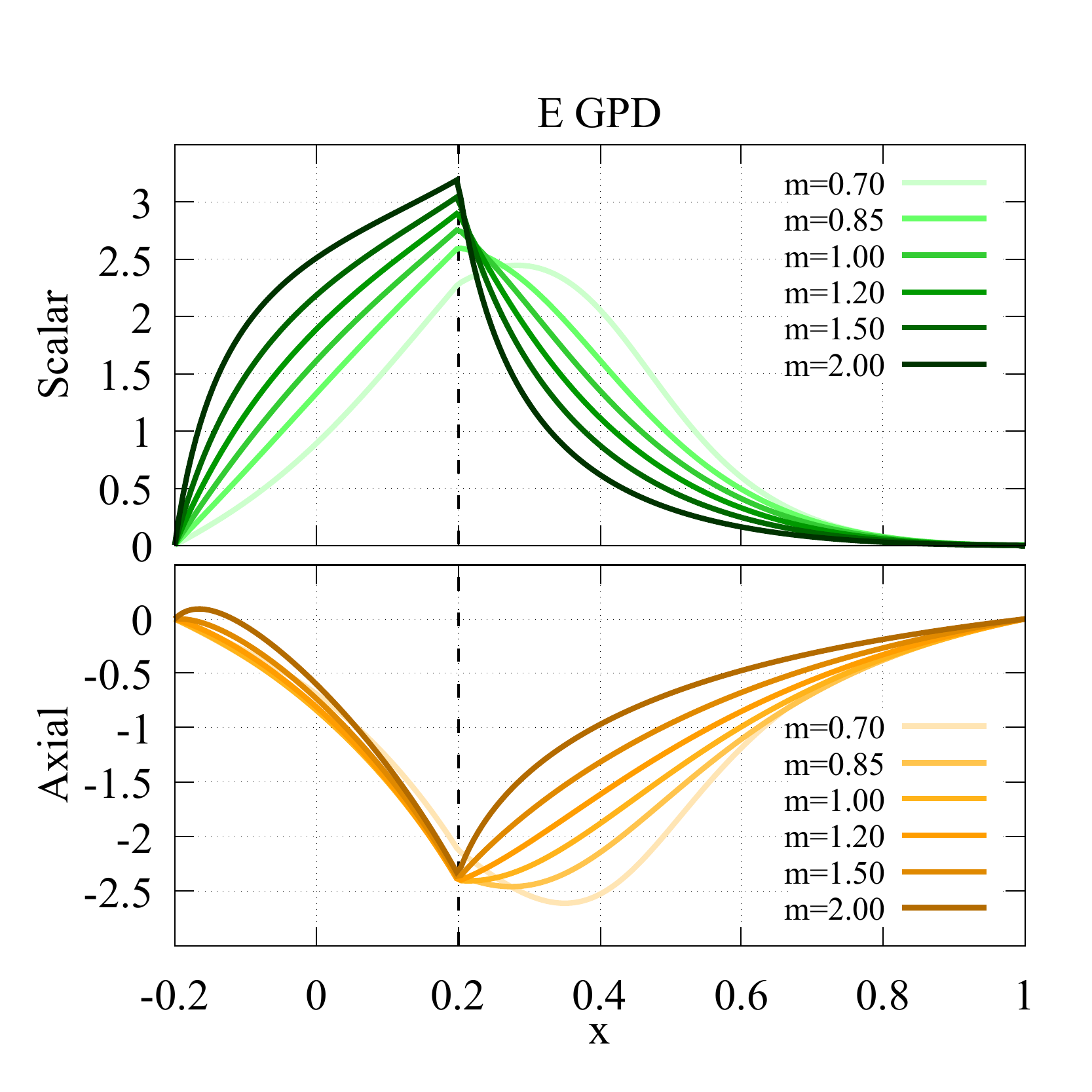}
    \end{subfigure}
    \vspace{-0.5cm}
    \caption{Zero transverse momentum transfer $E$ GPD as a function of $x$ for $\beta = 1$ GeV, $M = 0.939$ GeV and $\xi=0.2$. Left panel: results for different values of $m_q$ ($m_s=m_a=1$ GeV). Right panel: results for different values of $m_s=m_a=m$ ($m_q=0.35$ GeV). The mass values in the legend are given in GeV.}
    \label{fig:E_xi_FL_mqm}
\end{figure}

Before we show the convergence of the E axial QGPD to the E GPD, we will first present results illustrating the general properties of nonzero skewness GPDs in DSM. \Cref{fig:H_xi_FL_mqm} shows the results for H GPDs with various quark and diquark masses at $\xi=0.2$, while \Cref{fig:E_xi_FL_mqm} shows the corresponding results for E GPDs.

In the calculations for this paper, we used a function describing the $E$ GPD for an axial-vector diquark, a result that was previously not available in the literature. In addition, we also calculated the quasi-GPD (QGPD) for this diquark. Following \cite{Bhattacharya:2018zxi}, we calculated QGPDs for both $\Gamma = \gamma^0$ and $\Gamma = \gamma^3$ cases. For details on QGPD calculations in this model, we refer to \cite{Bhattacharya:2018zxi} and \cite{Bhattacharya:2019cme}. 
For $\Gamma = \gamma^0$, we get:
    \begin{align}
        \frac{2}{\delta}N^a_{E(0)}&= -2\frac{m_q^2}{m_a^2}M^2+2M\left[\frac{m_q}{m_a^2}\left(M^2\left\{ \frac{k^0}{\delta P^3}-1\right\}-\frac{t}{4}\left\{ 1+\frac{k^0}{\delta P^3}\right\}\right)-4k^0\frac{m_q}{\delta P^3}\right] \nonumber \\
        & \quad  - 4M^2x\left(\frac{k^0}{\delta P^3}+1\right) +2\frac{k^0}{\delta P^3}\frac{M^2}{m_a^2}\left[2 x (P^3)^2+x\left(M^2-\frac{t}{4}\right)  \right] \nonumber \\
        & \quad- 2\frac{M^2}{m_a^2}\left[(k^0)^2+x^2(P^3)^2+\vec{k}^2_\bot -M^2 x +\frac{t}{4}(x+1)\right] \nonumber \\
        & \quad+ \frac{4}{m_a^2}\left(M^2-\frac{t}{4} \right)\delta M^2\xi \frac{\vec{k}_\bot\cdot \vec{\Delta}_\bot}{\vec{\Delta}^2_\bot}\left(\frac{k^0}{\delta P^3}+1  \right)-8\xi M^2\delta \frac{\vec{k}_\bot\cdot \vec{\Delta}_\bot}{\vec{\Delta}^2_\bot} \left(\frac{k^0}{\delta P^3} +1\right)
    \end{align}
and the result for $\Gamma = \gamma^3$ reads:
    \begin{align}
        2 N_{E(3)}^a&= - 2\frac{m_q^2}{m_a^2}M^2+2m_q M \left[\frac{M^2}{m_a^2}(x-1)-\frac{t}{4 m_a^2}(x+1) -4x \right] - 4 M^2 \frac{k^0}{\delta P^3}(x+1) \nonumber \\
        &\quad +2k^0 \delta P^3 \frac{M^2}{m_a^2}\left[ -2 x +\frac{x}{(\delta P^3)^2}\left(M^2-\frac{t}{4}\right)\right] + 2\frac{M^2}{m_a^2}\Bigg[(k^0)^2+(P^3)^2x^2-\vec{k}^2_\bot\nonumber \\
        & \quad +M^2\frac{k^0}{P^3\delta}-\frac{t}{4}\left(\frac{k^0}{\delta P^3}+1 \right) \Bigg] 
        +4\frac{M^2}{m_a^2}\left(M^2-\frac{t}{4}\right)\frac{\xi}{\delta}  \frac{\vec{k}_\bot\cdot \vec{\Delta}_\bot}{\vec{\Delta}^2_\bot} (x+1)\nonumber \\
        & \quad- 8 \frac{M^2 \xi}{\delta} \frac{\vec{k}_\bot\cdot \vec{\Delta}_\bot}{\vec{\Delta}^2_\bot} (x+1),
    \end{align}
where the $E$ GPD can be calculated as follows:
\begin{equation}
\begin{split}
E_{Q(0(3))}(x,\xi,t;P^3)=&-\frac{g^2P^3}{(2\pi)^3}\int_0^{2\pi}d\theta \int_0^\Lambda \text{d}{k}_\bot \; k_\bot\\
&\cdot \Bigg[  \frac{N_{E(0(3))(k^0_{1-})}}{
				(k^0_{1-}-k^0_{1+})
				(k^0_{1-}-k^0_{2-})
				(k^0_{1-}-k^0_{2+})	
				(k^0_{1-}-k^0_{3-})
				(k^0_{1-}-k^0_{3+})	}    \\
				&+  \frac{N_{E(0(3))(k^0_{2-})}}{
				(k^0_{2-}-k^0_{1-})
				(k^0_{2-}-k^0_{1+})
				(k^0_{2-}-k^0_{2+})	
				(k^0_{2-}-k^0_{3-})
				(k^0_{2-}-k^0_{3+})	}     \\
								&+  \frac{N_{E(0(3))(k^0_{3-})}}{
				(k^0_{3-}-k^0_{1-})
				(k^0_{3-}-k^0_{1+})
				(k^0_{3-}-k^0_{2-})	
				(k^0_{3-}-k^0_{2+})
				(k^0_{3-}-k^0_{3+})	}    \Bigg] ,
\end{split}
\end{equation}
with:
\begin{equation}
k^0_{1\pm}=\xi P^3\pm \sqrt{(x-\xi \delta)^2(P^3)^2+\left(\vec{k}_\bot+\frac{\vec{\Delta}_\bot}{2}\right)^2+m_q^2-i\varepsilon} ,
\end{equation}
\begin{equation}
k^0_{2\pm}=-\xi P^3\pm \sqrt{(x+\xi \delta)^2(P^3)^2+\left(\vec{k}_\bot-\frac{\vec{\Delta}_\bot}{2}\right)^2+m_q^2-i\varepsilon} ,
\end{equation}
\begin{equation}
k^0_{3\pm}=\delta P^3\pm \sqrt{(1+x)^2(P^3)^2+\vec{k}^2_\bot+m_a^2-i\varepsilon} .
\end{equation}
\begin{figure}[h]
    \centering
    \includegraphics[width=\linewidth]{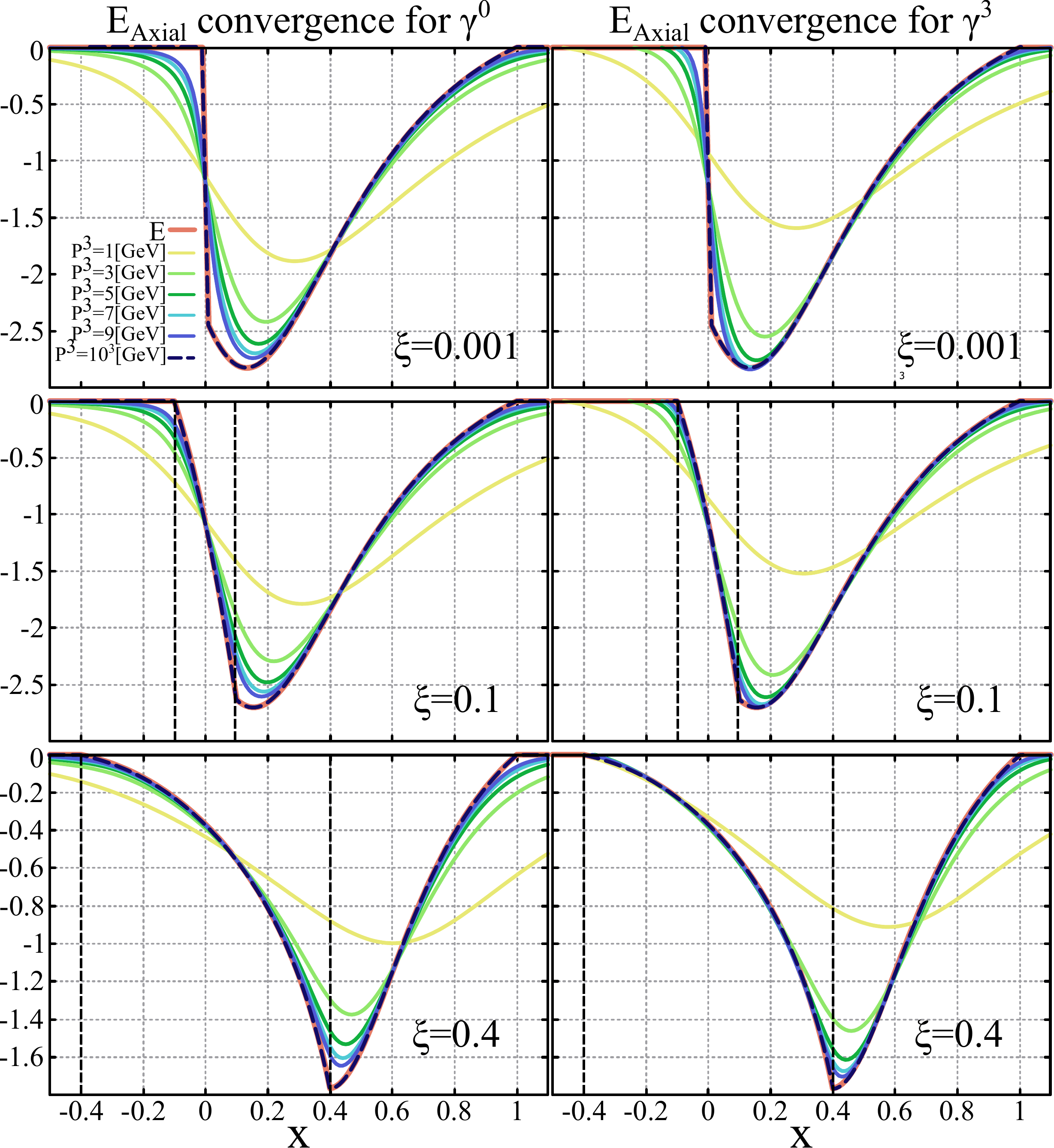}
    \caption{E QGPD for axial-vector diquark as a function of $x$ for different values of $\xi$ and $P^3$. Left panel: results for $\gamma^0$. Right panel: results
for $\gamma^3$. The standard axial-vector diquark $E$ GPD is shown for comparison. The limits of the ERBL region are indicated by vertical dashed
lines.}
    \label{fig:E_conv}
\end{figure}
We have analytically verified that, as $P^3 \to \infty$, these formulas reduce to the analytic expressions used for the standard axial-diquark $E$ GPD. For more information, see \cite{Bhattacharya:2019cme}. In \Cref{fig:E_conv}, we present numerical convergence of QGPDs to the light-cone GPD for three different values of skewness: $\xi=0.001$, $\xi=0.1$, and $\xi=0.4$. To verify the convergence, we have not used our calibrated parameters, but the values used in \cite{Bhattacharya:2019cme}, so in this case we omit the quasi-hard cutoff and switch to a hard cutoff. 
Thus, we use the following values:
\begin{equation*}
    g=1, \quad M=0.939\,[\text{GeV}], \quad m_a = 1\,[\text{GeV}], \quad m_q=0.35\,[\text{GeV}], \quad \Lambda = 1\, [\text{GeV}],
\end{equation*}
to facilitate comparisons with earlier results for the convergence of QGPDs to GPDs.

\section{Additional tomography results}
\label{app:additional}
In this appendix we present additional tomography results for $x = 0.5$,
$x = 0.7$, and $x = 0.9$, complementing the discussion in \Cref{Sec:Tomogrpahy}. For
each value of $x$ we show the cross-section at $b_x = 0$ and the
corresponding two-dimensional heatmap in the $(b_x, b_y)$ plane, for both
unpolarized and transversely polarized nucleons, and for both the
proton-only and the proton+neutron calibrations.
\begin{figure}[h]
    \centering
    \begin{subfigure}[b]{\linewidth}
        \centering
        \includegraphics[width=\linewidth]{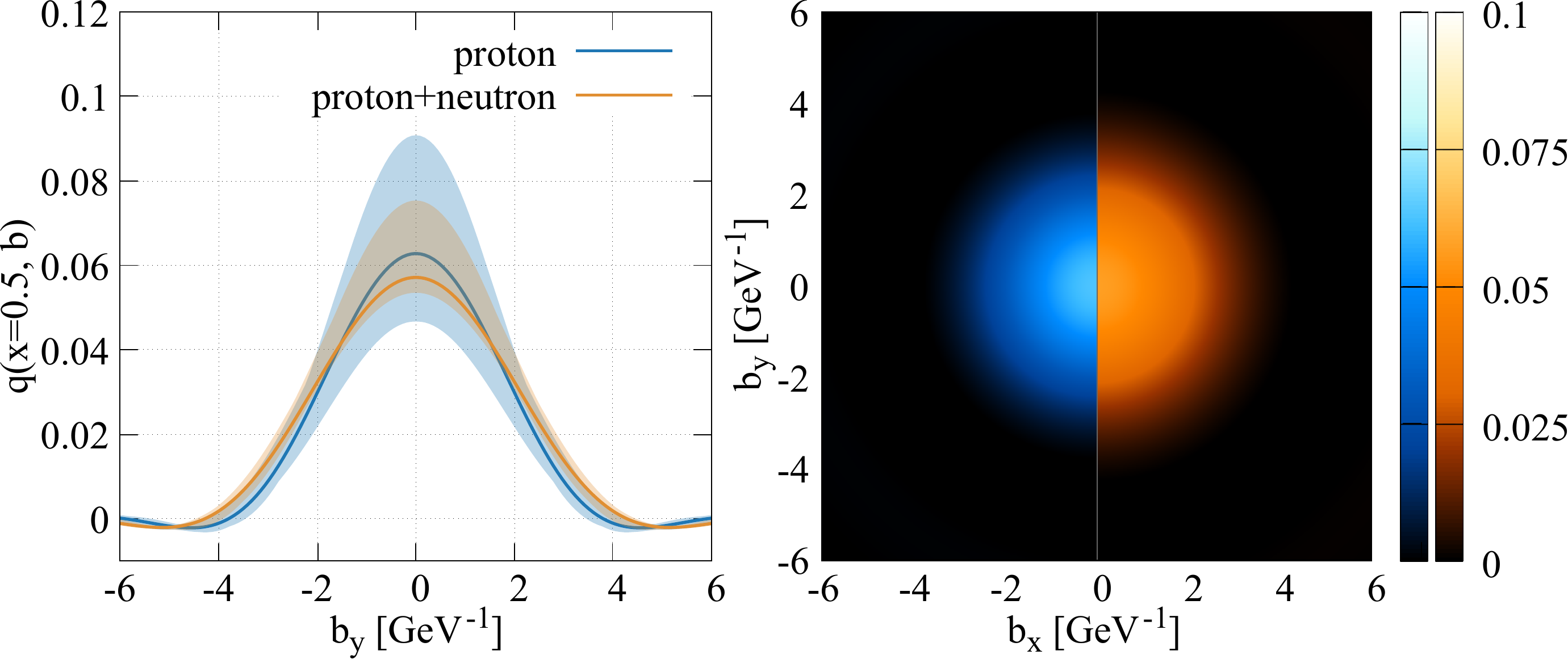}
        \caption{$x=0.5$, unpolarized up quark in proton.}
    \end{subfigure}
    \ContinuedFloat
    \centering
    \begin{subfigure}[b]{\linewidth}
        \centering
        \includegraphics[width=\linewidth]{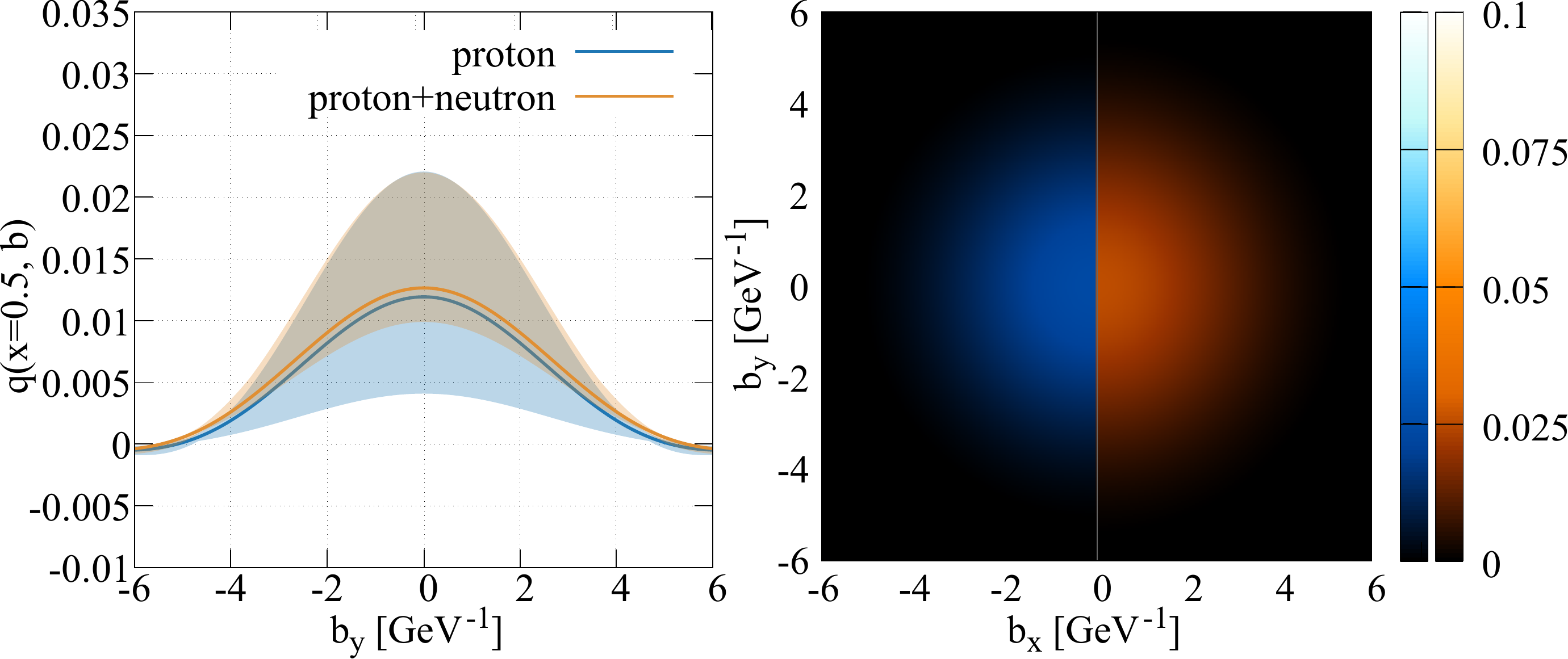}
        \caption{$x=0.5$, unpolarized down quark in proton (heatmap is multiplied by two).}
    \end{subfigure}
\end{figure}
\begin{figure}[p]
    \ContinuedFloat
    \centering
    \begin{subfigure}[b]{\linewidth}
        \centering
        \includegraphics[width=\linewidth]{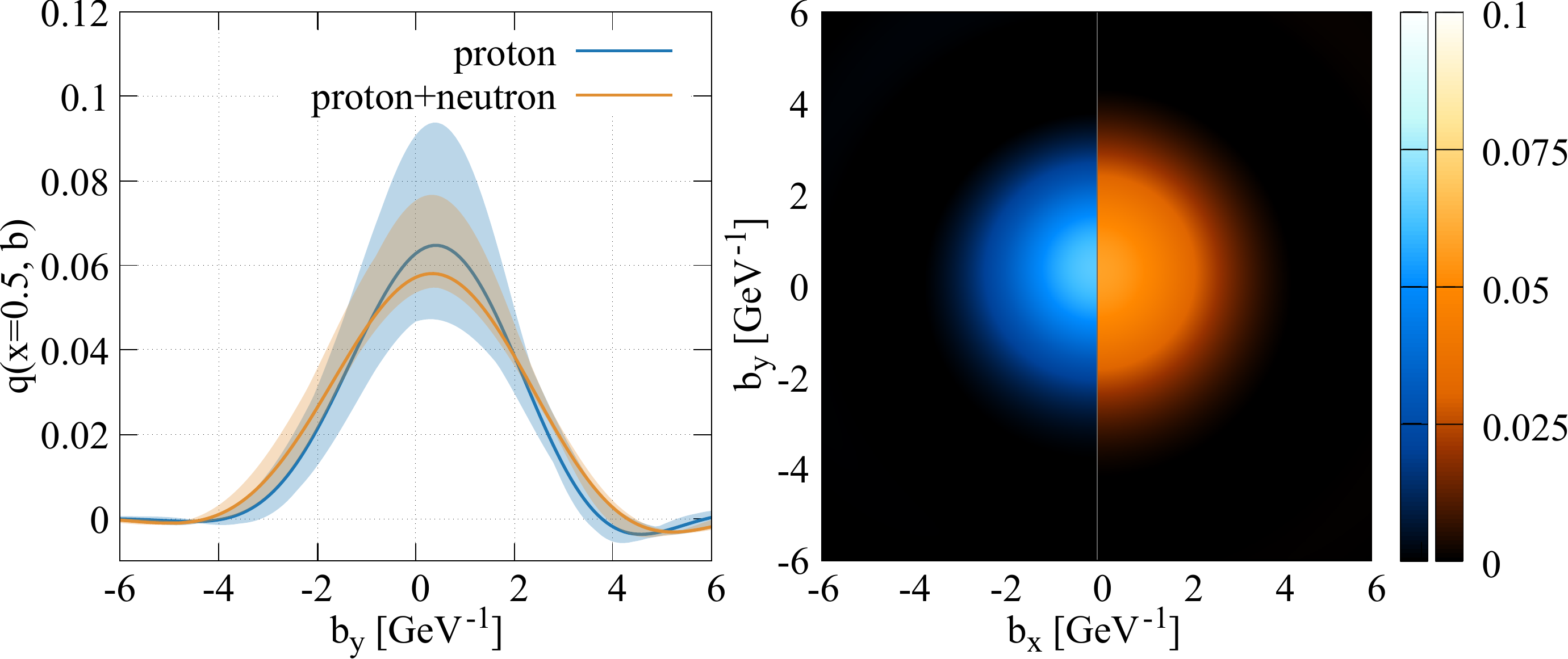}
        \caption{$x=0.5$, transversely polarized up quark in a proton.}
    \end{subfigure}
\end{figure}
\begin{figure}[p]
    \ContinuedFloat
    \centering
    \begin{subfigure}[b]{\linewidth}
        \centering
        \includegraphics[width=\linewidth]{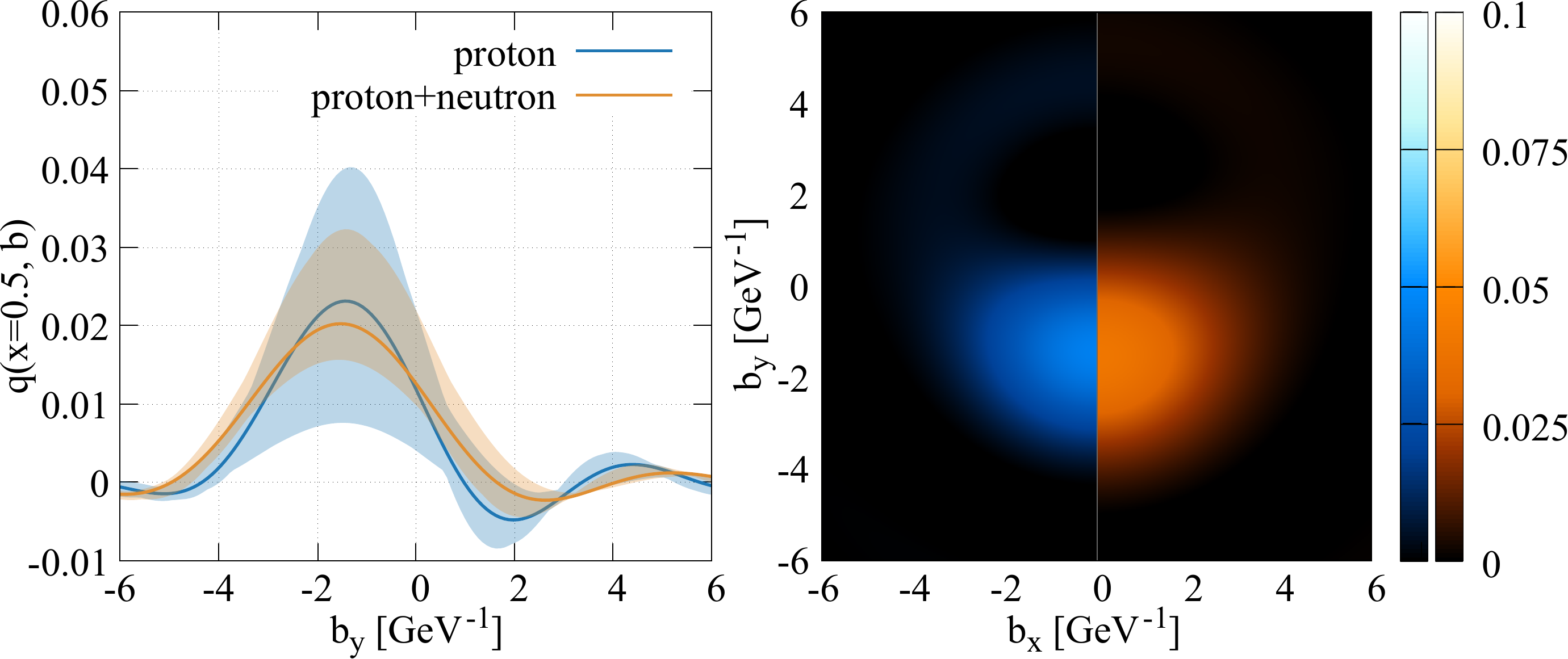}
        \caption{$x=0.5$, transversely polarized down quark in proton (heatmap is multiplied by two).}
    \end{subfigure}
        \caption{Nucleon tomography for unpolarized and transversely polarized proton for $x=0.5$. The left column shows the cross-section for $b_x=0$, of the heat map shown in the right column. We present both the result for the fit including only protons (blue) and both nucleons (orange). We show the central result on the heatmap.}
\end{figure}
\begin{figure}[p]
    \centering
    \begin{subfigure}[b]{\linewidth}
        \centering
        \includegraphics[width=\linewidth]{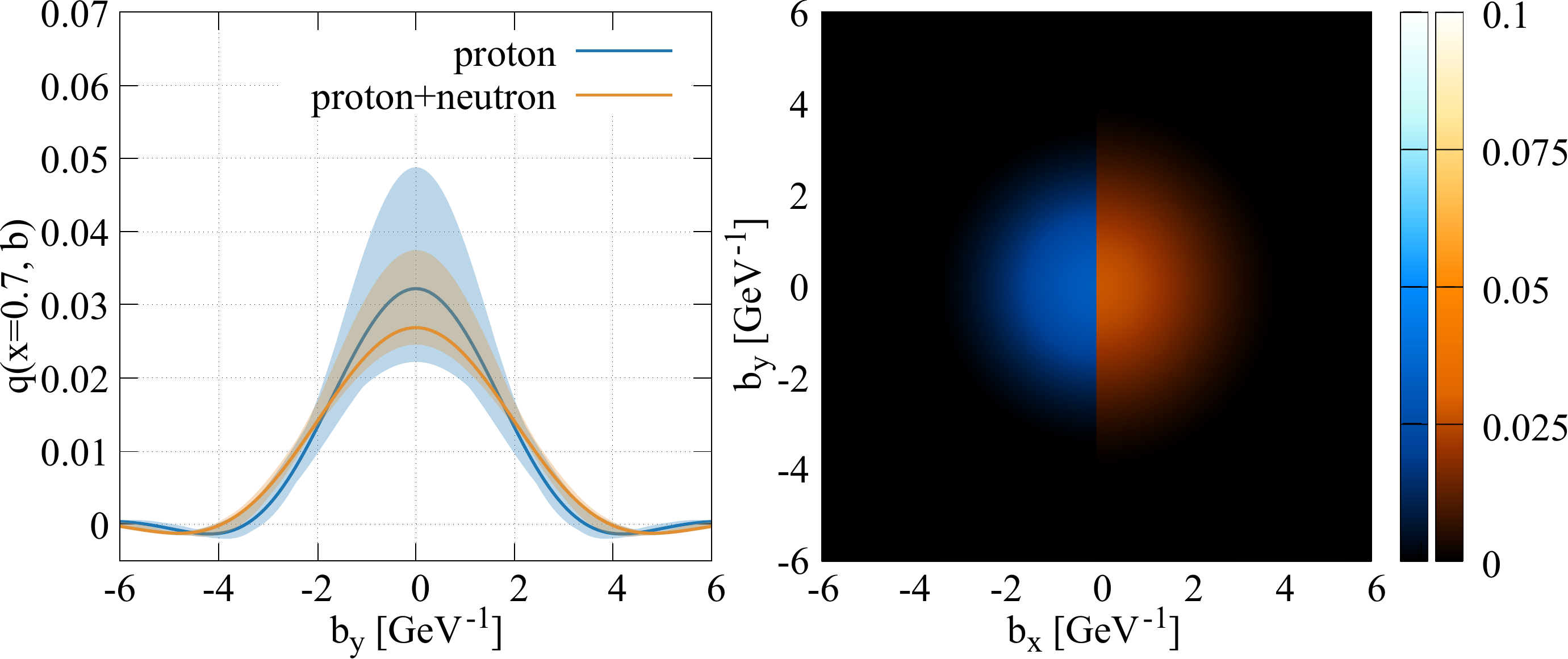}
        \caption{$x=0.7$, unpolarized up quark in proton.}
    \end{subfigure}
\end{figure}
\begin{figure}[p]
    \ContinuedFloat
    \centering
    \begin{subfigure}[b]{\linewidth}
        \centering
        \includegraphics[width=\linewidth]{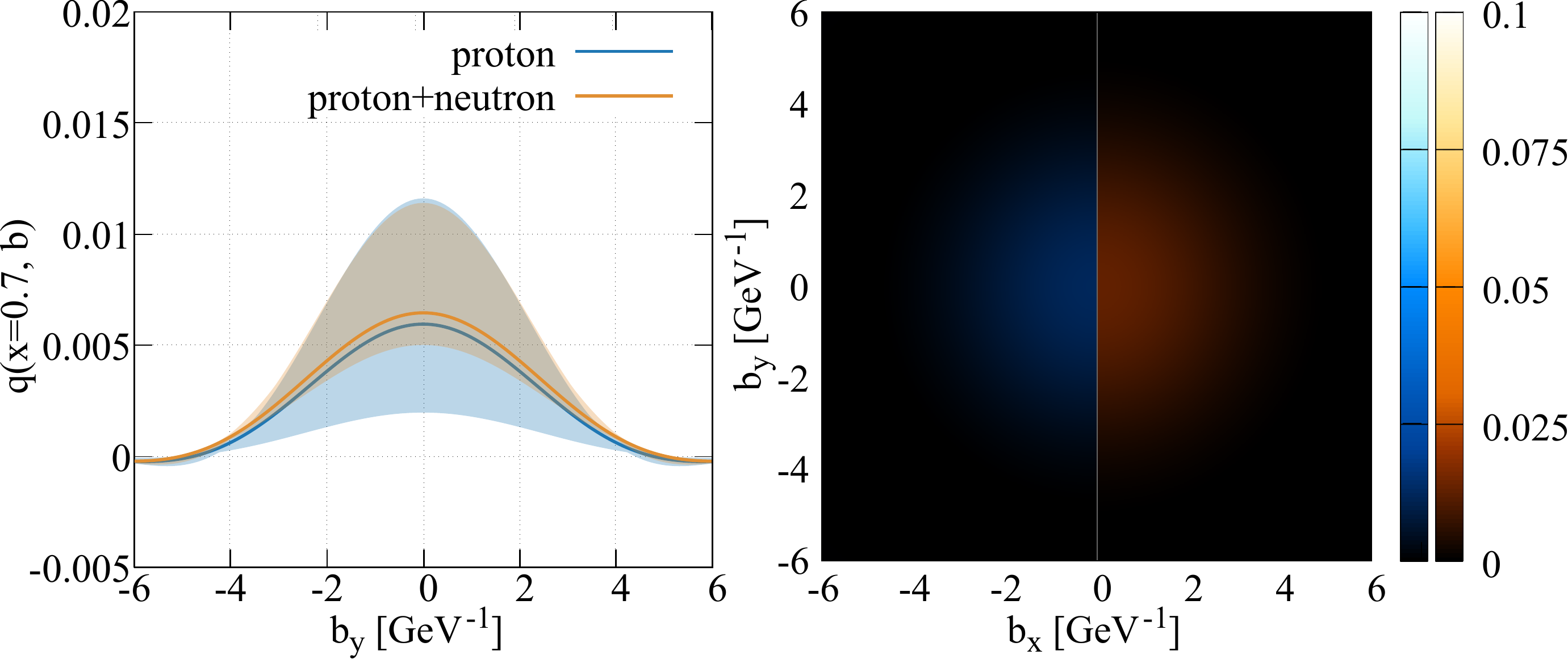}
        \caption{$x=0.7$, unpolarized down quark in proton (heatmap is multiplied by two).}
    \end{subfigure}
\end{figure}
\begin{figure}[p]
    \ContinuedFloat
    \centering
    \begin{subfigure}[b]{\linewidth}
        \centering
        \includegraphics[width=\linewidth]{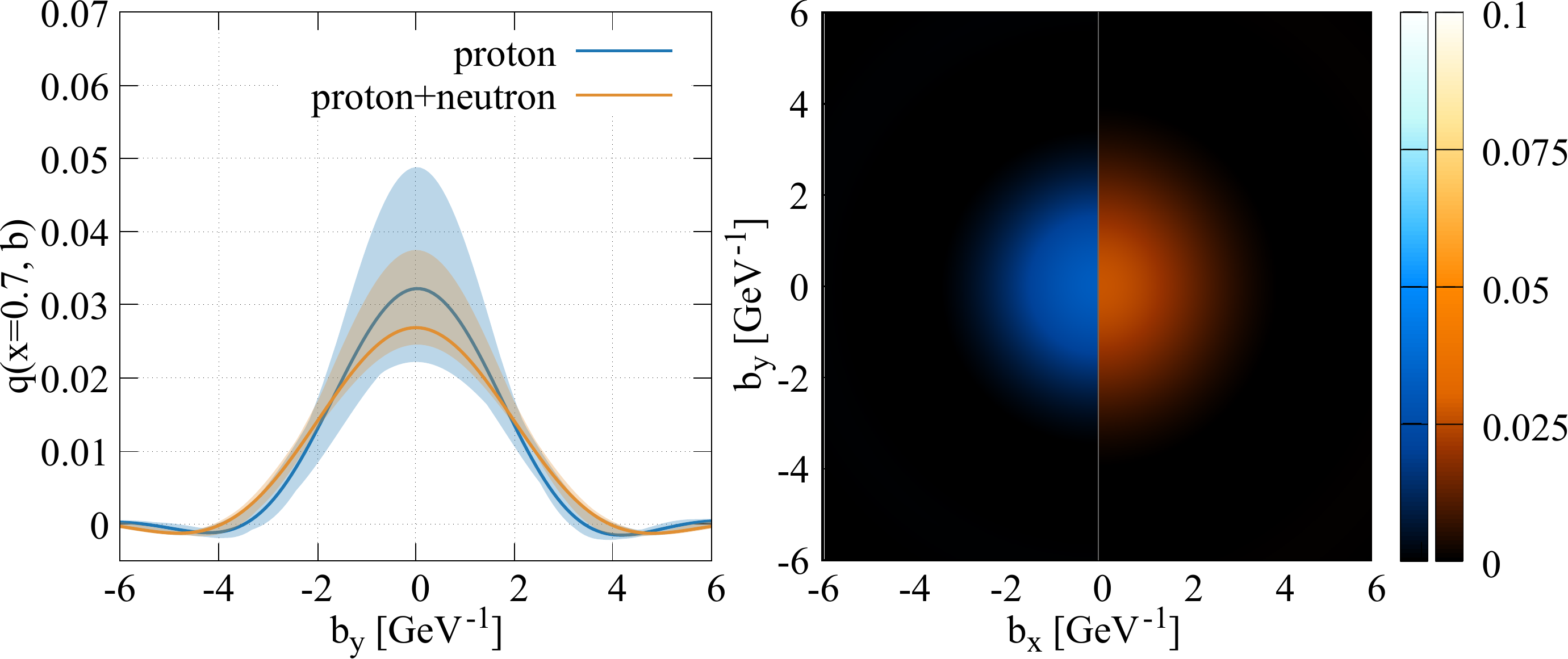}
        \caption{$x=0.7$, transversely polarized up quark in a proton.}
    \end{subfigure}
\end{figure}
\begin{figure}[p]
    \ContinuedFloat
    \centering
    \begin{subfigure}[b]{\linewidth}
        \centering
        \includegraphics[width=\linewidth]{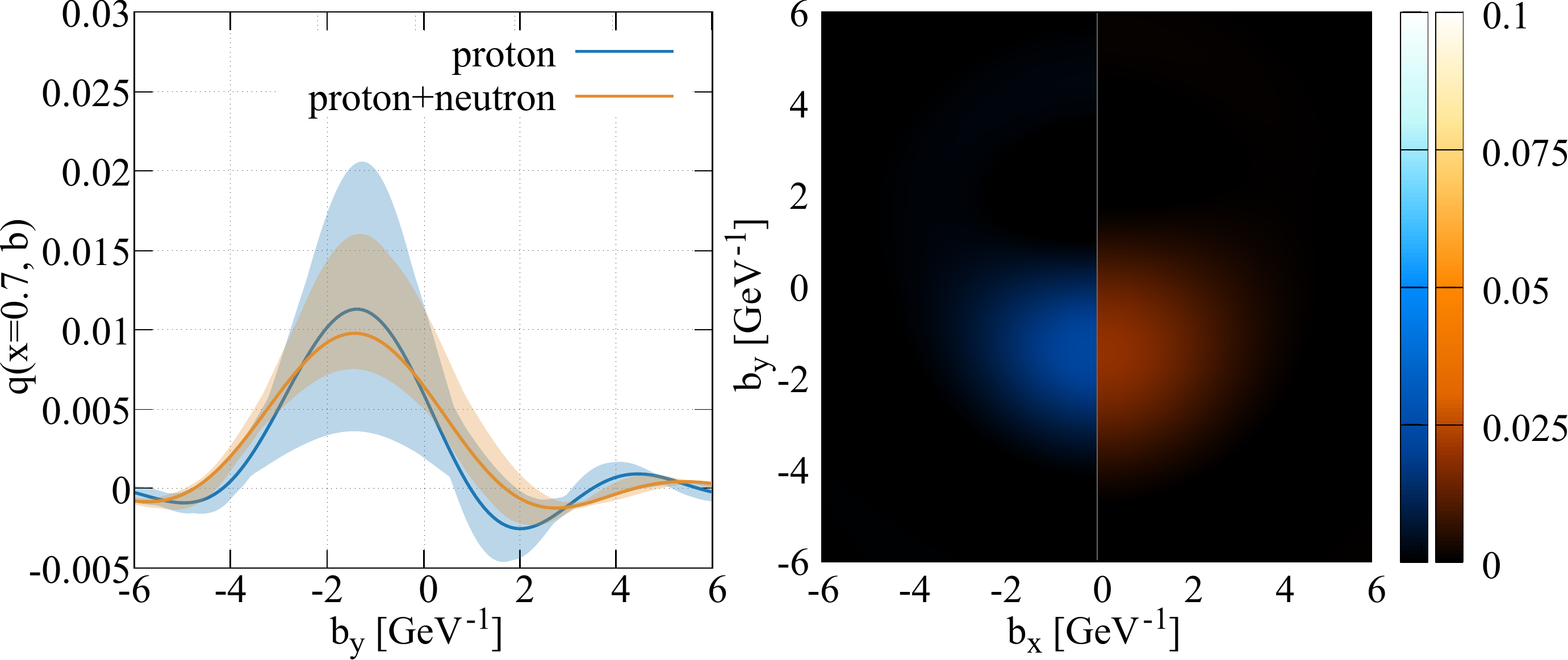}
        \caption{$x=0.7$, transversely polarized down quark in proton (heatmap is multiplied by two).}
    \end{subfigure}
        \caption{Nucleon tomography for unpolarized and transversely polarized proton for $x=0.7$. The left column shows the cross-section for $b_x=0$, of the heat map shown in the right column. We present both the result for the fit including only the proton (blue) and both nucleons (orange). We show the central result on the heatmap.}
\end{figure}
\begin{figure}[p]
    \centering
    \begin{subfigure}[b]{\linewidth}
        \centering
        \includegraphics[width=\linewidth]{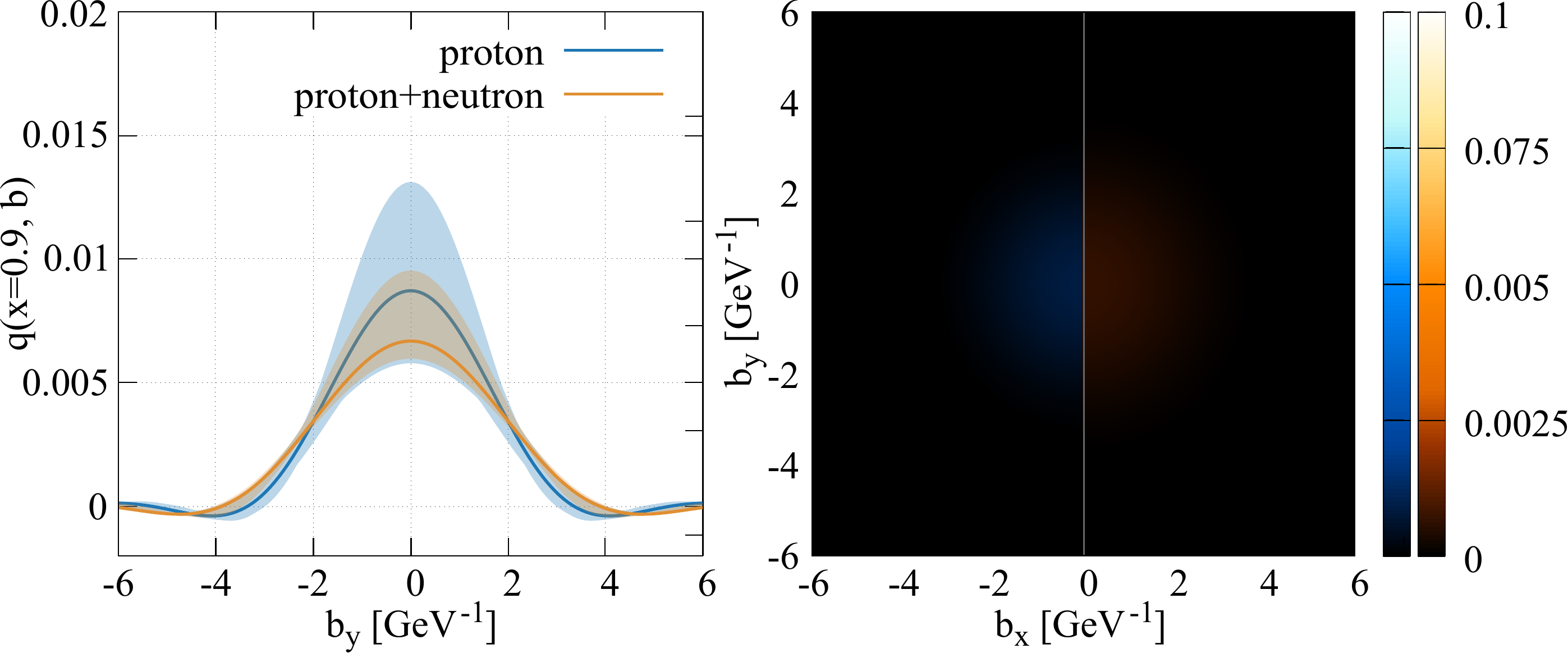}
        \caption{$x=0.9$, unpolarized up quark in proton.}
    \end{subfigure}
\end{figure}
\begin{figure}[p]
    \ContinuedFloat
    \centering
    \begin{subfigure}[b]{\linewidth}
        \centering
        \includegraphics[width=\linewidth]{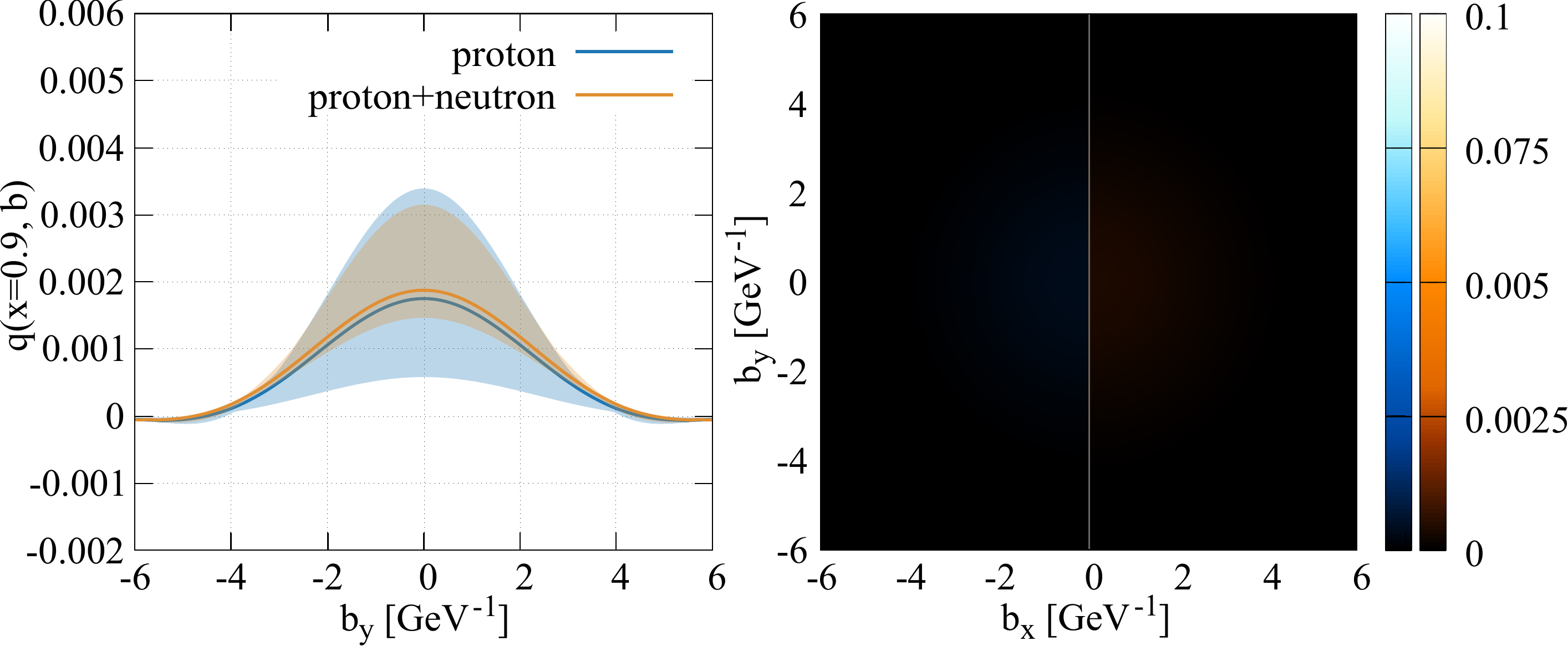}
        \caption{$x=0.9$, unpolarized down quark in proton (heatmap is multiplied by two).}
    \end{subfigure}
    \ContinuedFloat
    \centering
    \begin{subfigure}[b]{\linewidth}
        \centering
        \includegraphics[width=\linewidth]{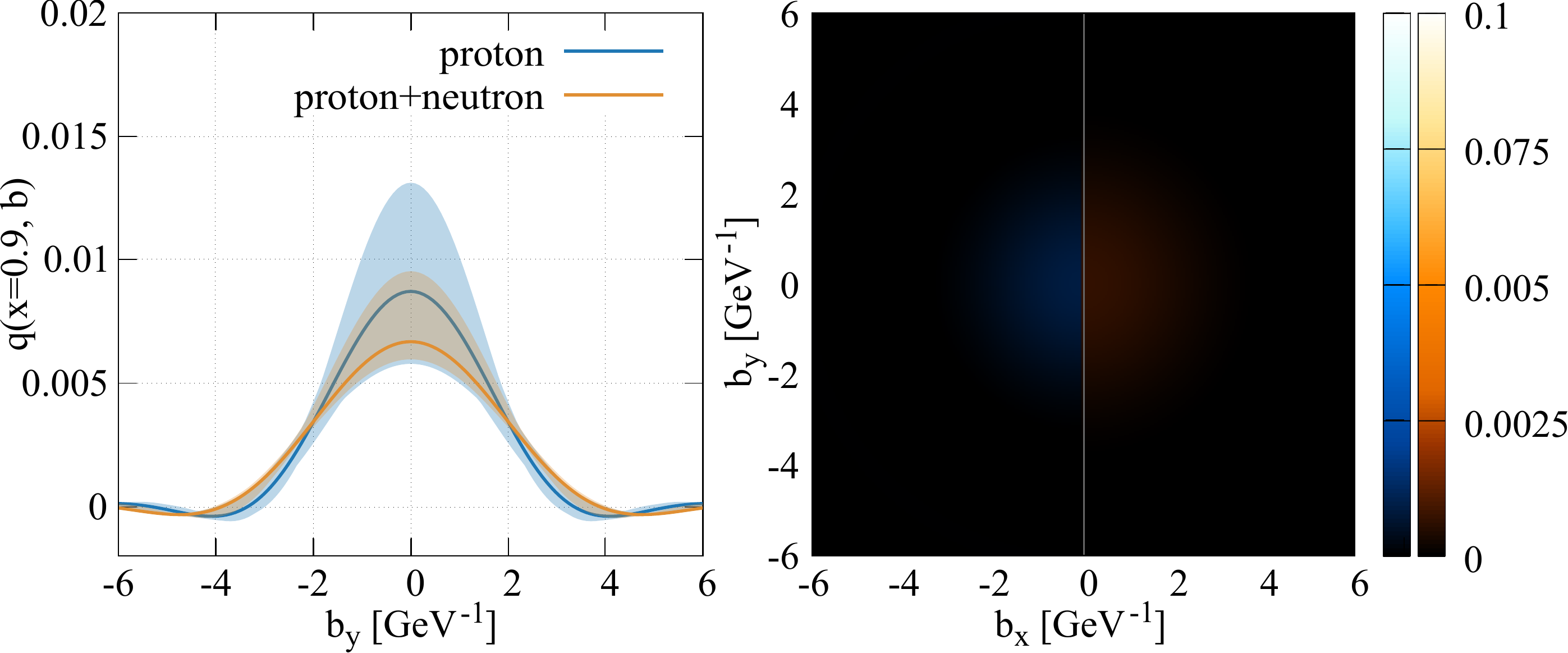}
        \caption{$x=0.9$, transversely polarized up quark in proton.}
    \end{subfigure}
    \ContinuedFloat
    \centering
    \begin{subfigure}[b]{\linewidth}
        \centering
        \includegraphics[width=\linewidth]{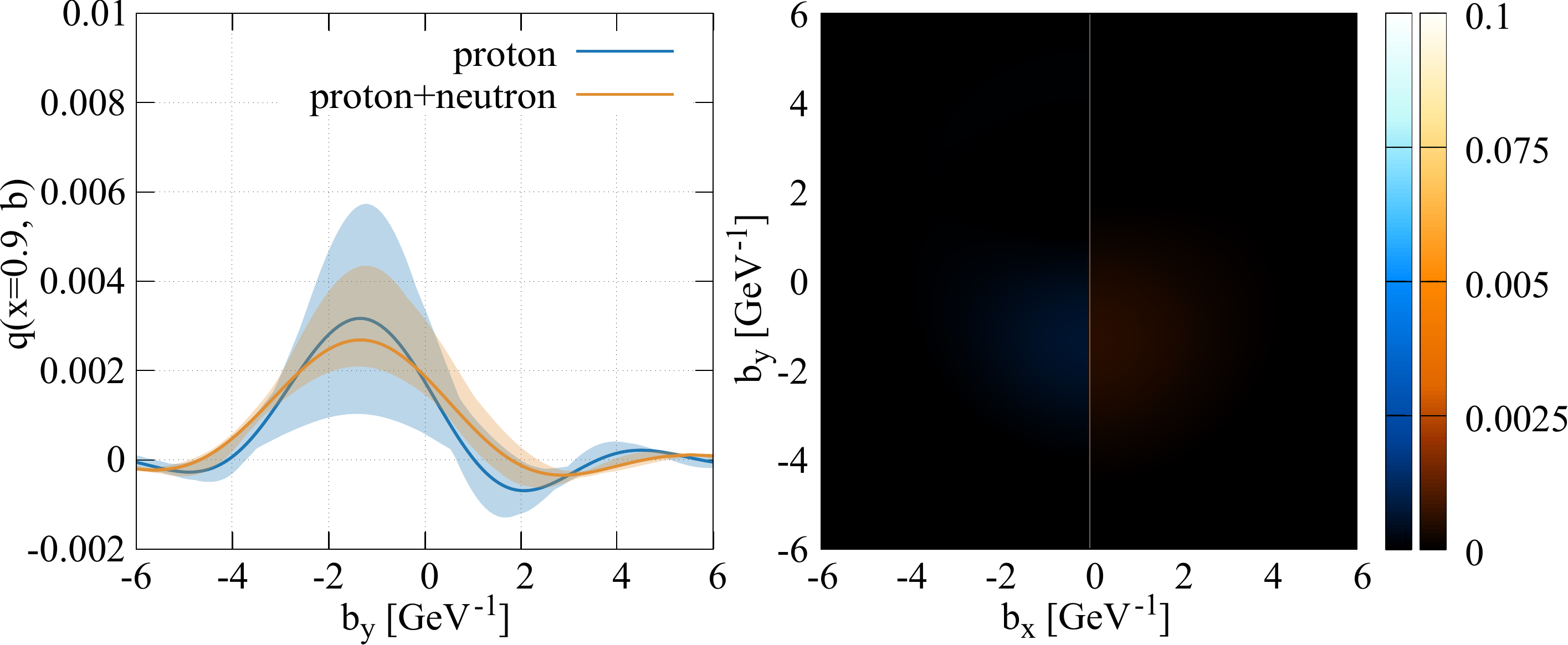}
        \caption{$x=0.9$, transversely polarized down quark in proton (heatmap is multiplied by two).}
    \end{subfigure}
        \caption{Nucleon tomography for unpolarized and transversely polarized proton for $x=0.9$. The left column shows the cross-section for $b_x=0$, of the heatmap shown in the right column. We present both the result for the fit including only protons (blue) and both nucleons (orange). We show the central result on the heatmap.}
\end{figure}

\end{document}